\documentclass[
12pt,
amsmath,amsfonts,
notitlepage,
nofootinbib,
longbibliography
]
{revtex4-1}
\usepackage[colorlinks,allcolors=blue,linktocpage]{hyperref}

\pdfoutput=1
\usepackage{graphicx}
\usepackage[dvipsnames]{xcolor}
\usepackage[utf8]{inputenc}

\usepackage{bm}
\usepackage{siunitx}
\usepackage{slashed}

\usepackage{feynmp-auto}

\DeclareMathOperator{\Tr}{Tr}

\renewcommand{\Re}{\operatorname{Re}}
\renewcommand{\Im}{\operatorname{Im}}

\begin{document}
\title{Reconciling resonant leptogenesis and baryogenesis via neutrino oscillations}
\author{Juraj Klaric}\email{juraj.klaric@epfl.ch}
\author{Mikhail Shaposhnikov}\email{mikhail.shaposhnikov@epfl.ch}
\author{Inar Timiryasov}\email{inar.timiryasov@epfl.ch}
\affiliation{Institute of Physics, Laboratory for Particle Physics and Cosmology,
	\'{E}cole Polytechnique F\'{e}d\'{e}rale de Lausanne, CH-1015 Lausanne,
Switzerland}

\begin{abstract}
	Right-handed neutrinos offer an elegant solution to two well established phenomena beyond the Standard Model (SM)---masses and oscillations of neutrinos, as well as the baryon asymmetry of the Universe.
	It is also a minimalistic solution since it requires only singlet Majorana fermions to be added to the SM particle content.
	If these fermions are nearly degenerate, the mass scale of right-handed neutrinos can be very low and accessible by the present and planned experiments.
	There are at least two well studied mechanisms of the low-scale leptogenesis: baryogenesis via oscillations and resonant leptogenesis.
	These two mechanisms were often considered separate, but they can in fact be understood as two different regimes of one and the same mechanism, described by a unique set of quantum kinetic equations.
	In this work we show, using a unified description based on quantum kinetic equations,
	that the parameter space of these two regimes of low-scale leptogenesis significantly overlap.
	We present a comprehensive study of the parameter space of the low-scale leptogenesis with the mass scale ranging from $0.1$~GeV to ${\sim 10^6}$~GeV. The unified perspective of this work reveals the synergy between intensity and energy frontiers in the quest for heavy Majorana neutrinos.
\end{abstract}

\maketitle
\tableofcontents

\begin{fmffile}{direct_rates}
\setlength{\unitlength}{0.5mm}
\fmfset{arrow_len}{3mm}

\section{Introduction}
\label{sec:introduction}
The origin of the light neutrino masses and the baryon asymmetry of the Universe remain some of the burning problems for physics beyond the Standard Model (SM).
Adding heavy neutrinos to the SM offers an economical solution to both problems (throughout this work we will use the terms heavy neutrinos and HNLs---Heavy Neutral Leptons interchangeably).
The masses of active neutrinos are coming from the type-I seesaw mechanism~\cite{Minkowski:1977sc,GellMann:1980vs,Mohapatra:1979ia,Yanagida:1980xy,Schechter:1980gr,Schechter:1981cv},
the baryon asymmetry of the Universe (BAU) is generated through
combined action of anomalous processes with fermion number non-conservation \cite{Kuzmin:1985mm} and lepton number and flavor violating reactions involving heavy neutrinos.
The scale of the heavy neutrino masses however remains an open question.
The seesaw mechanism on its own does not imply any specific scale for the heavy neutrino masses.
One may wonder if the answer can come from leptogenesis. And indeed, the specific
patterns of Majorana masses of HNLs single out different scales.
For example, if the spectrum of HNLs is hierarchical, $M_I \ll M_J$, a lower bound of $10^9~\si{\GeV}$ was found for the mass of the lightest heavy neutrino~\cite{Davidson:2002qv}. The mechanism leading to BAU in this case is known as \emph{thermal leptogenesis}. If the spectrum is nearly degenerate, i.e.\ there is a pair of HNLs such that $\Delta M \ll M$, the leptogenesis may take place for
$M$ as small as $\mathcal{O}(100)$~GeV or even few MeV~\cite{Canetti:2010aw}.
The corresponding leptogenesis mechanisms are called \emph{resonant} leptogenesis~\cite{Liu:1993tg,Flanz:1994yx,Flanz:1996fb,Covi:1996wh,Covi:1996fm,Pilaftsis:1997jf,Pilaftsis:1997dr,Pilaftsis:1998pd,Buchmuller:1997yu,Pilaftsis:2003gt}, and baryogenesis \emph{via oscillations}~\cite{Akhmedov:1998qx,Asaka:2005pn}.

In the recent years these last scenarios have received significant attention, from both the theoretical (see e.g.~\cite{Shaposhnikov:2006nn,Shaposhnikov:2008pf,Canetti:2010aw,Asaka:2010kk,Anisimov:2010gy,Asaka:2011wq,Besak:2012qm,Canetti:2012vf,Drewes:2012ma,Canetti:2012kh,Shuve:2014zua,Bodeker:2014hqa,Abada:2015rta,Hernandez:2015wna,Ghiglieri:2016xye,Hambye:2016sby,Hambye:2017elz,Drewes:2016lqo,Asaka:2016zib,Drewes:2016gmt,Hernandez:2016kel,Drewes:2016jae,Asaka:2017rdj,Eijima:2017anv,Ghiglieri:2017gjz,Eijima:2017cxr,Antusch:2017pkq,Ghiglieri:2017csp,Eijima:2018qke,Ghiglieri:2018wbs,Ghiglieri:2019kbw,Bodeker:2019rvr,Ghiglieri:2020ulj,Klaric:2020lov,Domcke:2020ety,Eijima:2020shs,DeSimone:2007edo,DeSimone:2007gkc,Garny:2009qn,Garny:2011hg,Iso:2013lba,Garbrecht:2011aw,Dev:2014laa,Dev:2014tpa,Garbrecht:2014aga,Dev:2015wpa,Hambye:2016sby,Jiang:2020kbt}) and experimental (see, e.g.~\cite{Liventsev:2013zz,Aaij:2014aba,Artamonov:2014urb,Aad:2015xaa,Khachatryan:2015gha,Antusch:2017hhu,CortinaGil:2017mqf,Izmaylov:2017lkv,Mermod:2017ceo,Drewes:2018gkc,Ballett:2019bgd,Sirunyan:2018mtv,SHiP:2018xqw,Boiarska:2019jcw,Bolton:2019pcu,NA62:2020mcv,Tastet:2020tzh,Bondarenko:2021cpc,CortinaGil:2021gga,Sirunyan:2018mtv,Boiarska:2019jcw,Aad:2019kiz,Wulz:2019lsz,NA62:2020mcv,Drewes:2018gkc,Alekhin:2015byh,SHiP:2018xqw,Curtin:2018mvb,Gligorov:2017nwh,Feng:2017uoz,Kling:2018wct,Hirsch:2020klk}) perspectives.
This interest is primary related to the potential testability of the model in the
present and near future experiments. HNLs are a canonical example of feebly interacting particles (see, e.g.~\cite{Beacham:2019nyx,Lanfranchi:2020crw}), which could have avoided discovery not because they are heavy but because their interactions are very weak.

For the observed BAU to be generated, the three  Sakharov conditions need to be met~\cite{Sakharov:1967dj}:
\begin{enumerate}
	\item[1)] Efficient baryon number violation.
\item[2)] Sizeable C and CP violation.
	\item[3)] Substantial deviation from thermal equilibrium.
\end{enumerate}
Heavy Majorana neutrinos can provide both CP violation via the complex Yukawa couplings, as well as a deviation from equilibrium.
In leptogenesis scenarios the asymmetry is generated in the lepton sector (hence the name) and transferred to the baryon sector by non-perturbaitve sphaleron processes which violate baryon number conservation~\cite{Kuzmin:1985mm}.
In the thermal leptogenesis the deviation from equilibrium takes place during the decays of HNLs, when the decay rate cannot catch up the expansion rate of the Universe.
The asymmetry between decays into leptons and anti-leptons comes about because of the interference between tree and one loop processes like $N \to \ell H$. See, e.g. \cite{Buchmuller:2004nz} for details. This \emph{decay asymmetry} is further enhanced if the two HNLs are degenerate in mass, as it is the case in the resonant leptogenesis.\footnote{In fact, the resonant enhancement has been already noted in~\cite{Kuzmin:1985mm}. } We will discuss this mechanism in greater detail in section~\ref{sub:resonant_leptogenesis}.
In the baryogenesis via neutrino oscillations, the deviation from equilibrium happens during freeze-in. As a result of the seesaw mechanism, the Yukawa couplings of light HNLs must be tiny. Thus the equilibration rate is much lower than the Hubble rate, so the HNLs remain out of equilibrium. The asymmetry generated in the processes such as scatterings, decays, and inverse decays of the HNLs is further enhanced by their oscillations.

Several studies of resonant leptogenesis suggested that the scale of heavy neutrinos can be as low as $\sim 100$~GeV.
At the same time, leptogenesis via oscillations was primarily studied in the few-\si{\GeV} regime.
In Ref.~\cite{Blondel:2014bra} it was suggested that the maximal HNL mass in this mechanism is around the $W$-boson mass, where decays into $W$ and $Z$ bosons become kinematically allowed, and enhance the equilibration of the lepton asymmetries.
The argument came about as follows. If the HNLs are kinematically allowed to decay into W (or W and Z), the  rate of this process will exceed the Hubble rate well before the moment of sphaleron freeze-out.
This means that the HNLs will be in thermal equilibrium and all asymmetries will be washed out.
However, this is not the end of the story.
For $T \lesssim M$, the heavy neutrinos begin to freeze-out, and their abundance, as well as the washout of the lepton asymmetries become Boltzmann suppressed.

Due to the different approximations that were applied to these mechanisms,
the overlap between them remained an open question.

These mechanisms may appear quite different at first glance---but, as we show in this work, it turns out that the same equations can be used to describe both mechanisms.

\begin{figure}[h]
	\centering
	\includegraphics[width=0.45\textwidth]{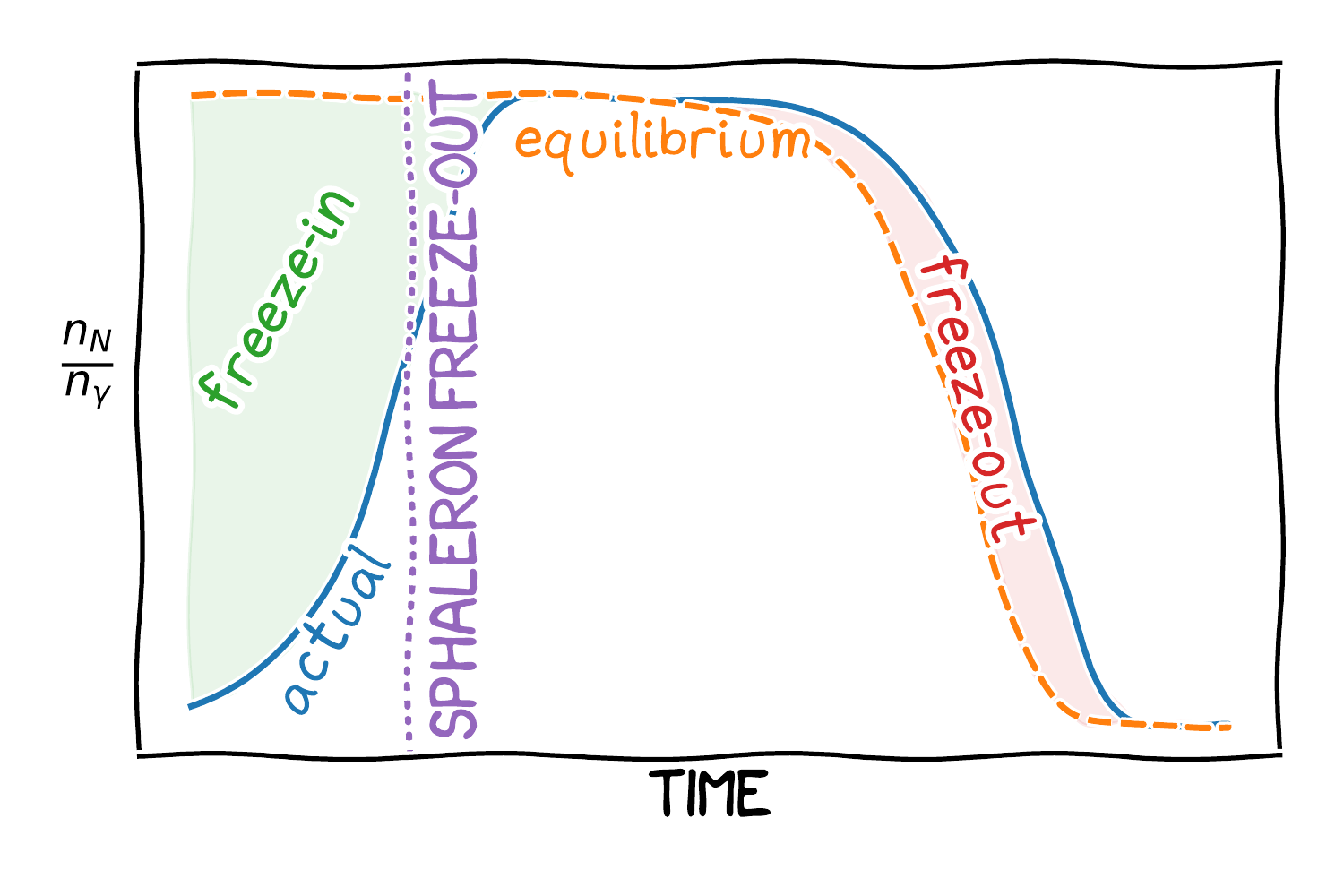}\quad \includegraphics[width=0.45\textwidth]{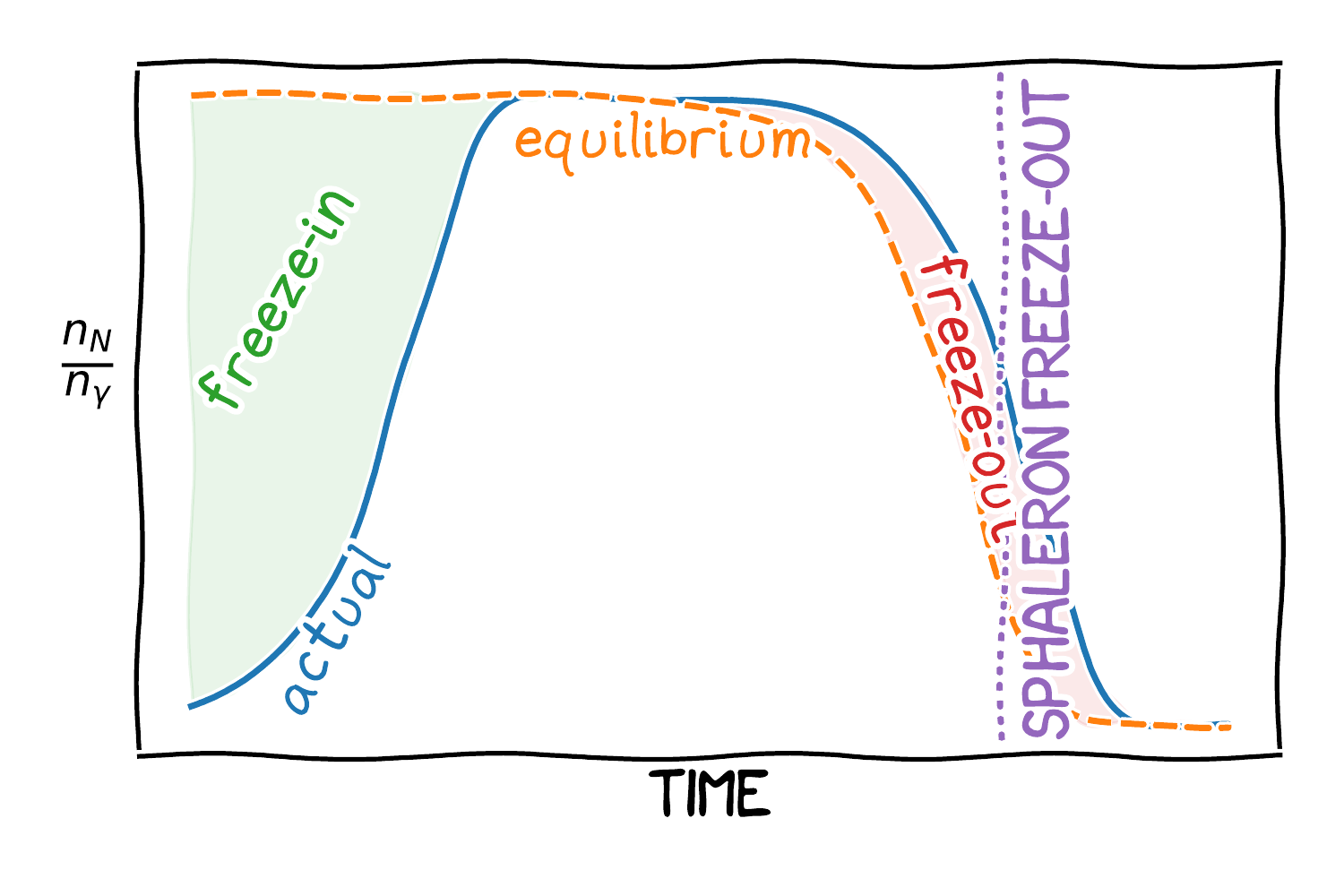}
	\caption{A sketch of the evolution of the HNL abundance in the early Universe. If we assume that the initial HNL abundance vanishes, there are two opportunities to generate the observed BAU.
	The first is during a period of freeze-in, while the first HNLs are being produced and they approach equilibrium. The second opportunity is when the Universe cools down to temperatures below the HNL mass, and the HNLs decay out-of-equilibrium simultaneously with a freeze-out of the SM lepton number caused by the Boltzmann-suppressed washout rates.}
	\label{fig:fifo}
\end{figure}

In this work we systematically study generation of the BAU in the model with two HNLs with masses in the range $10-10^4$~GeV.
We find that the parameter space of the \emph{baryogenesis via neutrino oscillations} is seamlessly connected with the parameter space of \textit{resonant leptogenesis}.
So there is just one mechanism, with different regimes. These regimes are characterized by whether the majority of the asymmetry is produced during the freeze-in or freeze-out of the HNLs (c.f. Fig.~\ref{fig:fifo}).
We identify three main reasons why leptogenesis remains viable for $M\sim M_{W,Z}$.
\begin{itemize}
	\item Although the equilibration rate of the heavy neutrinos generically exceeds the Hubble rate at the temperature of the sphaleron freeze-out, the lepton flavor washout rate can be suppressed, thereby preserving the BAU in one of the leptonic flavors. As was pointed out in~\cite{Garbrecht:2014bfa},
		this allows for freeze-in leptogenesis with HNL masses above the electroweak scale.\footnote{A more detailed study of the effect of freeze-in on hierarchical leptogenesis followed in~\cite{Garbrecht:2019zaa}, which confirmed the importance of freeze-in in the conventional leptogenesis scenario.}
	\item Due to the finite mass of the heavy neutrinos, their equilibrium distribution changes with temperature. This effect acts as a source for a deviation from equilibrium. We find that this effect remains important even for heavy neutrinos with masses at the GeV scale. This confirms the possibility of GeV-scale freeze-out leptogenesis as  suggested in~\cite{Hambye:2016sby}.
	\item We find that another factor preventing washout of the lepton asymmetries is approximate generalized lepton number conservation.
		\footnote{In principle there are several ways of assigning lepton number to the heavy neutrinos. We discuss them in section~\ref{sss:cons_chg}.}
		The processes that would erase the lepton asymmetry are suppressed by a factor $\sim \Delta M/\Gamma$, as identified in~\cite{Blanchet:2009kk,Deppisch:2010fr}.
\end{itemize}
The short account of the results was presented in \cite{Klaric:2020lov}.
The paper is organized as follows.
We start from an overview of the evolution of the leptogenesis calculations in section~\ref{sec:convergence_towards_the_unified_picture}.
Then we define our notations and introduce the seesaw mechanism in section~\ref{sec:seesaw}.
In section~\ref{sec:bau_bounds} we describe resonant leptogenesis and baryogenesis via neutrino oscillations.
We present the quantum kinetic equations for matrices of densities which at the core of our numerical study. We also show how the usual Boltzmann equations can be obtained as a limit of the quantum kinetic equations.
Section~\ref{sec:the_rates} is dedicated to the determination of the heavy neutrino production rates entering the kinetic equations. With all ingredients at hand, we perform the study of the parameter space in section~\ref{sec:production_of_the_baryon_asymmetry}.
This section also contains detailed discussion of the obtained results. Next, in section~\ref{sec:comparison_with_other_works} we discuss several other works which considered the HNLs with masses around $M_W$. We conclude in~\ref{sec:discussion_and_conclusions}. Technical details of the implementation, relation to the pseudo-Dirac basis, conserved lepton numbers, and fine tuning are discussed in the appendices.

\section{Convergence towards a unified picture}
\label{sec:convergence_towards_the_unified_picture}

In this section we present a brief overview of the development of the calculations and methodology in low-scale leptogenesis.

The importance of a resonant enhancement for leptogenesis~\cite{Liu:1993tg,Flanz:1996fb,Flanz:1994yx,Covi:1996wh,Covi:1996fm,Pilaftsis:1997jf,Buchmuller:1997yu} was realized soon after leptogenesis was proposed as a baryogenesis mechanism.\footnote{It is worth noting that resonant enhancement in baryogenesis predates idea of leptogenesis itself~\cite{Kuzmin:1970nx,Kuzmin:1985mm}.}
Such a resonantly enhanced decay asymmetry offered an exciting opportunity---the
mass scale of the HNLs could in principle be lowered to the electroweak scale~\cite{Pilaftsis:1997jf,Pilaftsis:2003gt,Pilaftsis:2005rv}
(it is worth noting that both the effects from a non-instantaneous freeze-out of sphalerons and spectator effects~\cite{Barbieri:1999ma,Buchmuller:2001sr,Davidson:2008bu}
were included in~\cite{Pilaftsis:2005rv}).

However, the fact that the finite-order perturbation theory breaks down in the limit of degenerate HNL masses was already clear in the earliest papers~\cite{Covi:1996fm,Pilaftsis:1997jf,Pilaftsis:1997dr},
and different approaches of resolving these issues followed soon thereafter~\cite{Roulet:1997xa,Covi:1997dr,Buchmuller:1997yu},
with the general conclusion that the decay asymmetry is regulated by the width of the HNLs.
This prompted further investigations into resonant leptogenesis using techniques of non-equilibrium Quantum Field Theory (see e.g.~\cite{Schwinger:1960qe,Keldysh:1964ud,Baym:1961zz,Danielewicz:1982kk,Niemi:1983nf,Landsman:1986uw,Calzetta:1986cq,Knoll:2001jx,Blaizot:2001nr,Calzetta:2008iqa,Berges:2015kfa}).
One of the most successful formalisms for in this framework is the CTP (closed-time-path) formalism, also known as the Schwinger-Keldysh formalism~\cite{Schwinger:1960qe,Keldysh:1964ud}.
In this formulation of non-equilibrium QFT one typically deals with $n$-point functions of the different particles and uses them to calculate observables, such as particle numbers and distributions,
and their time-evolution can be obtained by solving the Schwinger-Dyson equations on the CTP.
One of the biggest differences compared to equilibrium QFT is that time translation invariance is explicitly broken (both by the boundary condition and by the expansion of the Universe),
and each two-point function is a function of two time coordinates (and one momentum coordinate assuming translation invariance).
Consequentially, the Schwinger-Dyson equations are integro-differential equations which include both derivatives and integration over the time variables.

The main difference between the existing approaches is in the strategies used to solve these equations.
Perhaps the most straightforward way is to solve the Schwinger-Dyson equations directly.
However, this is only possible numerically or in specific limits.
Nonetheless, this can be used to cross-check the results of other methods~\cite{Garny:2011hg,Iso:2013lba}.
Another approach is to perform a Wigner transformation of the equations (it is often used to describe transport phenomena~\cite{Weinstock:2005xlg}),
which leads to density-matrix like equations in~\cite{Garbrecht:2011aw}.
This method was further developed to estimate the resonant enhancement in resonant leptogenesis~\cite{Iso:2014afa,Garbrecht:2014aga},
and was also used in~\cite{Drewes:2016gmt} to derive the quantum Boltzmann equations used in leptogenesis via oscillations.\footnote{
One should note that quantum Boltzmann equations are also used to describe decoherence effects in flavored leptogenesis~\cite{Blanchet:2011xq,Abada:2006ea,DeSimone:2006nrs,Beneke:2010dz}.}
One of the shortcomings of this approach is that the off-diagonal matrix elements lie on an unphysical ``average'' energy shell $E=(E_1+E_2)/2$,
and cannot be used for arbitrarily large HNL mass differences.\footnote{One should note that as soon as $\Delta M \gg \Gamma$, the standard Boltzmann equations may be used again, ensuring overlap between the two approximations since $M \gg \Gamma$ is typically satisfied in leptogenesis.}
Recently, equations valid for arbitrary HNL mass spectra were derived in~\cite{Bodeker:2019rvr}.
The formalism developed in~\cite{Millington:2012pf} instead works in the so-called two-momentum picture,
also leads to similar quantum-Boltzmann equations~\cite{Dev:2014laa,Dev:2014wsa},
however, with an important difference---an \emph{mixing} term that acts as an additional source of lepton asymmetries.
The differences between these results prompted further investigations in~\cite{Kartavtsev:2015vto,Racker:2020avp}.

Meanwhile, another line of research start to develop.
The idea of baryogenesis via oscillations was put forward in work~\cite{Akhmedov:1998qx}.
It was further developed in Ref.~\cite{Asaka:2005pn}.
In particular, it was shown there, that the lepton back-reactions (missed in~\cite{Akhmedov:1998qx}) play an important role and the BAU generation is possible with two singlet fermions.
This work  introduced the kinetic equations for three lepton chemical potentials and two $2\times 2$ HNL density matrices.
This equations were used in all consequent works on the baryogenesis via neutrino oscillations.
Further clarifications of the mechanism were presented in~\cite{Shaposhnikov:2008pf,Canetti:2012kh}.
In particular, the role of the process without helicity flip (we will refer to them as fermion number violating) was pointed out in Ref.~\cite{Shaposhnikov:2008pf}.
It was noted  that once such processes are included, the total asymmetry may be generated at the fourth order in Yukawa couplings.
However, the relevant rates were not estimated correctly at that time.
The neutrality of cosmic plasma has been accounted for in~\cite{Shuve:2014zua} by introducing the so-called susceptibility matrices relating the leptonic chemical potentials to the number densities.
The temperature dependence of the susceptibility matrices and corrections coming from fermion masses were introduced in~\cite{Ghiglieri:2016xye}.
An important step has been performed in refs.~\cite{Ghiglieri:2016xye,Eijima:2017anv}, where the role of the fermion number violating processes was systematically accounted for and the equations were extended to the Higgs phase.
Effects related to the non-instantaneous freeze-out of sphalerons were discussed in refs.~\cite{Eijima:2017cxr,Ghiglieri:2017csp}.
The equations of~\cite{Ghiglieri:2016xye} were further improved in~\cite{Ghiglieri:2018wbs,Ghiglieri:2019kbw,Ghiglieri:2020ulj}.

It is interesting that the oscillations found in~\cite{Dev:2014laa} were initially assumed to be a genuinely different source of asymmetry than the one in the scenario of leptognesis via oscillations,
since they already appeared at fourth order in the Yukawa couplings ($\mathcal{O}(F^4)$), instead of $\mathcal{O}(F^6)$ as reported in the initial papers~\cite{Akhmedov:1998qx,Asaka:2005pn}.
As discussed above, this apparent discrepancy was a result of the relativistic approximations, where LNV terms suppressed by a factor $(M/T)^2$ were neglected.
Early investigations~\cite{Hambye:2016sby} into non-relativistic corrections suggested that these terms are already important for GeV-scale HNLs,
and may allow for freeze-out leptogenesis with GeV-scale HNLs.

In spite of the gradual convergence of the methods and results, a conclusive study showing that the parameter space of the two regimes of leptogenesis are connected was missing.
Strong hints of such a possibility were already provided in~\cite{Garbrecht:2014bfa}, where it was shown that freeze-in leptogenesis remains important for HNL masses above the TeV scale.
Similarly, the study in~\cite{Hernandez:2016kel} suggests that leptogenesis via HNL oscillations extends to HNL masses as large as $100$ GeV.
On the other hand in~\cite{Hambye:2016sby}, it was suggested that already moderate decay asymmetries allow for freeze-out leptogenesis at the GeV scale.
More recently this was confirmed in a similar study including flavor effects~\cite{Granelli:2020ysj}.
In this work we perform a unified study including all the effects relevant for mass scales up to a few TeV.

\section{The seesaw formula and the light neutrino masses}
\label{sec:seesaw}
For completeness of the paper in this section we briefly review the type-I seesaw mechanism and how it constrains the properties of the heavy neutrinos. The Lagrangian of the model reads
\begin{equation}
	\mathcal{L}= \mathcal{L}_\mathrm{SM}
	+ i \bar{\nu}_{R_I} \slashed{\partial} \nu_{R_I}
	- F_{\alpha I} \bar{L}_\alpha \tilde{\Phi} \nu_{R_I}
	- \frac12 {M_M}_{IJ} \bar{\nu}^c_{R_I}\nu_{R_J},
	\label{Lagr}
\end{equation}
where $\mathcal{L}_{SM}$ is the SM Lagrangian, $\nu_{R_I}$ are right-handed neutrinos
labeled with the generation indices $I, J$, $L_\alpha$ are the left-handed lepton doublets labeled with
the flavor index $\alpha = e, \mu, \tau$ and $\tilde{\Phi} = i\sigma_2 \Phi$,
$\Phi$ is the Higgs doublet. $F_{\alpha I}$ is the matrix of
Yukawa couplings in the basis where charged lepton Yukawa couplings and the Majorana mass
term of the right-handed neutrinos $M_{M}$ are diagonal.
After electroweak symmetry breaking, the Higgs field in the Lagrangian~\eqref{Lagr} obtains a vacuum expectation value $\langle \Phi \rangle = (0\,,v)^T$, $v = 174.1$~GeV at zero temperature.
The interaction terms in~\eqref{Lagr} effectively become Dirac mass terms coupling the left and right chiral components of the neutrinos.
However, since the right-handed neutrinos also have a Majorana mass, the spectrum of the theory can only be obtained once we diagonalize the full mass matrix.
Introducing $m_D \equiv v F $ and assuming that the elements of $m_D$ are much smaller then the elements of $M_M$, we can approximately diagonalize the neutrino mass matrix
\begin{align}
	m_\nu = m_D M_M^{-1} m_D^T + \mathcal{O}(m_D^4)\,,
	\label{mass:neutrino}
\end{align}
which gives us three light mass eigenstates
and the heavy eigenstates with a mass matrix
\begin{align}
	M_N = M_M + m_D^T m_D M_M^{-1} + M_M^{-1} m_D^T m_D + \mathcal{O}(m_D^4)\,.
	\label{mass:heavyNeutrino}
\end{align}
The number of the heavy eigenstates is the same as the number of the right-handed fields $\nu_{R_I}$. At least two HNLs are need  to explain the two observed mass splittings in the active neutrino sector.
In this work we focus on the minimal scenario with two heavy neutrinos.
We label the physical states of heavy neutrinos as $N_2$ and $N_3$\footnote{
	We leave the label $N_1$ for a potential sterile neutrino dark matter candidate of the $\nu$MSM~\cite{Asaka:2005pn}.
}
and denote their masses $M_2$ and $M_3$.
Throughout this work we will be interested in the case when $N_{2,3}$ have close masses, i.e. $|M_2+M_3| \gg |M_2-M_3|$. Therefore it will be convenient to use the average mass $M$ and the mass splitting $\Delta M$. In order to match the notations of Ref.~\cite{Eijima:2018qke}, we define them through
\begin{equation}
	\begin{aligned}
		M_2 &= M - \Delta M,\\
		M_3 &= M + \Delta M.
	\end{aligned}
	\label{eq:M_12}
\end{equation}
So strictly speaking $\Delta M$ is a half of the mass splitting.

\subsection{Parametrization of the Yukawa couplings}
\label{sub:parametrization_of_the_yukawa_couplings}
The masses of the light neutrinos $m_\nu$ are  constrained by the neutrino oscillations experiments (we use the global fit~\cite{Esteban:2020cvm}).
Out of the $9$ parameters in the light neutrino mass matrix, $5$ are already measured: two mass differences, and three mixing angles.
The remaining unknown parameters are the mass of the lightest neutrino, two Majorana phases, and the $CP$-violating phase~$\delta$.\footnote{
	It is exciting that these parameters may be probed in the not so distant future,
	for inverted hierarchy, the next generation of neutrinoless double beta decay experiments may provide information on the Majorana phases~\cite{Giuliani:2019uno},
	the $CP$-violating phase $\delta$ is already constrained by T2K~\cite{Abe:2019vii}, with further improvements expected from the DUNE experiment~\cite{Abi:2018dnh}.
}
In the model with two HNLs the lightest neutrino is massless (up to tiny loop corrections~\cite{Davidson:2006tg}). Therefore it makes sense to speak about the neutrino mass hierarchy rather than ordering. In what follows we refer to normal (inverted) mass hierarchy as NH (IH).

The measured low-energy parameters mean that the choice of heavy neutrino masses $M$ and the Yukawa couplings $F$ is not completely free.
To take this into account, we can parametrize the neutrino Yukawa couplings using the Casas-Ibarra parametrization~\cite{Casas:2001sr}:
\begin{align}
	F = \frac{i}{v} U_\nu \sqrt{m_\nu^\mathrm{diag}} \mathcal{R} \sqrt{M_M}\,,
	\label{parametrization:CI}
\end{align}
where the matrix $m_\nu^\mathrm{diag}$ is the diagonal neutrino mass matrix ($M_M$ is already diagonal in our basis), $U_\nu$ is the Pontecorvo-Maki-Nakagawa-Sakata (PMNS) matrix, and $\mathcal{R}$ is a complex orthogonal matrix $\mathcal{R} \mathcal{R}^T=1$.
For the PMNS matrix we use the standard parametrization~\cite{Zyla:2020zbs}:
\begin{align}
	U_\nu
	= V^{(23)} U_\delta V^{(13)} U_{-\delta} V^{(12)} \mathrm{diag}(1,\: e^{i \alpha_{21}/2},\: e^{i \alpha_{31} /2})
	\ , \label{PMNS}
\end{align}
where $U_{\pm \delta} = \mathrm{diag}(1,\: e^{\mp i \delta /2},\: e^{\pm i \delta /2})$, and the non-vanishing entries of $V^{(\alpha \beta)}$ for $\alpha = e,\:\mu,\:\tau$ are
\begin{align}
	V^{(\alpha \beta)}_{aa}
	&= V^{(\alpha \beta)}_{bb}
	= \cos \theta_{\alpha \beta}
	\ ,
	&  V^{(\alpha \beta)}_{\alpha \beta}
	&= -V^{(\alpha\beta)}_{\beta\alpha}
	= \sin \theta_{\alpha \beta}
	\ ,
	&  V^{(\alpha \beta)}_{\gamma\gamma}|_{\gamma\neq\alpha\,,\beta}
	&= 1 \ .
\end{align}
In the case of two heavy neutrinos there is only one relevant Majorana phase in the PMNS matrix.
We parametrize it as $\eta = \frac12 (\alpha_{21} - \alpha_{31} )$ for normal, and $\eta = \frac12 \alpha_{21} $ for inverted neutrino mass hierarchy with $\eta \in [0,2\pi]$.
The light neutrino mass matrix $m_\nu^\mathrm{diag}=\mathrm{diag}(m_1,m_2,m_3)$ with $m_1=0$ for NH, and $m_2=0$ for IH.

In the model with two right-handed neutrinos the matrices $\mathcal{R}$ depend on the neutrino mass hierarchy are given by
\begin{align}
	\mathcal{R}^{\rm NH}=
	\begin{pmatrix}
		0 && 0\\
		\cos \omega && \sin \omega \\
		-\xi \sin \omega && \xi \cos \omega
	\end{pmatrix}\,,\quad \quad
	\mathcal{R}^{\rm IH}=
	\begin{pmatrix}
		\cos \omega && \sin \omega \\
		-\xi \sin \omega && \xi \cos \omega \\
		0 && 0
	\end{pmatrix}
	\,.
\end{align}
with a complex angle $\omega=\Re\omega+i \Im\omega$, and the discrete parameter $\xi=\pm1$.
The change of the sign of $\xi$ can
be compensated by $\omega \to - \omega$
along with $ N_3 \to - N_3$~\cite{Abada:2006ea}, so we  fix  $\xi = +1$.
It is sufficient to constrain, $\Re \omega \in [0,\pi]$, as larger angles only change the overall sign of the Yukawa couplings.

To summarize, there are six free parameters of the theory.
They are listed in table~\ref{table_parameters} along
with their ranges considered in this work.
\begin{table}[htb!]
	\begin{center}
		\begin{tabular}{| c | c | c | c | c | c |}
			\hline
			$M$, GeV & $\log_{10} (\Delta M/M)$  & $\Im \omega$ & $\Re \omega$ & $\delta$ &
			$\eta$\\ \hline
			$[0.1 - 7000]$  & $[-19,-0.5]$  & $[-7,7]$ & $[0, \pi]$ &$[0, 2\pi]$ & $[0, 2\pi]$\\
			\hline
		\end{tabular}
	\end{center}
	\caption{\label{table_parameters} Parameters of the theory:
		average mass $M$; mass splitting $\Delta M$; $\Im \omega$; $\Re \omega$;
		Dirac $\delta$ and Majorana $\eta$ phases.
	In the second line we indicate the ranges of these parameters which were considered in this work.}
\end{table}
The upper boundary of the mass range table~\ref{table_parameters} is somewhat arbitrary. In section~\ref{sub:approximate_scaling_between_the_tev_scale_and_the_davidson_ibarra_bound} we show that a specific scaling law exists above $M\sim2$~TeV, so our results are applicable for heavier HNLs as well.

\subsection{Heavy neutrino mixing}
As a consequence of the seesaw mechanism, the heavy neutrino mass eigenstates are mixed with the doublet neutrinos, and can interact with the rest of the standard model,
the interaction states $\nu_{L \alpha}$ are superpositions of the mass eigenstates $\nu_i$ and $N_I$:
\begin{align}
	\nu_{L \alpha} = \left[ U_\nu \right]_{\alpha i} \nu_i + \theta_{\alpha I} N_I^c\,.
	\label{mixing}
\end{align}
The mixing angle between a heavy neutrino $N_I$ and the active neutrinos $\nu_{L\,\alpha}$ is expected to be small, and is approximately given by
\begin{align}
	\label{eq:theta}
	\theta_{\alpha I} \approx \frac{v F_{\alpha I}}{ M_{I}}\,.
\end{align}
This mixing angle appears in amplitudes for heavy neutrino production, which means that the heavy neutrino production rate is suppressed by
\begin{align}
	U^2_{\alpha I} \equiv |\theta_{\alpha I}|^2\,.
\end{align}
For phenomenological applications it is convenient to introduce the quantities
\begin{align}
	U_I^2 \equiv \sum_\alpha U_{\alpha I}^2\,,\quad
	U_\alpha^2 \equiv \sum_I U_{\alpha I}^2, \quad \text{and }\quad
	U^2 \equiv \sum_{\alpha I} U_{\alpha I}^2\,,
	\label{U2:def}
\end{align}
which quantify the total mixing angle of a particular heavy neutrino, the total mixing to a particular flavor, and the overall mixing between the heavy and light neutrinos.
\begin{align}
	U^2 = \frac{\sum_i m_i}{M} \cosh \left( 2 \Im \omega \right)
	\label{eq:U2Imomega}
\end{align}
This total mixing is useful to characterize the overall suppression of the interactions of the HNLs.

\section{Baryogenesis through leptogenesis}
\label{sec:bau_bounds}
As stated in the introduction, any leptogenesis mechanism relies on HNLs to satisfy two of the Sakharov conditions which are not met in the SM---$CP$ violation, and a sufficient deviation from thermal equilibrium.
In thermal leptogenesis the $CP$ violation arises via the loop corrections to the HNL decay rates:
\begin{align}
	\Gamma \sim
	\left\vert
	\parbox{20mm}{
		\begin{fmfgraph*}(30,20)
			\fmfpen{thin}
			\fmfleft{i1} \fmfright{o1,o2}
			\fmf{plain,width=thick}{i1,v1}
			\fmf{plain}{v1,o2}
			\fmf{dashes}{v1,o1}
		\end{fmfgraph*}
	}
	+
	\parbox{20mm}{
		\begin{fmfgraph*}(30,20)
			\fmfpen{thin}
			\fmfleft{i1} \fmfright{o1,o2}
			\fmf{plain,width=thick,tension=1.2}{i1,v1}
			\fmf{plain,width=thick}{v2,v3}
			\fmf{plain}{v1,v2}
			\fmf{plain,tension=1.5}{v3,o2}
			\fmf{dashes}{v1,v3}
			\fmf{dashes,tension=1.5}{v2,o1}
		\end{fmfgraph*}
	}
	+
	\parbox{20mm}{
		\begin{fmfgraph*}(30,20)
			\fmfpen{thin}
			\fmfleft{i1} \fmfright{o1,o2}
			\fmf{plain,width=thick,tension=3}{i1,v1}
			\fmf{plain,width=thick,tension=3}{v2,v3}
			\fmf{phantom,tension=3}{v1,v2}
			\fmf{dashes,left,tension=0}{v1,v2}
			\fmf{plain,right,tension=0}{v1,v2}
			\fmf{plain}{v3,o2}
			\fmf{dashes}{v3,o1}
		\end{fmfgraph*}
	}
	\right\vert^2
	\,.
	\label{fdiag:rlg}
\end{align}
These decay rates can differ for decays into leptons and anti-leptons, which leads to an overall lepton asymmetry.
This difference is often parametrised by the \emph{decay asymmetry}:
\begin{align}
	\epsilon^\alpha_I =
	\frac{\Gamma(N_I \rightarrow \bar{\phi} \ell_\alpha) - \Gamma(N_I\rightarrow \phi \bar{\ell_\alpha}) }{\Gamma(N_I\rightarrow \bar{\phi} \ell_\alpha) + \Gamma(N_I \rightarrow \phi \bar{\ell_\alpha}) }\,.
	\label{def:decayAsymm}
\end{align}
The evolution of the lepton asymmetries and the HNL number densities is an out-of-equilibrium process which can be described by
the following Boltzmann equations (see e.g.~\cite{Nardi:2006fx,Blanchet:2011xq,Davidson:2008bu}):
\begin{align}
	T_\mathrm{ref} \frac{d Y_{N_I}}{d z} &= - \Gamma_{N_I} (Y_{N_I}-Y_{N_I}^\mathrm{eq})\,,\\
	T_\mathrm{ref} \frac{d Y_{\Delta_\alpha}}{d z} &= 2 \epsilon^\alpha_I \Gamma_{N_I} (Y_{N_I}-Y_{N_I}^\mathrm{eq}) - W_\alpha \mu_\alpha \,.
	\label{eqs:Boltzmann}
\end{align}
We used the notation $z=T_\mathrm{ref}/T$ (where $T_\mathrm{ref}$ is some reference temperature - either the HNL mass $M$ or the sphaleron temperature $T_\mathrm{sph}$),
$\Gamma_{N_I}$ is the HNL decay rate,
$W_\alpha$ is the lepton number washout rate,
$\mu_\alpha$ are the chemical potentials in the lepton flavors,
and $Y_X = n_X/s$ are the yields---the ratios of the number density to the entropy density.
In the case of hierarchical HNL masses, the decay asymmetry $\epsilon^\alpha_I$ is bounded from above~\cite{Davidson:2002qv}, which translates to a lower bound on the mass of the lightest HNL:
\begin{align}
	M_I \gtrsim 10^9~\si{\GeV}\,,
	\label{bound:DI}
\end{align}
known as the Davidson-Ibarra bound.

Leptogenesis with HNLs below this mass scale can either be realized through resonant leptogenesis~\cite{Pilaftsis:2003gt} or through leptogenesis via neutrino oscillations~\cite{Akhmedov:1998qx,Asaka:2005pn}.
These two mechanisms were discovered independently for the different possible masses of the heavy neutrinos, leptogenesis via oscillations for $\mathrm{GeV}$-scale HNLs and resonant leptogenesis for HNLs with masses from the $\mathrm{TeV}$ scale to the Davidson-Ibarra bound.
In the following sections we briefly overview the main features of the two mechanisms and discuss the similarities and differences between them.

\subsection{Resonant Leptogenesis}
\label{sub:resonant_leptogenesis}
The  Davidson-Ibarra  bound on the decay asymmetry from~\cite{Davidson:2002qv} relaxes  (as does the bound on the HNL mass) if the HNL masses are degenerate~\cite{Liu:1993tg,Flanz:1996fb,Flanz:1994yx,Covi:1996wh,Covi:1996fm,Pilaftsis:1997jf,Buchmuller:1997yu}.
This is a consequence of the resonant enhancement in the $CP$-violating decays, which was discussed even before the idea of leptogenesis in~\cite{Kuzmin:1970nx}.

This \emph{resonant} contribution to the decay asymmetry comes from the interference of the tree level and wave-function amplitudes (the first and the last diagram in~\eqref{fdiag:rlg}).
\begin{align}
	\epsilon^\mathrm{w.f.}_I = \frac{\Im [ (F^\dagger F)_{32}]^2}{16 \pi (F^\dagger F)_{II}} \frac{M_2 M_3}{M_3^2-M_2^2}
	\label{eq:decayAsymmetry}
\end{align}
which is enhanced when $M_3 \rightarrow M_2$.
In the exactly degenerate limit the decay asymmetry diverges in the one-loop approximation.
This apparent divergence is as an artifact of applying the $S$-matrix theory to unstable HNLs, which cannot be asymptotic states.
To find the correct size of such a decay asymmetry, in the past two decades there were several studies of leptogenesis using out-of-equilibrium QFT methods~\cite{Buchmuller:2000nd,DeSimone:2007edo,DeSimone:2007gkc,Garny:2009rv,Garny:2009qn,Anisimov:2010aq,Beneke:2010wd,Garny:2011hg,Iso:2013lba,Garbrecht:2011aw,Herranen:2010mh,Fidler:2011yq, Herranen:2011zg, Millington:2012pf,Millington:2013isa,Dev:2014laa,Dev:2014tpa,Garbrecht:2014aga,Dev:2015wpa}.

The decay asymmetry then obtains a finite regulator of order of the decay width of the heavy neutrinos~$A\sim M_I \Gamma_I$, and
\begin{align}
	\epsilon^\text{w.f.} = \frac{\Im [ (F^\dagger F)_{32}]^2 [(F^\dagger F)_{22} + (F^\dagger F)_{33}]}{16 \pi (F^\dagger F)_{22} (F^\dagger F)_{33}} \frac{M_2 M_3 (M_2^2-M_3^2)}{(M_2^2-M_3^2)^2 + A^2}\,,
	\label{eq:decayAsymmetryR}
\end{align}
The different approaches of deriving the Boltzmann equations and the decay asymmetries from first principles do not always agree and can lead to $\mathcal{O}(1)$ different results in the degenerate limit, i.e.\ when $\Delta M \sim \Gamma_I$.\footnote{
	One should however note that the \emph{two-time} formulation~\cite{Anisimov:2010dk,Garny:2011hg,Iso:2013lba} (which relies on very few approximations),
	and the so-called \emph{Wigner space} approach~\cite{Garbrecht:2011aw,Garbrecht:2014aga} (where all interaction rates are evaluated on the same HNL average mass shell)
	lead to excellent agreement when the masses are not hierarchical~\cite{Dev:2017wwc}.
}
In practice the decay asymmetries $\epsilon$ combined with the Boltzmann equations (even if derived using out-of-equilibrium QFT methods) do not accurately describe the HNL dynamics.
Instead one has to rely on quantum-kinetic equations that include the HNL coherence terms as dynamical degrees of freedom as we discuss in the following section.

\subsection{Leptogenesis via neutrino oscillations}
\label{sub:kinetic_equations}
In leptogenesis via HNL oscillations~\cite{Akhmedov:1998qx,Asaka:2005pn} the asymmetry is not produced in the decays of the HNLs, but instead while the HNLs are produced and approach equilibrium in the early Universe.
The HNLs generated via the SM particle interactions are produced in their interaction basis, which does not necessarily coincide with their mass basis.
Due to this misalignment, the HNLs begin to oscillate and through these $CP$ violating oscillations they generate the lepton asymmetry.

Since the GeV-scale HNLs in this mechanism remain relativistic at temperatures relevant for baryogenesis $T\geq T_{sph} \approx 130\,\mathrm{GeV}$,
the caluclation of the BAU does not rely on calculating their decay rates as in resonant leptogenesis.
Instead, a more apropriate picture is that of neutrino oscillations.
To correctly take these processes into account, in~\cite{Akhmedov:1998qx,Asaka:2005pn},
the equations used to describe the kinetic evolution of neutrinos due to Raffelt and Sigl~\cite{Sigl:1992fn} were modified to include oscillations between HNLs.

\paragraph{Evolution equations for baryogenesis via neutrino oscillations.}

Compared to the initial developments~\cite{Akhmedov:1998qx,Asaka:2005pn}, there have been several systematic improvements to the kinetic equations for HNLs.
One of the most significant improvements in recent years is the inclusion of corrections caused by the finite HNL mass~\cite{Eijima:2017anv,Ghiglieri:2017gjz}.
On the other hand, it was also found that the same equations arise in the non-equilibrium formulation of QFT~\cite{Drewes:2016gmt,Antusch:2017pkq}.
The key difference compared to the Boltzmann equations~\eqref{eqs:Boltzmann} is that besides the HNL number density $Y_{N_I}$,
one also has to keep track of the HNL correlations.
This information is encoded in the \emph{density matrix} $(\rho_N)_{IJ}$, where $(\rho_N)_{II}\sim f_{N_I}$ in the mass basis.
As we will show in section~\ref{sec:boltzmannFromDnsMat}, the off-diagonal correlations $(\rho_N)_{IJ}$ become negligibe in the limit of fast oscillations,
and we recover the usual Boltzmann equations.
The equations governing the HNL densities\footnote{The equations in the current form---with lepton chemical potentials---are valid as long as leptons stay in equilibrium.
	In particular, for temperatures above $T>85$~TeV , the right-handed electrons are not in equilibrium~\cite{Bodeker:2019ajh}, and  susceptibility matrices relating chemical potentials with number densities need to be modified accordingly.
} (modified from~\cite{Garbrecht:2011aw,Drewes:2016gmt,Eijima:2017anv,Ghiglieri:2017gjz,Antusch:2017pkq} to be valid in both relativistic and non-relativistic limits, c.f.~\cite{Ghiglieri:2019kbw,Ghiglieri:2020ulj,Bodeker:2019rvr})
including both the positive (negative) HNL $\rho_N$ ($\bar{\rho}_N$) helicities, and the leptonic asymmetries $n_{\Delta_\alpha}$ are given by:
\begin{subequations}
	\begin{alignat}{3}
		i \frac{d n_{\Delta_\alpha}}{dt}
		&= -&&2 i \frac{\mu_\alpha}{T} \int \frac{d^{3}k}{(2 \pi)^{3}} \Tr [\Gamma_{\alpha} ] f_{N} (1-f_{N})  \,
		+ i \int \frac{d^{3}k}{(2 \pi)^{3}} \, \text{\text{Tr}}[\tilde{\Gamma}_\alpha \, (\delta \bar{\rho}_{N} - \delta \rho_N)]\,,\label{kin_eq_a}
		\\
		i \, \frac{d \delta \rho_{N}}{dt}&= - &&i \, \frac{d\rho_{N}^{eq}}{dt}
		+[H_N, \rho_N]
		- \frac{i}{2} \, \{ \Gamma , \delta \rho_{N} \}
		- \frac{i}{2} \, \sum_\alpha \tilde{\Gamma}_\alpha \, \left[ 2 \frac{\mu_\alpha}{T} f_{N} (1-f_{N}) \right],\label{kin_eq_b}
		\\
		i \, \frac{d \delta \bar{\rho}_{N}}{dt}&= -&&i \, \frac{d\rho_{N}^{eq}}{dt}
		-[H_N, \bar{\rho}_{N}]
		- \frac{i}{2} \, \{ \Gamma , \delta \bar{\rho}_{N}\}
		+ \frac{i}{2} \, \sum_\alpha \tilde{\Gamma}_\alpha \, \left[ 2 \frac{\mu_\alpha}{T} f_{N} (1-f_{N}) \right].
		\label{kin_eq_c}
	\end{alignat}\label{KE_1a}
\end{subequations}
Where we introduced $\delta \rho_N = \rho_N - \rho_N^{eq}$, and $\rho_N^{eq} \approx 1_{2\times2} f_N$.
The function $f_N = 1/\left( e^{\omega_k/T}+1 \right) $ is the equilibrium
distribution function of the massive fremions, $\omega_k = \sqrt{M^2+k^2}$.
This distribution function is temperature-dependent. Its time derivative acts as a source of the deviation from equilibrium, therefore in what follows we will refer to  $ d\rho^{eq}/d t$ as the \emph{source term}.
Note that we omit any Hubble expansion terms as we implicitly consider the \emph{comoving densities}.
The effective Hamiltonian describing the
coherent oscillations of the HNLs is
\begin{equation}
	H_N = H_0 + H_I, \quad
	H_0 = \frac{M^2}{2 E_N}, \quad
	H_I = h_{+} Y_{+} + h_{-} Y_{-}\,,
	\label{H_N}
\end{equation}
where $E_{N}=\sqrt{k_{N}^{2} + M^{2}}$. The quadratic combinations of Yukawa couplings $Y_\pm$ appearing in Eq.~\eqref{H_N} and in the rates are defined below.
The damping rates are
\begin{align}
	\Gamma = \gamma_+ Y_+ + \gamma_- Y_-\,, &&
	\Gamma_\alpha = \gamma_+ Y_+^\alpha + \gamma_- Y_-^\alpha\,,
	\label{damping}
\end{align}
The communication terms, describing the transitions from HNLs to active neutrinos, are
\begin{align}
	\tilde{\Gamma} = - \gamma_+ Y_+ + \gamma_- Y_-\,, &&
	\tilde{\Gamma}_\alpha = - \gamma_+ Y_+^\alpha + \gamma_- Y_-^\alpha\,.
	\label{communication}
\end{align}
In the expressions above the subscripts $+$ and $-$
refer to the fermion number conserving and violating quantities correspondingly.
The functions $h_\pm$ and $\gamma_\pm$
depend only on kinematics (\emph{i.e.} on the common mass of HNLs).
These functions have to be determined over the whole temperature region of the interest,
which includes both symmetric and Higgs phases.

The dependence on the Yukawa coupling constants factorizes out
from the rates~\eqref{H_N}--\eqref{communication}, and we have
\begin{align}
	Y_+ = F^\dagger F \,,&&
	Y_- = G F^T F^* G^*\,,\\
	(Y_+^\alpha)_{IJ} = F^\dagger_{I \alpha}F_{\alpha J} \,,&&
	(Y_-^\alpha)_{IJ} = (G F^T)_{I \alpha} (F^* G^*)_{\alpha J}\,,
	\label{YukawaMatrices}
\end{align}
where the matrix $G$ encodes the generalized Majorana condition between the Majorana fields $N_I \equiv \nu_{R_I} + \nu_{R_I}^c$, with $N_I = G_{IJ} N^C_J$.
If we transform $N_I$ from the basis where the Majorana condition is given by $\tilde{N}_I=\tilde{N}_I^C$ by a unitary transformation
$N_I = U_{I J} \tilde{N}_J$, this matrix is given by $G = U^\dagger U^*$.
The relation between the number densities and the chemical potentials to leptons
in Eq.~\eqref{KE_1a}
has to take into account the neutrality of plasma.
When the system is in equilibrium with respect to sphaleron processes,
this relation reads
\begin{equation}
	\mu_\alpha = \omega_{\alpha \beta}(T) n_{\Delta_\beta},
	\label{susceptibilities}
\end{equation}
where $\omega_{\alpha \beta}(T)$ is the so-called susceptibility matrix,
see, e.g.~\cite{Ghiglieri:2016xye,Eijima:2017cxr}.

\subsection{Boltzmann equations as a limit of the quantum kinetic equations}
\label{sec:boltzmannFromDnsMat}
As mentioned above, one of the key quantities in thermal leptogenesis are the decay asymmetries $\epsilon$.
These quantities are typically calculated via Feynman diagrams as shown in Eq.~\eqref{fdiag:rlg}, but can also be obtained as a limit of the full density matrix equations.
In this section we show how these quantities arise from the density-matrix approach, and discuss when such approximations can be applied.
To find approximate expressions, it is instructive to explicitly write the equations governing the HNL correlations $\rho_{IJ}$ for $I\neq J$
(note that we omitted the subscript $N$ for brevity):
\begin{equation}
	\begin{aligned}
		\frac{d \delta \rho_{IJ}}{d t}
		&=-i (H_{II} - H_{JJ}) \delta \rho_{IJ} - \frac12 (\Gamma_{II} +\Gamma_{JJ}) \delta \rho_{IJ}\\& - \frac12 \Gamma_{IJ} (\delta \rho_{II} - \delta \rho_{JJ}) - i H_{IJ} (\delta \rho_{II} - \delta \rho_{JJ})\,.
	\end{aligned}
\end{equation}
If the diagonal elements $\delta \rho_{II}, \; \delta \bar{\rho}_{JJ}$ change on a time-scale much slower than the oscillation period, we can approximate the off-diagonal correlations from~\eqref{KE_1a} by
the steady state limit (obtained by setting $\frac{d \delta \rho_{IJ}}{d t} \rightarrow 0$):
\begin{align}
	\text{for } I\neq J :\quad \delta \rho_{IJ} &\approx \frac{\frac{i}{2} \Gamma_{IJ} (\delta \rho_{II} + \delta \rho_{JJ}) + H_{IJ} (\delta \rho_{II} - \delta \rho_{JJ})}
	{(H_{II}-H_{JJ}) - \frac{i}{2} (\Gamma_{II} + \Gamma_{JJ})}\,,
	\label{rhoij:quasistatic}
\end{align}
where the analogous solutions for $\delta \bar{\rho}$ are be obtained by replacing $H \rightarrow -H$, and $\rho \rightarrow \bar{\rho}$.
We may insert this solution into the source term for the lepton asymmetry from Eq.~\eqref{kin_eq_a}:
\begin{align}
	&\Tr \left[ \tilde{\Gamma}_\alpha (\bar{\rho}_N - \rho_N) \right] \approx\\\notag
	&\sum_I \delta \rho_{II} \frac{H_{22}-H_{33}}{(H_{22}-H_{33})^2 + \left( \frac{\Gamma_{22} + \Gamma_{33}}{2} \right)^2} \times \\\notag
	&\times \left\{ 2(\gamma_+^2 + \gamma_-^2) \Im \left[ (F^\dagger F)_{23} F_{\alpha 2} F^*_{\alpha 3} \right] +
	4 \gamma_+ \gamma_- \Im \left[ (F^\dagger F)_{23} F^*_{\alpha 2} F_{\alpha 3} \right] \right\} \,,
\end{align}
where we neglect higher order terms in the Yukawa couplings, including the helicity asymmetry $\bar{\rho}_{II} - \rho_{II}$.
The two terms can be identified with the  \emph{lepton flavor violating}
and \emph{lepton number violating} source terms for the asymmetry.

Substituting this solution into Eq.~\eqref{kin_eq_a}, and inserting the vacuum expressions for $H$ gives us an approximate expression for the two decay asymmetries
\begin{align}
	\epsilon_I^{\alpha\,\mathrm{LNV}} =
	- \frac{(M_2^2-M_3^2)\Im \left[ (F^\dagger F)_{23} F^*_{\alpha 2} F_{\alpha 3} \right]}{(M_2-M_3)^2 + E_N^2 (\gamma_+ + \gamma_-)^2 ([F^\dagger F]_{22} + [F^\dagger F]_{33})^2}
	\frac{4 E_N \gamma_+ \gamma_-}{(\gamma_+ + \gamma_-)[F^\dagger F]_{II}}\,,
	\label{effDecAsymm:LNV}
\end{align}
and
\begin{align}
	\epsilon_I^{\alpha\,\mathrm{LFV}} =
	\frac{(M_2^2-M_3^2)\Im \left[ (F^\dagger F)_{23} F_{\alpha 2} F^*_{\alpha 3} \right]}{(M_2-M_3)^2 + E_N^2 (\gamma_+ + \gamma_-)^2 ([F^\dagger F]_{22} + [F^\dagger F]_{33})^2}
	\frac{2 E_N (\gamma_+^2 + \gamma_-^2)}{(\gamma_+ + \gamma_-)[F^\dagger F]_{II}}\,.
	\label{effDecAsymm:LFV}
\end{align}
Notice that $\epsilon_I^{\alpha\,\mathrm{LFV}}$ only violates lepton flavor,
as the sum over $\alpha$ leads to a vanishing lepton asymmetry.

A few comments on these expressions are in order:
\begin{itemize}
	\item in contrast to the asymmetries used in~\cite{Hambye:2016sby, Granelli:2020ysj}, the decay rates entering~\eqref{effDecAsymm:LNV} and~\eqref{effDecAsymm:LFV}
		include explicitly the helicity-dependence of the rates, and lead to a result similar to~\cite{Garbrecht:2019zaa} in the hierarchical limit,
	\item in the non-relativistic limit $\gamma_\pm \rightarrow M/(16 \pi)$ and $|k_0| \rightarrow M$ we recover the relation~\eqref{eq:decayAsymmetryR},
	\item in the same limit these expressions qualitatively agree with the regulators from~\cite{Garny:2011hg}, as well as those from~\cite{Garbrecht:2014aga} if $\Re[F^\dagger F]_{23} \ll [F^\dagger F]_{II}$,
	\item although the decay asymmetry from Eqs.~\eqref{effDecAsymm:LNV} and~\eqref{effDecAsymm:LFV} remain finite when ${\Delta M^2_N \ll |k_0| \gamma_\pm F^2}$,
		the approximation of fast oscillations which we used here breaks down, and one has to rely on quantum kinetic equations~\eqref{KE_1a}.
\end{itemize}

\begin{figure}[h]
	\centering
	\includegraphics[width=0.9\textwidth]{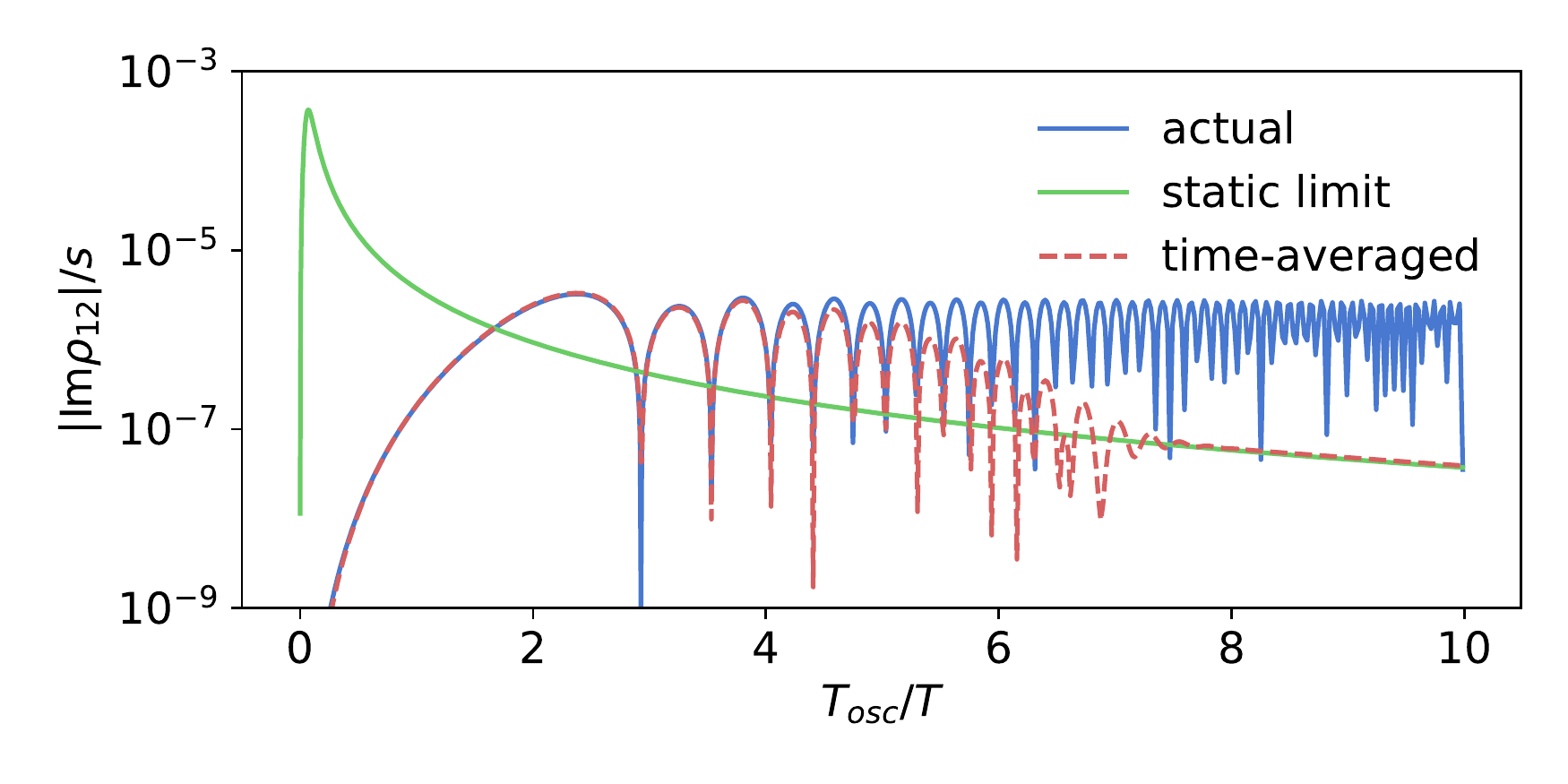}
	\caption{An example time evolution of the off-diagonal correlations which enters the decay asymmetry (blue, full).
		In the limit of fast oscillations, this highly oscillatory behavior is well described by the static limit (green, full).
		The fast oscillations lead to a numerically stiff system. In (red, dashed) we show how the full solution approaches the static limit once we average it over some a finite time interval.
		This example shows that using the static solution before a single oscillation period can lead to a significant overestimate of the BAU, even when it is regulated by a finite width.
	}
	\label{fig:rho12}
\end{figure}

\paragraph{Applicability of the decay asymmetries.}
We conclude this section with a short comment on the different ways of regulating the decay asymmetry $\epsilon^\mathrm{LFV}$.
To explore the breakdown of Boltzmann equations in the GeV-mass regime, we recover the result from~\cite{Akhmedov:1998qx,Asaka:2005pn},
following the procedure described in~\cite{Drewes:2012ma}.
Taken at face value, Eqs.~\eqref{effDecAsymm:LNV} and~\eqref{effDecAsymm:LFV} appear well-behaved in the limit of $\Delta M_N \rightarrow 0$,
but once we take the full picture of heavy neutrino oscillations into account we find that this is not the whole story.
In the relativistic limit $T \gg M$, the rate $\gamma_+ \sim T$, and $\gamma_- \approx 0$, equation~\eqref{effDecAsymm:LFV} simplifies to:
\begin{align}
	\epsilon_I^{\alpha\,\mathrm{LFV}} &\approx \frac{(M_2^2-M_3^2)\Im \left[ (F^\dagger F)_{23} F_{\alpha 2} F^*_{\alpha 3} \right]}{(M_2^2-M_3^2)^2 + E_N^2 \gamma_+^2 ([F^\dagger F]_{22} + [F^\dagger F]_{33})^2}
	\frac{2 E_N \gamma_+}{[F^\dagger F]_{II}}
	\,.
\end{align}
In contrast to the temperature independent decay asymmetries often used for resonant leptogenesis,
this decay asymmetry has a dramatic temperature dependence, and becomes enhanced when $M/T\rightarrow 0$:
\begin{align}
	\label{eq:LFVsrc}
	{\epsilon_I^\alpha}^\text{LFV} \sim
	\frac{1}{z^2} \frac{\Im \left[ (F^\dagger F)_{23} F_{\alpha 2} F^*_{\alpha 3} \right]}{M_2^2-M_3^2}
	\frac{2 E_N \gamma_+}{[F^\dagger F]_{II} T^2}\,,
\end{align}
where $z\equiv M/T$. The question is what prevents the initial lepton asymmetries from diverging, we might expect that the asymmetry is already made finite by the width in Eq.~\eqref{effDecAsymm:LFV}.
On the other hand, if we solve the kinetic equations, we find that another effect suppresses the decay asymmetries much sooner: the finite time required for HNL oscillations.
As shown in figure~\ref{fig:rho12}, the approximate solution obtained in the limit of fast oscillations misses the size of the off-diagonal correlations by several orders of magnitude,
since the requirement of fast oscillations is not satisfied until several oscillations occurred.
The fact that the oscillation time scale regulates the divergences of Eq.~\eqref{eq:LFVsrc} is exactly what leads to the exponent $(M_2-M_3)^{3/2}$ in expression for the total lepton asymmetry~\cite{Asaka:2005pn}.
As we have shown in this section, a systematic treatment of the low-scale leptogenesis in all corners of the parameter space is only possible within the framework of quantum kinetic equations~\eqref{KE_1a}. The crucial ingredient of these equations are the rate coefficients entering equations \eqref{H_N}--\eqref{communication}. The next section is dedicated to the determination of these coefficients in all relevant regimes.

\section{The heavy neutrino production rates}
\label{sec:the_rates}

In recent years significant progress has been made towards determining the production rate of the heavy neutrinos in the early universe~\cite{Anisimov:2010gy,Besak:2012qm,Ghisoiu:2014ena,Garbrecht:2013urw,Garbrecht:2019zaa, Biondini:2017rpb}.
In particular, the production rate of GeV-scale heavy neutrinos at temperatures $T\gg M$ has been studied in great detail~\cite{Ghiglieri:2017gjz,Ghiglieri:2017csp,Ghiglieri:2016xye}.
For GeV-scale neutrinos important effects arise at temperatures below the electroweak crossover, where the mixing between the heavy and light neutrinos can significantly affect the production rate~\cite{Eijima:2017anv,Ghiglieri:2017gjz}.
The active neutrino interaction rate---which affects the heavy neutrino production at these regime---has even been calculated at  NLO~\cite{Jackson:2019tnr}.

At present, there are no readily available estimates of the full heavy neutrino production rate for $T\sim M$.
In~\cite{Garbrecht:2019zaa}, the rate was calculated only including the \emph{naive} leading order $1\leftrightarrow 2$, and $2\leftrightarrow 2$ processes, thereby neglecting the $1+n\rightarrow 2+n$ processes enhanced by multiple soft scatterings, also known as the LPM (Landau-Pomeranchuk-Migdal) effect.
On the other hand, earlier calculations~\cite{Anisimov:2010gy,Besak:2012qm,Ghisoiu:2014ena}, only provide the helicity-averaged rate, which is not sufficient to track the full evolution of heavy neutrinos as the universe cools down from $T\gg M$ to $T\sim M$.
Below we describe our approach to the calculation of the heavy neutrino production rate.

The production rate of the heavy neutrinos can be expressed through the spectral (\textit{antihermitian}\footnote{In literature this is also known as the \textit{imaginary} part of the self-energy, which is strictly speaking not accurate for heavy neutrinos whose self-energy is a complex matrix in Dirac and flavor spaces.}) part of their self-energy $\Sigma^\mathcal{A}_N$.
To simplify the calculation, we factor out the Yukawa couplings from the self energies and introduce
\begin{align}
	\slashed{\Sigma}_N = g_w (  \hat{\slashed{\Sigma}} P_R F^\dagger F + \hat{\slashed{\Sigma}} P_L G F^T Y^* G^*)\,,
\end{align}
where $g_w=2$ counts the $SU(2)$ degrees of freedom, and the symmetric matrix $G$ encodes the generalized Majorana condition $N=G N^c$, and must be included to ensure flavor covariance of the equations.

The rate coefficients $\gamma_\pm$ are related to the self-energy as
\begin{align}
	\gamma_\pm =\frac{g_w}{|k_0|} (\hat{\Sigma}^\mathcal{A}_{N 0} \pm \hat{\Sigma}^\mathcal{A}_{N i}\hat{k}_i ) (k_0 \pm |k|) \Biggr|_{k_0=\omega_k}
	= \frac{g_w}{|k_0|} \hat{\Sigma}^\mathcal{A}_{N \mp} \cdot k_\pm \Biggr|_{k_0=\omega_k}\,,
\end{align}
where we define $\Sigma_\pm$ and $k_\pm$ as $a_\pm = a_0 \pm a_i k_i/|k|$.
The results of~\cite{Ghisoiu:2014ena}, as well as~\cite{Ghiglieri:2016xye} provide a calculation of the spin-averaged heavy neutrino production rate, which is related to the rates $\gamma_\pm$ as
\begin{align}
	\Im \Pi = (\gamma_+ + \gamma_-)|k_0|\,,
\end{align}
where $\Im \Pi$---in the notations of~\cite{Ghiglieri:2016xye}---is the antihermitian part of self-energy.

For $M \ll T$, when neutrinos are relativistic, and $k_-\sim \frac{M^2}{2k}$ the $\gamma_-$ rate is suppressed by $M^2/T^2$, and $\Im \Pi \approx \gamma_+$ corresponds to the total heavy neutrino production rate.
In~\cite{Ghiglieri:2017gjz}, these rates were expressed in terms of $Q_\pm$ with
\begin{align}
	\gamma_+(k) \approx Q_+(k) T\,, && \gamma_-(k) \approx Q_-(k) \frac{M^2}{T}\,.
\end{align}

On the other hand, for $M\gg T$, the two rates $\gamma_\pm$ become equal, as the heavy neutrinos become non-relativistic and both $k_+ \approx k_- \approx M$, and $\hat{\Sigma}_{N \pm}\approx \frac{M}{32 \pi}$.

In the intermediate regime, for temperatures $T\sim M$ we estimate the full rate by extrapolating the relativistic rate from~\cite{Ghiglieri:2017gjz},
and combining it with the rate of heavy neutrino decays above the ``thresholds''\footnote{
	We note that this threshold is not completely accurate, as there are other channels allowing the heavy neutrino to decay, furthermore, the thermal masses should be taken with other processes at the same order in the couplings - namely the Landau-Pomeranchuk-Migdal contributions. Nonetheless we use it to distinguish the regimes where the production rate is dominated by decays $M_N>T$ and the regime where the scatterings dominate $M_N < T$.
} corresponding to decays into the Higgs, $W\pm$ and $Z$ bosons for $M_N > M_{H,Z,W}$.

We perform the extrapolation as follows, we note that the heavy neutrino production rate $\Gamma_N = \gamma_+ + \gamma_-$
appears to have a rather weak dependence on the heavy neutrino mass in the results from~\cite{Besak:2012qm,Ghisoiu:2014ena}, for masses $M\lesssim T$.
The fermion number conserving (FNC) rate has the same behavior for $M\ll T$, where it becomes independent of $M$, whereas for $M \gg T$ the heavy neutrino decays dominate.
In contrast to this, the fermion number violating (FNV) rate $\gamma_-$ has a strong dependence on the heavy neutrino mass $M$ in the relativistic regime.
However, we note that the dependence does not directly appear in the self-energy $\hat\Sigma$, but mainly through the prefactor $k_-/|k_0| \sim M^2/ k^2$.

To summarize, we estimate the self-energy of the heavy neutrino as the sum of a $M$-independent term, and a $M$-dependent term for masses above the heavy neutrino decay threshold:

\begin{align}
	\hat{\Sigma}(k,M) \approx
	\hat{\Sigma}(k,0) + \theta(M-m_\phi) \hat{\Sigma}^{1\leftrightarrow2}(k,M)\,.
\end{align}

If we compare this approximation to the naive $1\leftrightarrow2$ decay rate, we find that there is a clear $M$-dependence in the $1\leftrightarrow2$ rate even for $M<m_\phi$.
This dependence is caused by the finite size of the effective lepton mass $m_\ell$, and the fact that the decay of the heavy neutrino would be forbidden for $m_\phi - m_\ell < M < m_\phi+m_\ell$.
Note however that this kinematically forbidden region is an artifact of our approximation when we treat the effective mass of the lepton as a physical mass term. Such kinematically forbidden regions in reality disappear once the $2\leftrightarrow2$ scatterings, as well as the LPM effect are included.

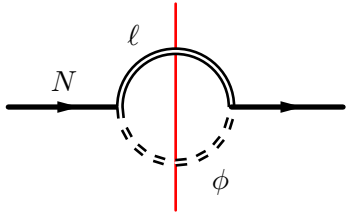
\begin{figure}
	\begin{align*}
		&\begin{gathered}
			\begin{fmfgraph*}(89,55)
				\fmfleft{i1} \fmfright{o1}
				\fmftop{c1}
				\fmfbottom{c2}
				\fmf{fermion,label=$N$,width=thick,label.side=left,tension=2}{i1,v1}
				\fmf{dbl_plain,left=1,tag=1}{v1,v2}
				\fmf{dbl_dashes,right=1,tag=2}{v1,v2}
				\fmf{fermion,width=thick,tension=2}{v2,o1}
				\fmfposition
				\fmfipath{p[]}
				\fmfiset{p1}{vpath1(__v1,__v2)}
				\fmfiset{p2}{vpath2(__v2,__v1)}
				\fmfiv{label=$\ell$}{point length(p1)/3 of p1}
				\fmfiv{label=$\phi$}{point length(p2)*2/3 of p2}
				\fmfiv{}{point length(p1)/2 of p1}
				\fmfiv{}{point length(p2)/2 of p2}
				\fmf{plain,foreground=red}{c1,c2}
			\end{fmfgraph*}
		\end{gathered}
	\end{align*}
	\caption{Heavy neutrino production before electroweak symmetry breaking.
		The double lines represent the resummed propagators for the Higgs and lepton doublets. Note that both the $1\leftrightarrow 2$ and $2\leftrightarrow2$ processes can be described by this diagram.
		At first approximation we can understand the ressumation of the propagators as the Higgs and the lepton doublet obtaining effective masses $m_\phi$ and $m_\ell$ respectively.
		An important soft process not included in this diagram is the Landau-Pomeranchuk-Migdal effect, which corresponds to soft gauge boson exchanges between the lepton and Higgs.
	}
	\label{fig:productionSP}
\end{figure}

The contributions from the $1 \leftrightarrow 2$ processes to the antihermitian part of the heavy neutrino self-energy (c.f. Fig~\ref{fig:productionSP}) are given by
\begin{align}
	\hat{\Sigma}_{N}(k) =
	\int \frac{\mathrm{d}^4 p}{(2\pi)^4} 2 \Delta_\phi^\mathcal{A}(p-k) \hat{S}_\ell^\mathcal{A}(p)
	[1-f_F(p) + f_B(p-k)],
	\label{integral:sigmahat}
\end{align}
where $ \Delta_\phi^\mathcal{A}(k)$ and $\hat{S}_\ell^\mathcal{A}(k)$ are the spectral functions.
To obtain the naive result for the $1\leftrightarrow 2$ heavy neutrino production rate, we replace them by the tree-level approximation
\begin{align}
	\Delta_\phi^\mathcal{A}(k) \approx \pi \delta(k^2-m_\phi^2) \mathrm{sign}(k^0)\,, &&
	\hat{S}_\ell^\mathcal{A}(k) \approx \pi \delta(k^2-m_\ell^2) \mathrm{sign}(k^0) (\slashed{k}-m_\ell)\,,
	\label{2ptf:3lvl}
\end{align}
where $m_\ell^2 = T^2 (g_1^2 + 3 g_2^2)/16$ and $m_\phi^2 = T^2 (g_1^2 + 3 g_2^2 + 4 h_t^2 + 8 \lambda )/16$.
Note that we ommit the chiral projector in $\hat{S}_\ell$, as we factored it out in the definition of $\hat{\Sigma}_N$.

When the on-shell condition is imposed through the delta functions from~\eqref{2ptf:3lvl}, the integral in Eq.~\eqref{integral:sigmahat} reduces to a one-dimensional integral with a well known result
\begin{align}
	\hat{\Sigma}_0^\mathcal{A} &= \frac{T^2}{16 \pi |\mathbf{k}|} [I_1(-\omega_+) - I_1(-\omega_-)]\,,
	\label{self-energy:spectral1to2}
	\\\notag
	\hat{\Sigma}_i^\mathcal{A} &= \frac{T^2 \hat{\mathbf{k}}_i}{16 \pi |\mathbf{k}|}
	\left[
		\frac{k^0}{|\mathbf{k}|} [I_1(-\omega_+) - I_1(-\omega_-)] -
		\frac{M^2 + m_\ell^2 - m_\phi^2}{2 |\mathbf{k}| T} [I_0(-\omega_+) - I_0(-\omega_-)]
	\right]\,,
\end{align}
where the functions $I_n$ correspond to the integrals
\begin{align}
	I_n(y)=\int^y \mathrm{d} x x^n [1-f_F(x)+f_B(y-x)]\,,
\end{align}
with the integration boundaries
\begin{align}
	\omega_\pm = \frac{- k_0 (m_\phi^2-m_\ell^2-M^2) \pm k \sqrt{[M^2-(m_\phi+m_\ell)^2][M^2-(m_\phi-m_\ell)^2]}}{2 M^2}\,.
\end{align}

\begin{figure}
	\centering
	\includegraphics[width=0.9\textwidth]{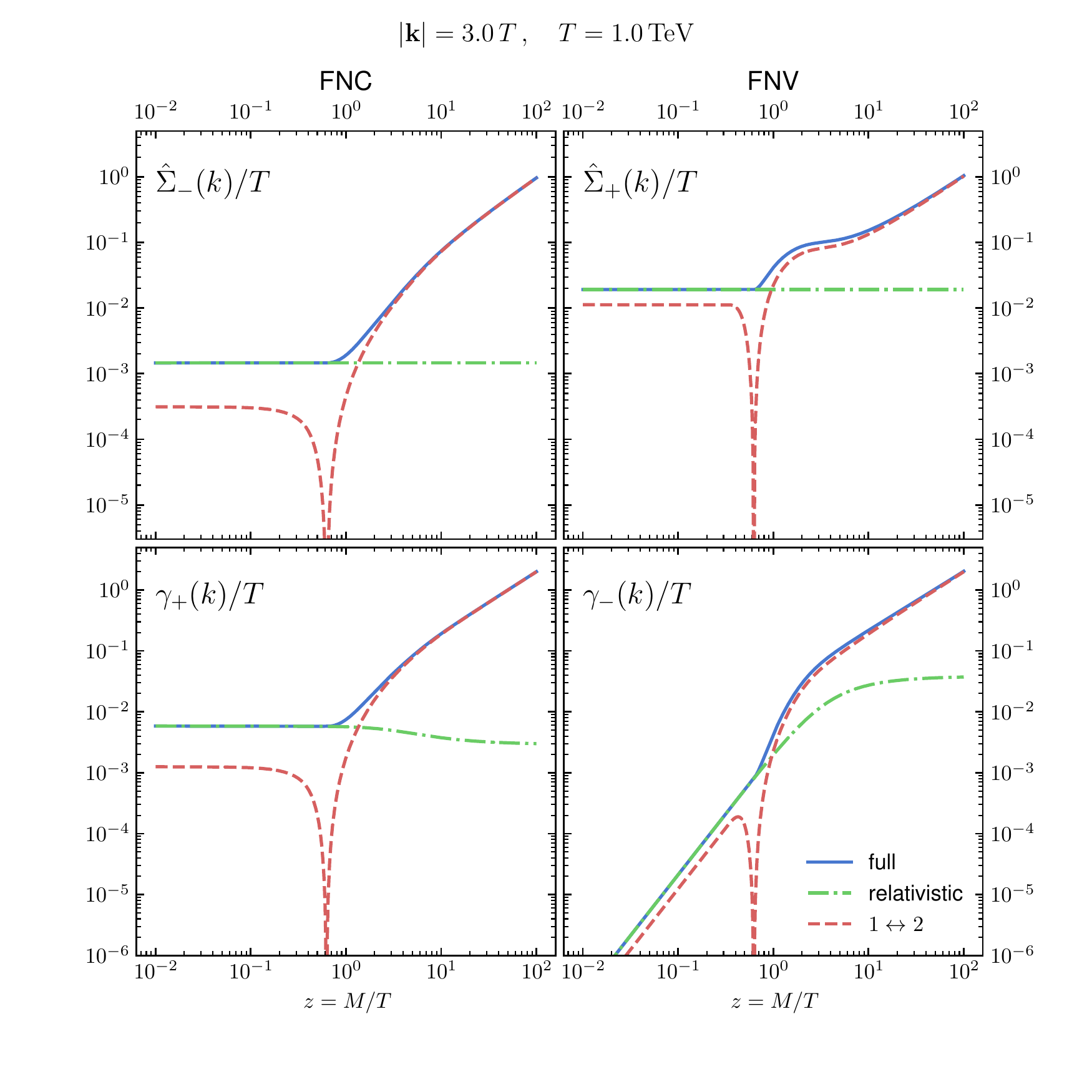}
	\caption{The extrapolation used to estimate the FNV and FNC rates.
		The full result (blue, full) is obtained by adding the extrapolated relativistic self-energy (green, dot-dashed)
		and the $1\leftrightarrow 2$ self-energy (red, dashed) for $M>M_H$.
		The upper panel shows the self-energies $\hat{\Sigma}_\pm$, which have a weaker dependence on the HNL mass $M$,
		compared to the interaction rates $\gamma_\pm$ which include a part of the HNL phase space suppression $k_\pm/|k_0|$.
	}
	\label{fig:sigmapmApp}
\end{figure}

\begin{figure}
	\centering
	\includegraphics[width=0.9\textwidth]{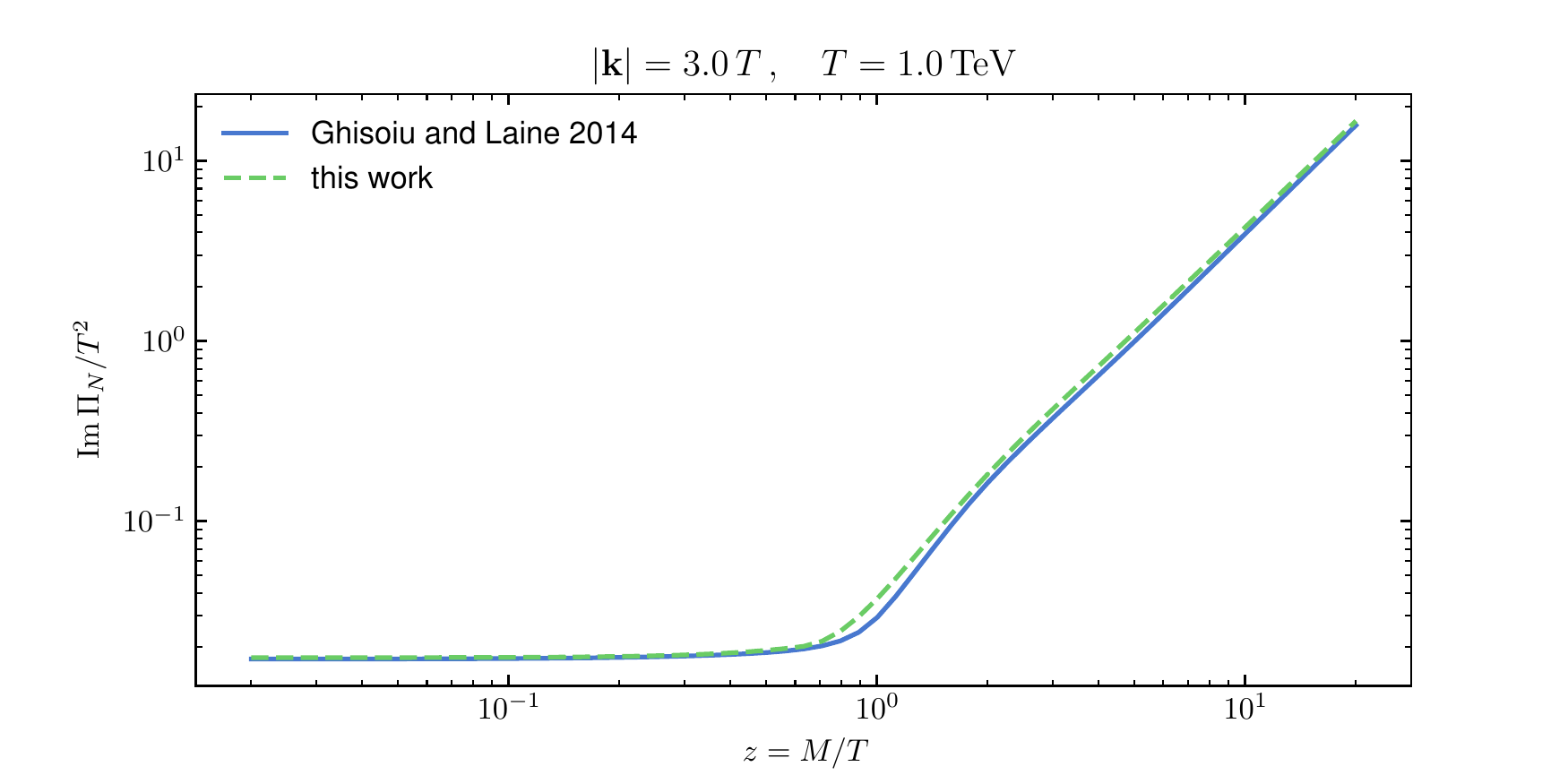}
	\caption{Comparison of the helicity-averaged HNL production rate from this work (green, dot-dashed) with a previous calculation from~\cite{Ghisoiu:2014ena}.
		In spite of using an extrapolation, we manage to reproduce the main features of the full calculation.
	}
	\label{fig:rate}
\end{figure}

\subsection{Production rates after electroweak symmetry breaking}
Following Ref.~\cite{Ghiglieri:2016xye}, we can identify two classes of processes that lead to the production of the heavy neutrinos.
\textit{Direct} processes, where the heavy neutrino interacts with Higgs and lepton doublets, as well as \textit{indirect} processes in which the heavy neutrino interacts through the mixing with the light neutrinos.

One of the biggest differences compared to the symmetric phase arises in the decay of the heavy neutrino, as the Higgs doublet is replaced by a real scalar field. At the same time, decay channels into the $Z$ and $W^\pm$ bosons open up.
There is a gauge dependence in how we separate the direct and indirect processes, but the total rate in their sum remains gauge invariant.
This gauge invariance can be lost if we do not keep all the direct and indirect diagrams at the same order in perturbation theory.
As the resonant mixing between the heavy and light neutrinos is best described by resumming the lepton propagators, we implicitly include diagrams that are not matched by the direct rate, and therefore introduce gauge dependence to our estimate of the heavy neutrino production rate.
We resolve this issue by choosing a specific gauge, in this case the 't Hooft-Feynman gauge, due to the fact that it does not introduce any new scales.

\subsubsection{Direct Heavy Neutrino Production}
In this section we perform a simple estimate of the direct heavy neutrino production rate. As discussed at the beginning of this section, there are various contributions to the self-energy of the heavy neutrinos.

\begin{figure}
	\begin{align*}
		\begin{gathered}
			\begin{fmfgraph*}(55,34)
				\fmfleft{i1} \fmfright{o1}
				\fmftop{c1}
				\fmfbottom{c2}
				\fmf{fermion,width=thick,label.side=left,tension=2}{i1,v1}
				\fmf{plain,left=1,tag=1}{v1,v2}
				\fmf{dashes,right=1,tag=2}{v1,v2}
				\fmf{fermion,width=thick,tension=2}{v2,o1}
				\fmfposition
				\fmfipath{p[]}
				\fmfiset{p1}{vpath1(__v1,__v2)}
				\fmfiset{p2}{vpath2(__v2,__v1)}
				\fmfiv{label=$\nu$}{point length(p1)/3 of p1}
				\fmfiv{label=$h$}{point length(p2)*2/3 of p2}
				\fmfiv{}{point length(p1)/2 of p1}
				\fmfiv{}{point length(p2)/2 of p2}
				\fmf{plain,foreground=red}{c1,c2}
			\end{fmfgraph*}
		\end{gathered}+
		\begin{gathered}
			\begin{fmfgraph*}(55,34)
				\fmfleft{i1} \fmfright{o1}
				\fmftop{c1}
				\fmfbottom{c2}
				\fmf{fermion,width=thick,label.side=left,tension=2}{i1,v1}
				\fmf{plain,left=1,tag=1}{v1,v2}
				\fmf{dashes,right=1,tag=2}{v1,v2}
				\fmf{fermion,width=thick,tension=2}{v2,o1}
				\fmfposition
				\fmfipath{p[]}
				\fmfiset{p1}{vpath1(__v1,__v2)}
				\fmfiset{p2}{vpath2(__v2,__v1)}
				\fmfiv{label=$\nu$}{point length(p1)/3 of p1}
				\fmfiv{label=$z$}{point length(p2)*2/3 of p2}
				\fmfiv{}{point length(p1)/2 of p1}
				\fmfiv{}{point length(p2)/2 of p2}
				\fmf{plain,foreground=red}{c1,c2}
			\end{fmfgraph*}
		\end{gathered}+
		\begin{gathered}
			\begin{fmfgraph*}(55,34)
				\fmfleft{i1} \fmfright{o1}
				\fmftop{c1}
				\fmfbottom{c2}
				\fmf{fermion,width=thick,label.side=left,tension=2}{i1,v1}
				\fmf{plain,left=1,tag=1}{v1,v2}
				\fmf{dashes,right=1,tag=2}{v1,v2}
				\fmf{fermion,width=thick,tension=2}{v2,o1}
				\fmfposition
				\fmfipath{p[]}
				\fmfiset{p1}{vpath1(__v1,__v2)}
				\fmfiset{p2}{vpath2(__v2,__v1)}
				\fmfiv{label=$e^\mp$}{point length(p1)/3 of p1}
				\fmfiv{label=$w^\pm$}{point length(p2)*2/3 of p2}
				\fmfiv{}{point length(p1)/2 of p1}
				\fmfiv{}{point length(p2)/2 of p2}
				\fmf{plain,foreground=red}{c1,c2}
			\end{fmfgraph*}
		\end{gathered}
	\end{align*}
	\caption{Direct heavy neutrino production in the broken phase of the electroweak theory. The dashed lines represent the Higgs field $h$ and the goldstone modes corresponding to the $W^\pm$ and $Z$ boson. The light (heavy) neutrinos are represented by the thin (thick) solid lines.}
	\label{fig:direct_rate}
\end{figure}
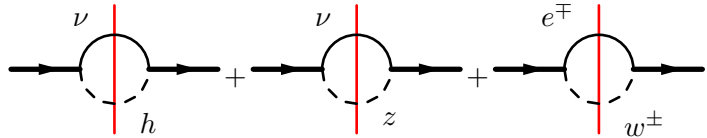

The situation slightly changes in the broken phase of the standard model. In the 't Hooft-Feynman gauge, the goldstone bosons also contribute to the heavy neutrino production rate.
We can approximate this rate by the same $1\leftrightarrow 2$ integrals from the previous section
however, with $m_\phi$ replaced by the mass of the appropriate goldstone boson (c.f. Fig.~\ref{fig:direct_rate}) i.e. the gauge boson to which the goldstone mode corresponds
\begin{align}
	\hat{\Sigma}_N^{(1\leftrightarrow2)} =
	\frac{1}{2 g_w} \hat{\Sigma}_N^{(1\leftrightarrow2)} \Biggr|_{m_\phi=m_H} +
	\frac{1}{2 g_w} \hat{\Sigma}_N^{(1\leftrightarrow2)} \Biggr|_{m_\phi \rightarrow m_Z} +
	\frac{1}{g_w} \hat{\Sigma}_N^{(1\leftrightarrow2)} \Biggr|_{m_\phi \rightarrow m_W}\,.
\end{align}
The individual rates were divided by factors of $2 g_w$, to take into account that only one isospin runs in the loop, as well as the $1/\sqrt{2}$ factor that appears in the coupling to the field $h$ compared to $\phi$.

Despite the fact that our calculation of the $1\leftrightarrow2$ rate misses the $1+n\leftrightarrow 2+n$ processes which are included in the LPM resummation, our results---after averaging over helicities---show a nice agreement with the full helicity averaged computation, see figure~\ref{fig:rate}.

The remaining direct processes, such as $2\leftrightarrow2$ scatterings can also be included in such diagrams if we replace the tree-level spectral functions from~\eqref{2ptf:3lvl} by their resummed counterparts.

\paragraph{Production in decays and inverse decays}
The heavy neutrinos can be produced in the decays of the Higgs particle. During the electroweak crossover, the heavy neutrino mixes with the light neutrinos, and can therefore also be interact with the $W$ and $Z$ bosons.
In the Unitary gauge, this mixing would appear only as a indirect contribution.
We use the 't Hooft-Feynman gauge, where the longitudinal modes of the massive gauge bosons manifest themselves as the goldstone modes $w$, $z$ and $q$.

An accurate estimate of the non-perturbative contribution to masses of the Higgs and vector bosons typically requires a lattice calculation~\cite{Kajantie:1996mn,DOnofrio:2015gop}. To ensure consistency with the rate calculation from~\cite{Ghiglieri:2018wbs}, we follow their approach and rely on a 1-loop calculation.

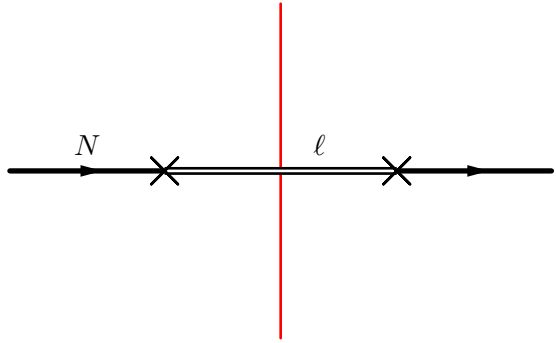
\begin{figure}
	\begin{align*}
		\begin{gathered}
			\begin{fmfgraph*}(144,89)
				\fmfleft{i1} \fmfright{o1}
				\fmftop{c1}
				\fmfbottom{c2}
				\fmf{fermion,label=$N$,width=thick,label.side=left,tension=1.5}{i1,v1}
				\fmf{dbl_plain,tag=1}{v1,v2}
				\fmf{fermion,width=thick,tension=1.5}{v2,o1}
				\fmfposition
				\fmfipath{p[]}
				\fmfiset{p1}{vpath1(__v1,__v2)}
				\fmfiv{label.angle=90,label.distance=1,label=$\ell$}{point length(p1)*2/3 of p1}
				\fmf{plain,foreground=red}{c1,c2}
				\fmfv{d.sh=cross}{v1,v2}
			\end{fmfgraph*}
		\end{gathered}
	\end{align*}
	\caption{Indirect heavy neutrino production. The double lines represent the resummed propagator for lepton doublet.}
	\label{fig:indirect_rate}
\end{figure}

\subsubsection{Indirect Heavy Neutrino Production}

At tree level the sum of the direct and indirect rates can be shown to be gauge invariant. However, if we resum the propagator of the doublet leptons, we implicitly include a whole set of gauge-dependent diagrams, which do not have their match in the tree-level diagrams of the direct processes.
The gauge dependence of resummed propagators is well documented in the literature (see e.g.~\cite{Carrington:2003ut}), and can be alleviated by using e.g.\ higher order effective actions.
Note that in $R_\xi$ gauges the mass of the goldstones is given as $\xi \times m_\mathrm{gb}$, where $m_\mathrm{gb}$ is the mass of the corresponding gauge boson.
Therefore, if we choose $\xi=1$, i.e.\ the 't Hooft-Feynman gauge, we avoid introducing new scales into the problem.

\begin{figure}
	\centering
	\includegraphics[width=0.9\textwidth]{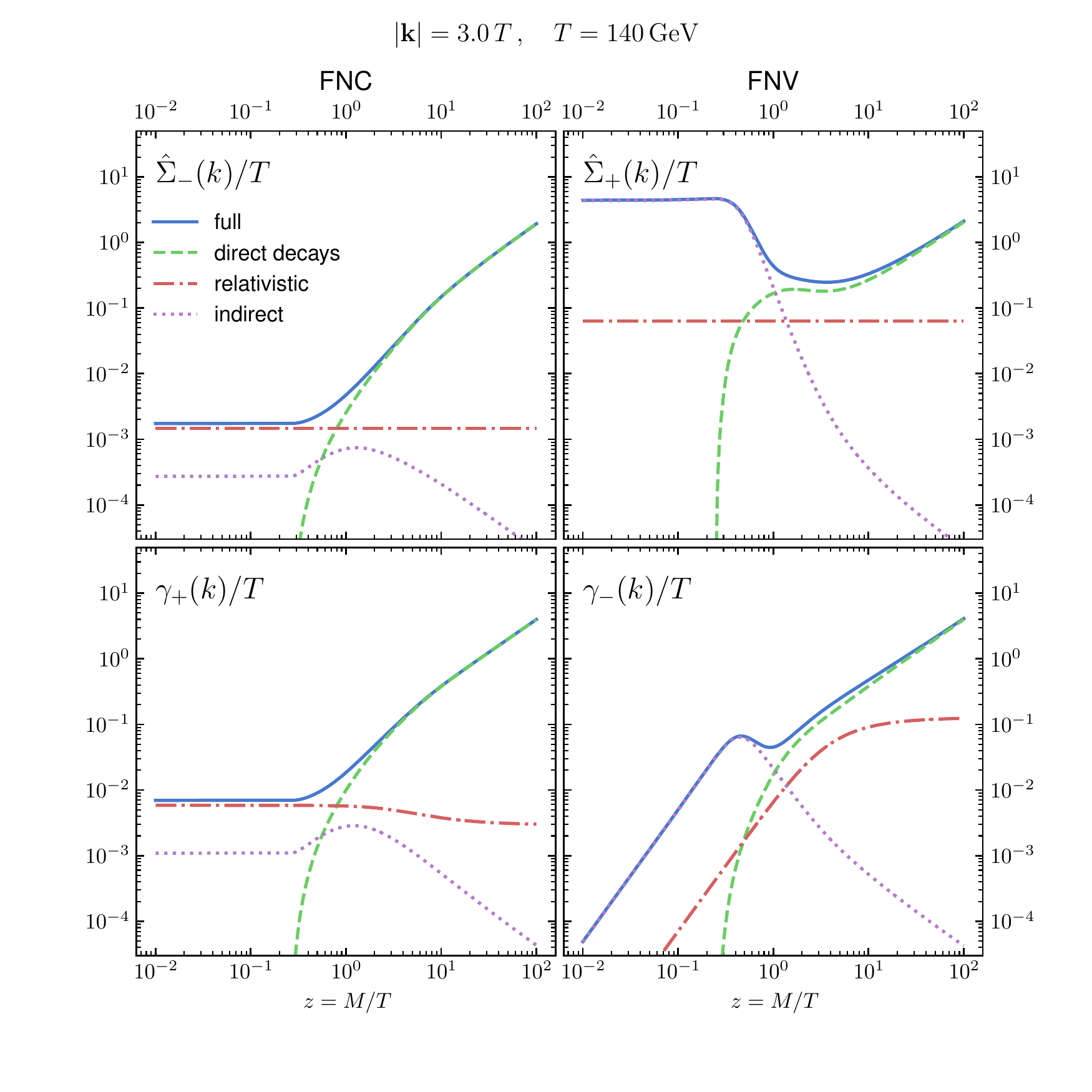}
	\caption{The FNV and FNC rates in the broken phase of the SM.
		The full rate (blue, full) is a sum of the relativistic direct contribution (red, dot-dashed), the direct $N$ decay (green, dashed),
		and the indirect contribution from the HNL mixing with the active neutrinos (purple, dotted).
		For completeness we show both the self-energies (upper panels) and the interaction rates (lower panels).
	}
	\label{fig:sigmapmAppBroken}
\end{figure}

The resummed lepton propagators are given by (see e.g.~\cite{Garbrecht:2008cb}):
\begin{align}
	S^\mathcal{A}_\ell =
	P_L \frac{
		2 (\slashed{k} - \slashed{\Sigma}^H_\ell) \Sigma^\mathcal{A}_\ell \cdot (k - \Sigma^H_\ell) -
		\slashed{\Sigma}^\mathcal{A}_\ell (\slashed{k} - \slashed{\Sigma}^H_\ell)^2
		+ {\slashed{\Sigma}^\mathcal{A}_\ell}^3
	}{
		\left[ (\slashed{k} - \slashed{\Sigma}^H_\ell)^2 - {\Sigma^\mathcal{A}_\ell}^2 \right]^2 +
		4 \left[ \Sigma^\mathcal{A}_\ell \cdot (k - \Sigma^H_\ell) \right]^2
	} P_R\,,
	\label{Sl:spectral}
	\\
	S^H_\ell =
	P_L \frac{
		2 \slashed{\Sigma}^\mathcal{A}_\ell \Sigma^\mathcal{A}_\ell \cdot (k - \Sigma^H_\ell)+
		(\slashed{k} - \slashed{\Sigma}^H_\ell)  \left[ (k - \Sigma^H_\ell)^2 - {\Sigma^\mathcal{A}_\ell}^2 \right]
	}{
		\left[ (\slashed{k} - \slashed{\Sigma}^H_\ell)^2 - {\Sigma^\mathcal{A}_\ell}^2 \right]^2 +
		4 \left[ \Sigma^\mathcal{A}_\ell \cdot (k - \Sigma^H_\ell) \right]^2
	} P_R\,,
	\label{Sl:hermitian}
\end{align}
where the superscripts $H$ and $\mathcal{A}$ stand for the hermitian (dispersive) and antihermitian (dissipative) parts of the lepton self-energy.
These expressions significantly simplify if we use light-cone coordinates, where the denominator factorizes, which effectively gives us two propagators, one for the particles, and the other for the holes in the plasma. For the retarded propagator this gives us:
\begin{align}
	S^R_\ell &= P_L
	\left( \frac{\gamma_-}{k_+ - \Sigma_+^R} + \frac{\gamma_+}{k_- - \Sigma_-^R} \right) P_R
	\label{prop:ParticleHole}
\end{align}
where the indices $\pm$ indicate the light-cone coordinates $a_\pm = (a_0 \pm a_i k_i/|k|)$.
The indirect FNV and FNC rates are then given by~\cite{Ghiglieri:2017gjz,Eijima:2017anv}:
\begin{align}
	\gamma^\mathrm{ID}_\pm = g_w \frac{k_\mp \hat{\Sigma}_\pm^\mathrm{ID}}{|k_0|} =
	v^2 \frac{k_\pm \Sigma_{\ell \pm}^\mathcal{A}}{	{\Sigma_{\ell \pm }^\mathcal{A}}^2 + ( k_\pm - \Sigma_{\ell \pm}^H )^2}\,,
	\label{gammaID}
\end{align}
where $\Sigma_\ell$ are the self-energies of the neutrinos.
Interestingly, for large HNL masses, these self-energies have to be evaluated on the HNL mass shell,
which leads to further uncertainties, as the active neutrino interaction rates have only been calculated for relativistic neutrinos.
To estimate this rate we use the same approach as for the rates in the symmetric phase, namely, we add the $1\rightarrow 2$ contribution to the relativistic estimate
when the HNL mass exceeds the decay thresholds.
Fortunately, any such uncertainties are not important for larger HNL masses,
since the off-shell active neutrino width does not grow faster than the HNL mass in the denominator itself, and the indirect contribution becomes negligible for $M\gg T$,
as shown in Fig.~\ref{fig:sigmapmAppBroken}.

By combining the estimates of the rates we have a consistent set of rates in both the broken and symmetric phases for any HNL mass.
In spite of the uncertainties related to our extrapolation, our rates remain consistent with the spin-averaged rates from the literature as shown in Fig.~\ref{fig:rate}.

\section{The parameter space of leptogenesis}
\label{sec:production_of_the_baryon_asymmetry}

\subsection{Baryogenesis through freeze-in and freeze-out}
As we have seen in the previous chapters, the same equations can be used to describe both leptogenesis through neutrino oscillations, and resonant leptogenesis.
Because there is no clear cut between the two mechanisms, we may instead look at whether the majority of the asymmetry is produced during freeze-in or freeze-out of the heavy neutrinos.

For a clear separation between the two regimes, we perform three parameter scans:
\begin{enumerate}
	\item Parameter scan with vanishing initial abundance of the heavy neutrinos --- \emph{both freeze-in and freeze-out contribute} to the BAU generation.
	\item Parameter scan with thermal initial conditions --- \emph{only freeze-out contributes}. Indeed, by freeze-in we understand the period during which HNLs reach the equilibrium. By artificially setting the thermal initial conditions we eliminate this period.
	\item Parameter scan with vanishing initial abundance of the heavy neutrinos and without the deviation from equilibrium caused by the expansion of the Universe --- \emph{only freeze-in contributes}. Technically this is achieved by putting to zero the source term $d \rho_N^\mathrm{eq}/ dt$ from Eqs.~\eqref{KE_1a}.
\end{enumerate}

Because of the approximate linearity of the evolution equations, the three parameter scans are not independent, and the baryon asymmetry from both freeze-in and freeze-out can be obtained by summing the other two parameter scans.
However, to avoid numerical errors due to cancellations of large numbers, we perform all three parameter scans independently.

\subsection{The range of allowed masses and mixing angles}
\label{subsec:paramspace}

The parameter space of leptogenesis is quite large, as we have discussed in section \ref{sub:parametrization_of_the_yukawa_couplings}, in total we have six unknown parameters. These parameters are: the average mass $M$; the mass splitting $\Delta M$; one Dirac and one Majorana phases of the PMNS matrix; the real and imaginary parts of the complex angle $\omega$.
Instead of constraining these parameters directly, it is instructive to see how they are related to the experimentally observable quantities. Among these the most relevant  are the masses of the heavy neutrinos and their mixing angles.
This is particularly important for direct searches as the size of the mixing angle determines the number of heavy neutrinos that can be produced.

The condition of reproducing both the baryon asymmetry of the Universe and the light neutrino masses imposes constraints on the allowed mixing angles of the heavy neutrinos.
The lower bound on the \emph{total} mixing angle $U^2$ is mainly coming from the requirement of reproducing the light neutrino masses, while the upper bound comes from leptogenesis.

The mixing angle of the heavy neutrinos is tied to the size of their Yukawa couplings, see Eq.~\eqref{eq:theta}. For large values of the Yukawa couplings, the mixing angle itself will be large. The main issue for leptogenesis is that a large value of Yukawa couplings at the same time leads to a large washout strength.

To perform the study of the parameter space we choose several specific benchmark points which extremize the ratios of the Yukawa couplings
\begin{align}
	\frac{\sum_i |F_{\alpha i}|^2}{\sum_{i \beta} |F_{\beta i}|^2}\,.
\end{align}
It is interesting that these points correspond to special values of the PMNS phases,
with\footnote{We should note however, that these are only effective low-energy parameters, which do not quantify the full amount of $CP$ violation in the theory.}
\begin{align}
	\delta = n \pi/2\,, && \eta = m \pi/2\,, && \Re \omega = l \pi/4 \quad \text{with } m,n,l \in \mathbb{Z}\,.
\end{align}
The same phases have been show to maximize the total mixing $U^2$ in Ref.~\cite{Eijima:2018qke,Drewes:2016jae}.
These choices of parameters are particularly favorable for leptogenesis as they allow for a hierarchy in the washout strengths. This means that even in the presence of a large overall washout parameter, the asymmetry may survive hidden in a particular lepton flavor.

\begin{figure}[h]
	\centering
	\includegraphics[width=0.45\textwidth]{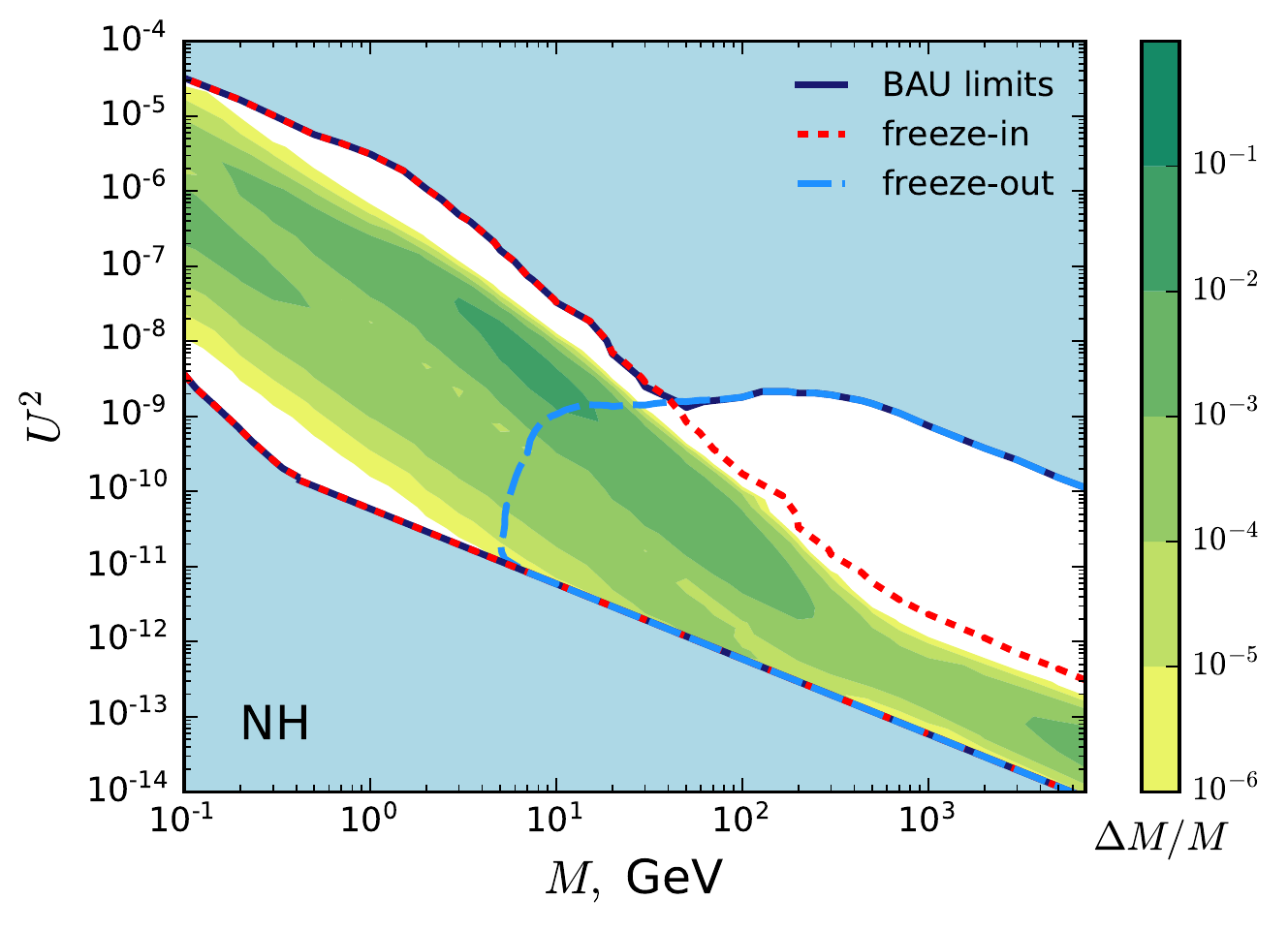}
	\includegraphics[width=0.45\textwidth]{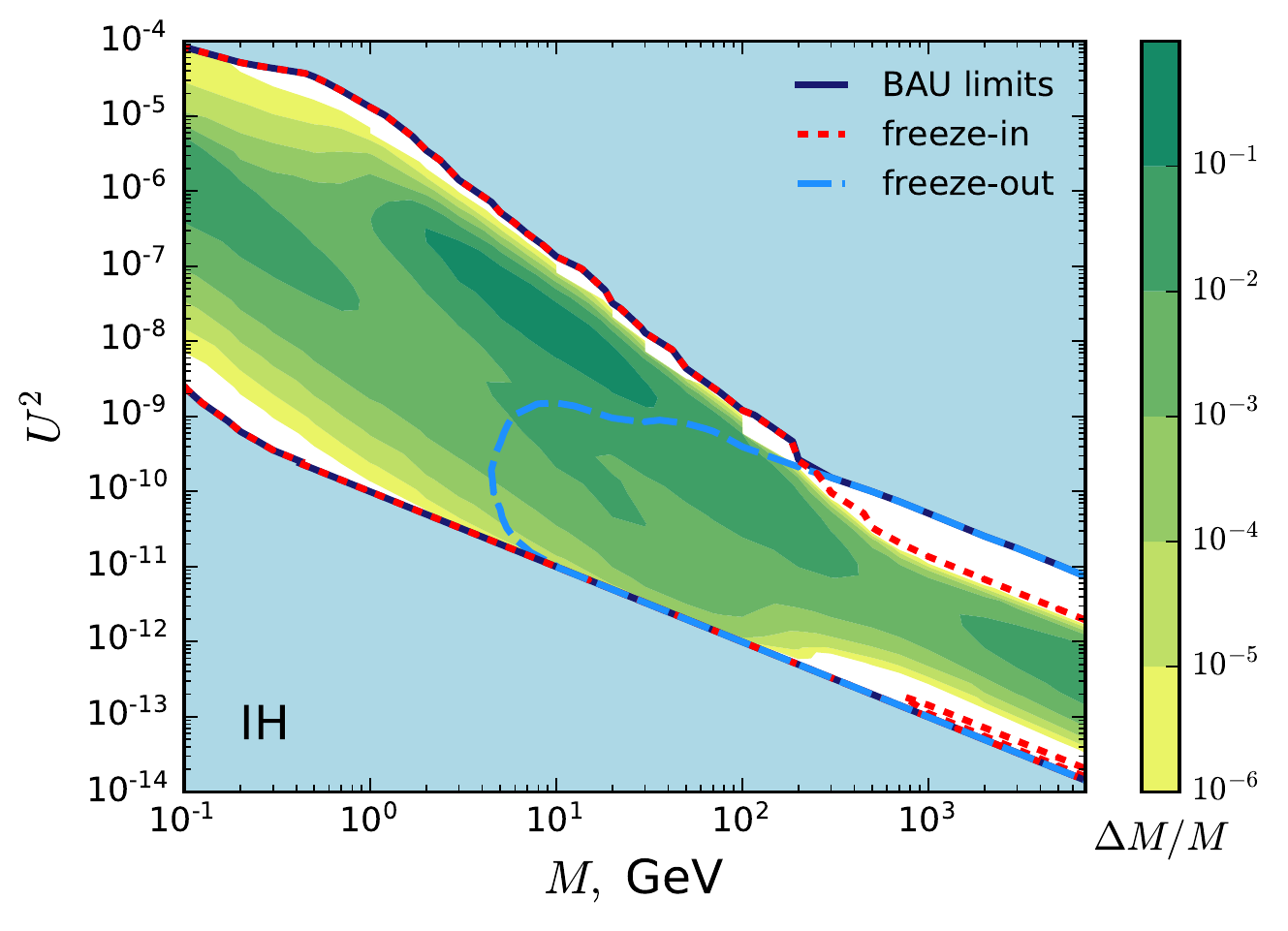}
	\caption{The range of the total mixing angle $U^2$ consistent with both the seesaw mechanism and leptogenesis as a function of HNLs' mass $M_N$. The black solid lines show the results obtained with the full kinetic equations and vanishing initial conditions for HNLs. The blue, dashed lines correspond to thermal initial conditions. In this regime the freeze-in does not contribute to the asymmetry generation. The red, dotted line corresponds to neglecting the effect of the expansion of the Universe on the distribution of the heavy neutrinos. In this case the freeze-out cannot contribute and the asymmetry is generated during freeze-in.
		The color contours represent the largest allowed value of the mass splitting $\Delta M /M$. Within the white regions the mass splitting is smaller than $10^{-6}$.
	The left (right) panel shows the case of normal (inverted) hierarchy.}
	\label{fig:MU2}
\end{figure}

Once the phases $\delta, \eta,$ and $\Re \omega$ are fixed, we scan over the remaining parameters. The details of the numerical procedure are specified in~\ref{App:Numerical}.
We limit the HNL mass to the range $[0.1, \;7\times 10^3]$~GeV.
The results are shown in fig.~\ref{fig:MU2}.
As one can see from fig.~\ref{fig:MU2}, the leptogenesis is efficient for the whole range of masses which we have considered.
Moreover, it extends beyond the considered mass range. In particular, as we show in section~\ref{sub:approximate_scaling_between_the_tev_scale_and_the_davidson_ibarra_bound}, there exists a simple scaling which allows to extend our results all the way till the Davidson-Ibarra bound.

\subsection{Constraints on the heavy neutrino mass splitting}
\label{sub:constraints_on_the_heavy_neutrino_mass_splitting}

The mass splitting between the heavy neutrinos is one of the most important parameters for both leptogenesis scenarios.
The main reason why leptogenesis is so sensitive to the mass splitting $\Delta M$ between the heavy neutrinos is that this parameter sets the scale for the oscillations that violate $CP$ and lead to a lepton asymmetry.
The temperature corresponding to the onset of oscillations depends on the Hubble rate and is given as~\cite{Asaka:2005pn}
\begin{align}
	T_\mathrm{osc} \approx (M_0 M \Delta M)^{1/3}\quad \text{if} \; T_\mathrm{osc}\gg M_N,
\end{align}
where $M_0=\sqrt{90 /\left(8 \pi^{3} g_{*}\right)} M_{\mathrm{Pl}}$ and $ g_{*}$ is the effective number of relativistic degrees of freedom.
For heavier neutrinos, it is possible that the oscillations begin when they are already non-relativistic, which gives us a different temperature since the typical HNL energy is $M$ instead of $T$
\begin{align}
	T_\mathrm{osc} \approx \left(M_0 \Delta M\right)^{1/2}\quad \text{if} \; T_\mathrm{osc} \lesssim M_N.
\end{align}
This oscillation temperature scale can be lowered once we take the Yukawa-induced thermal masses into account, as the thermal masses are aligned with the HNL interaction basis, and do not lead to oscillations themselves.
To understand the parameter space it is instructive to consider ``slices'' where we vary the mass splitting and the magnitude of the Yukawa couplings (\emph{i.e.} the parameter $\Im \omega$),
and keep the remaining parameters fixed.

\begin{figure}
	\centering
	\includegraphics[width=0.4\textwidth]{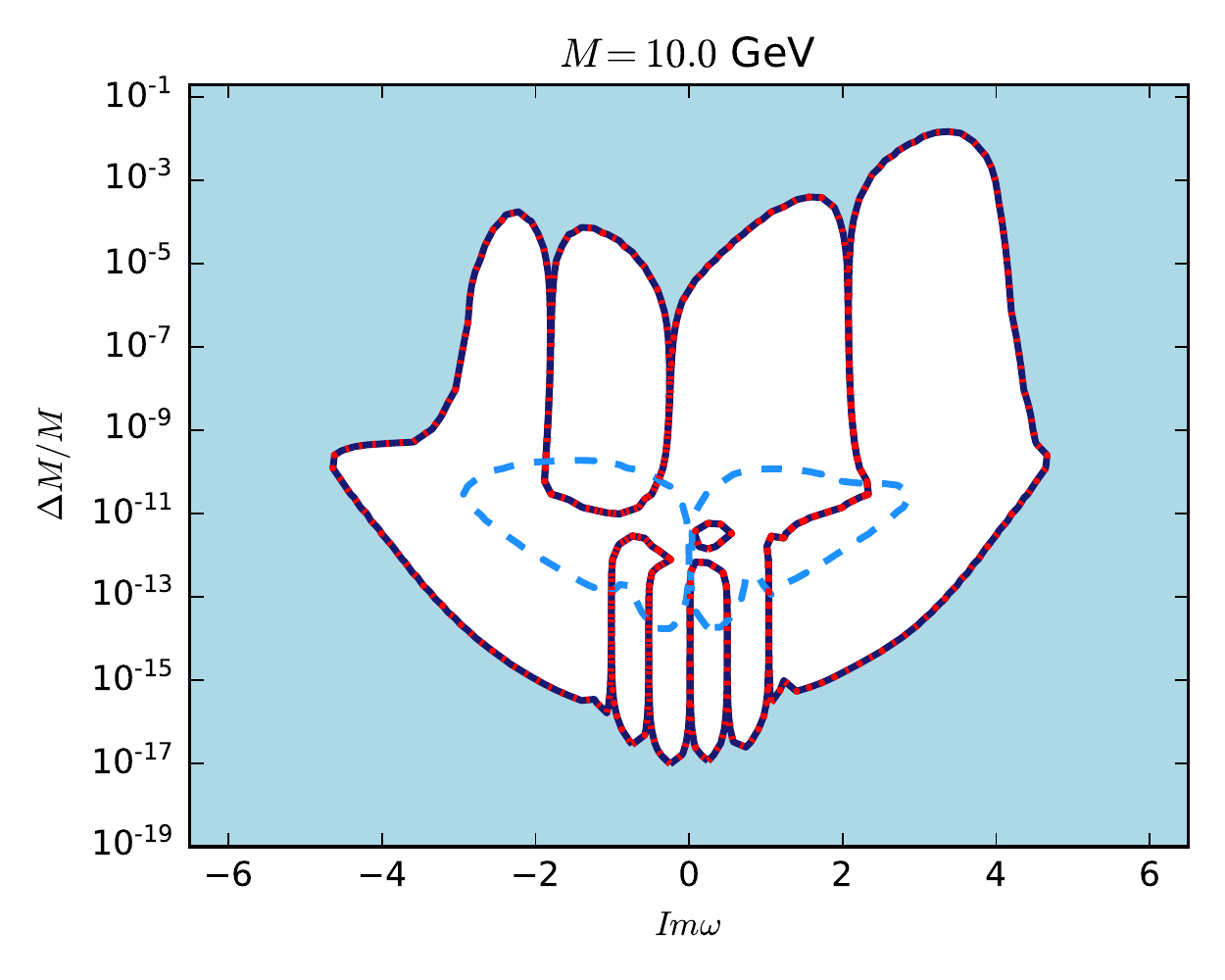} \includegraphics[width=0.4\textwidth]{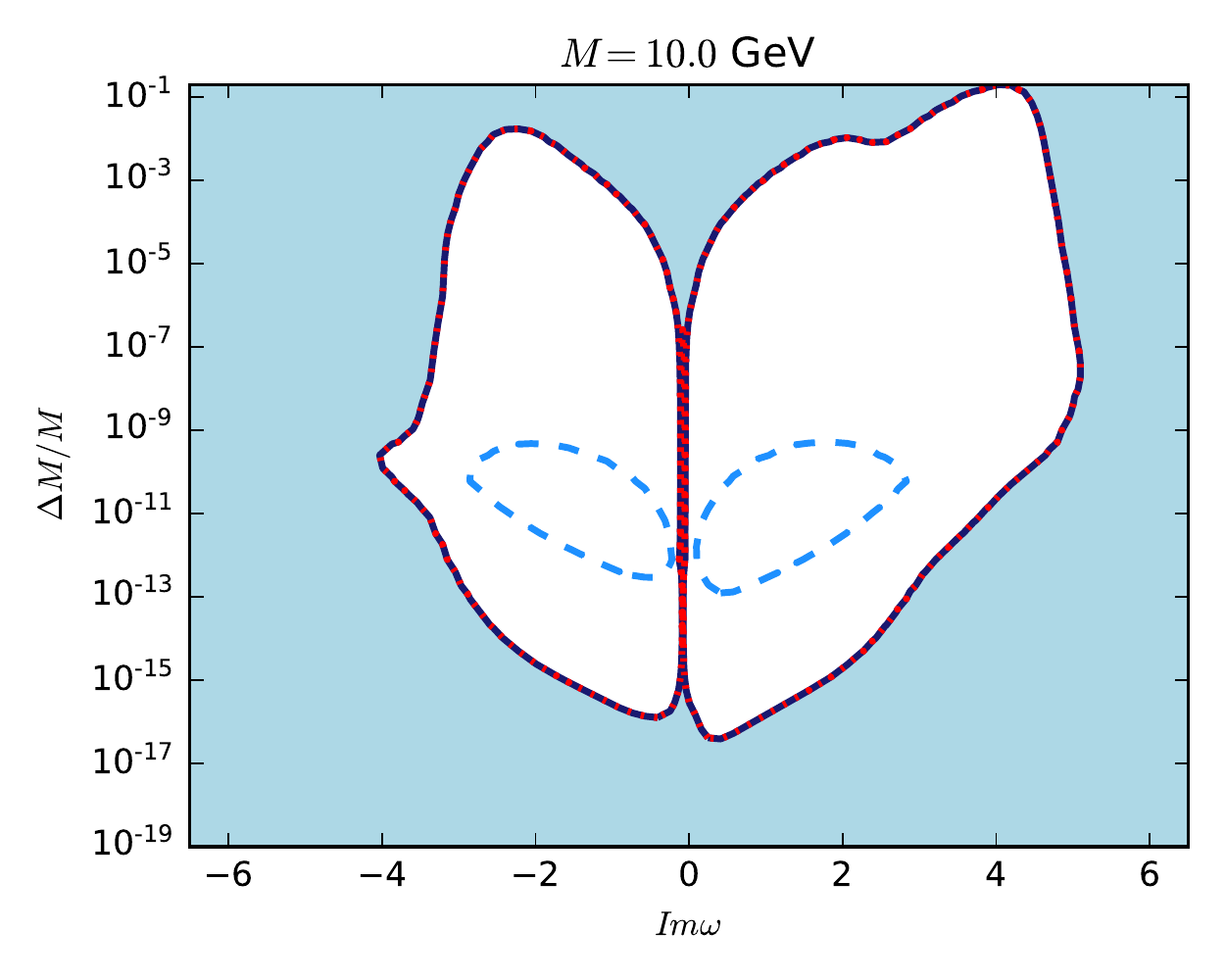}\\
	\includegraphics[width=0.4\textwidth]{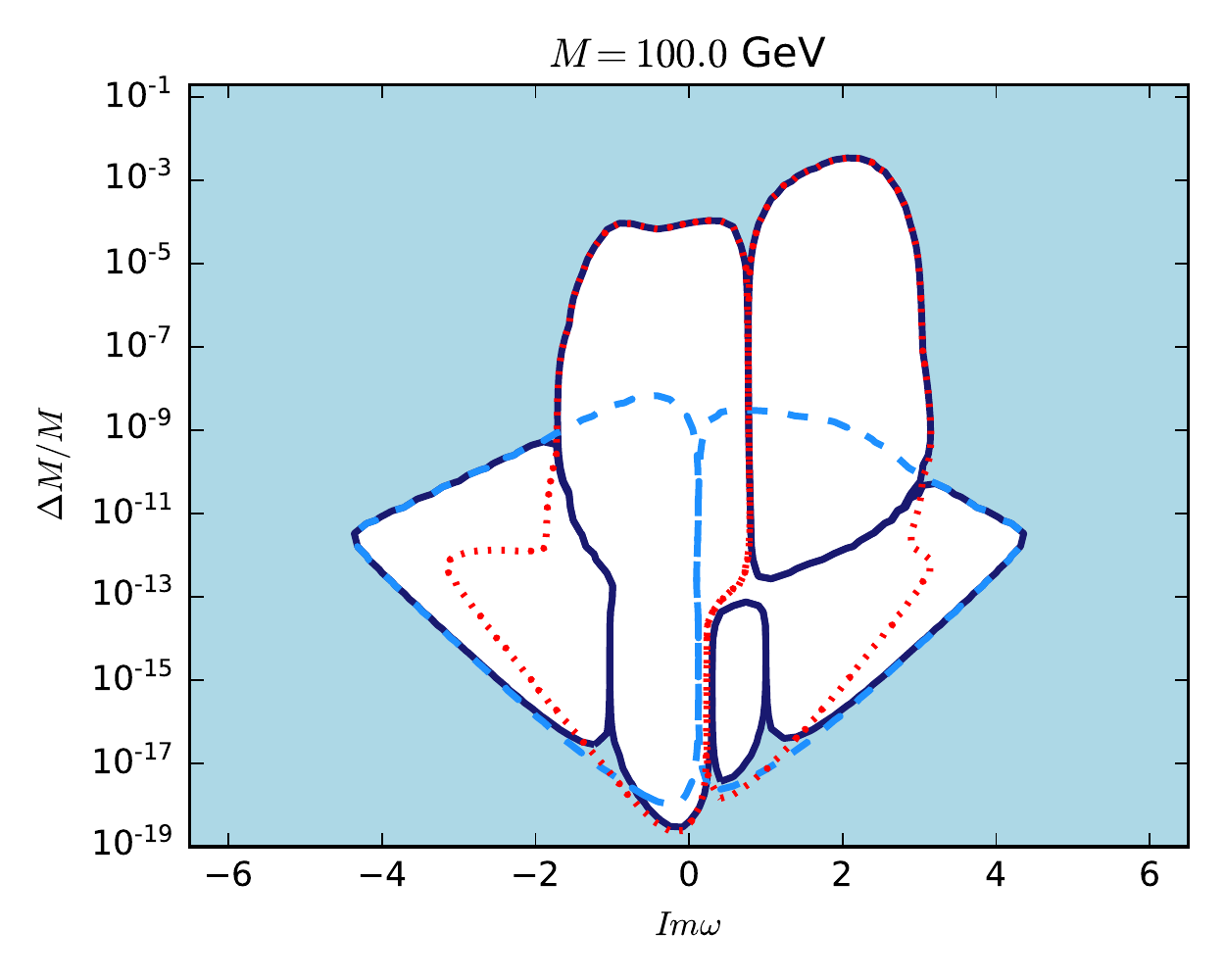} \includegraphics[width=0.4\textwidth]{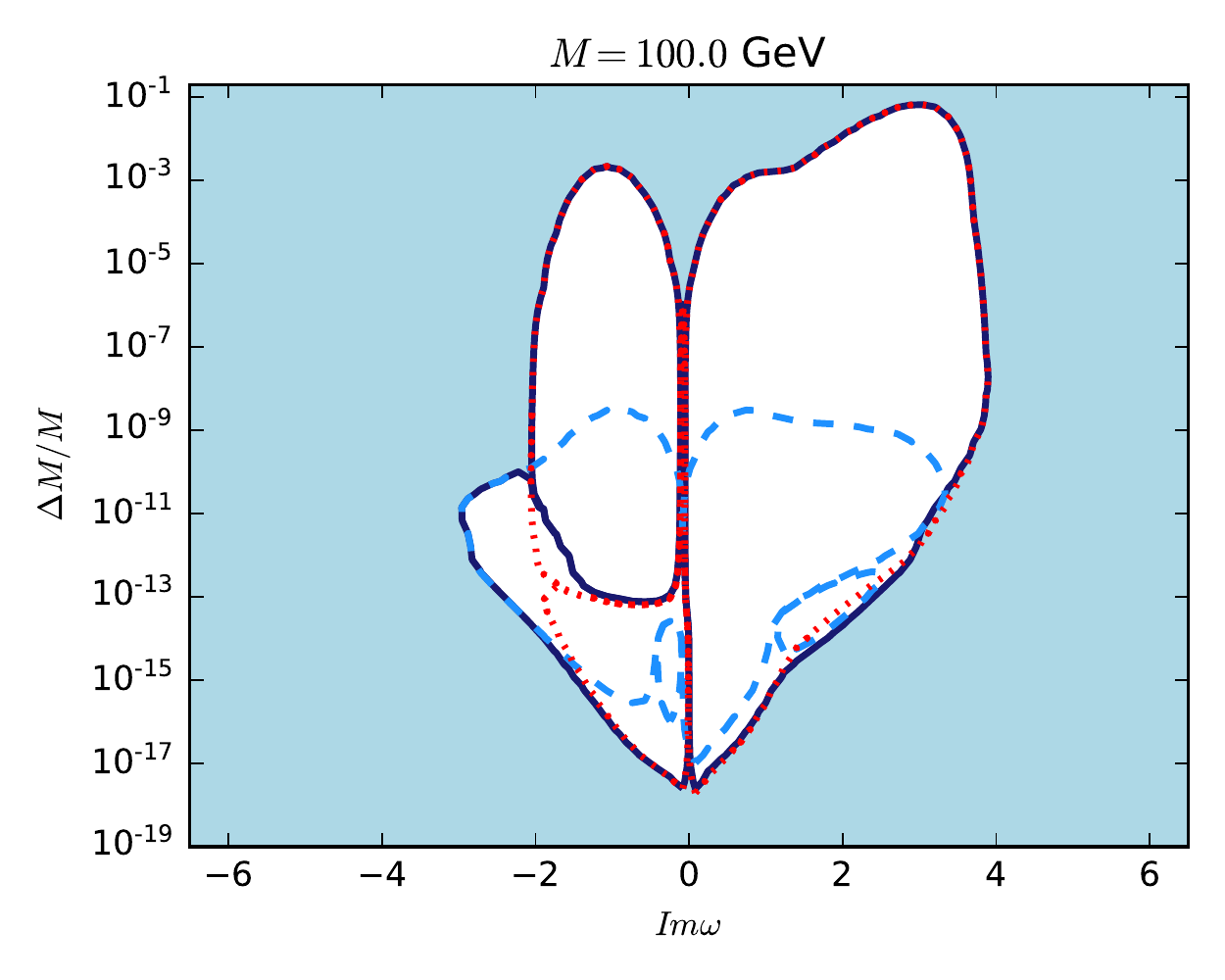}\\
	\includegraphics[width=0.4\textwidth]{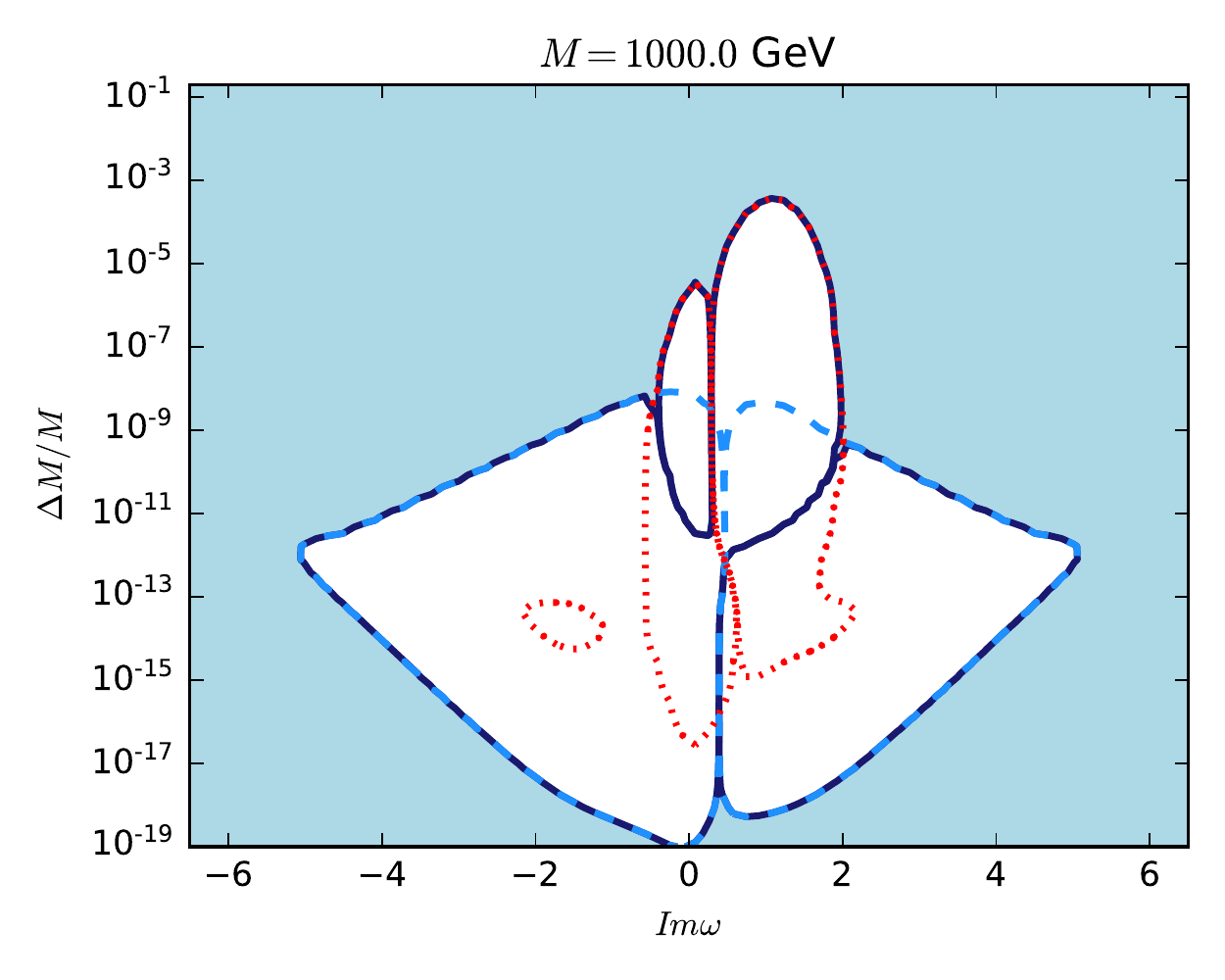} \includegraphics[width=0.4\textwidth]{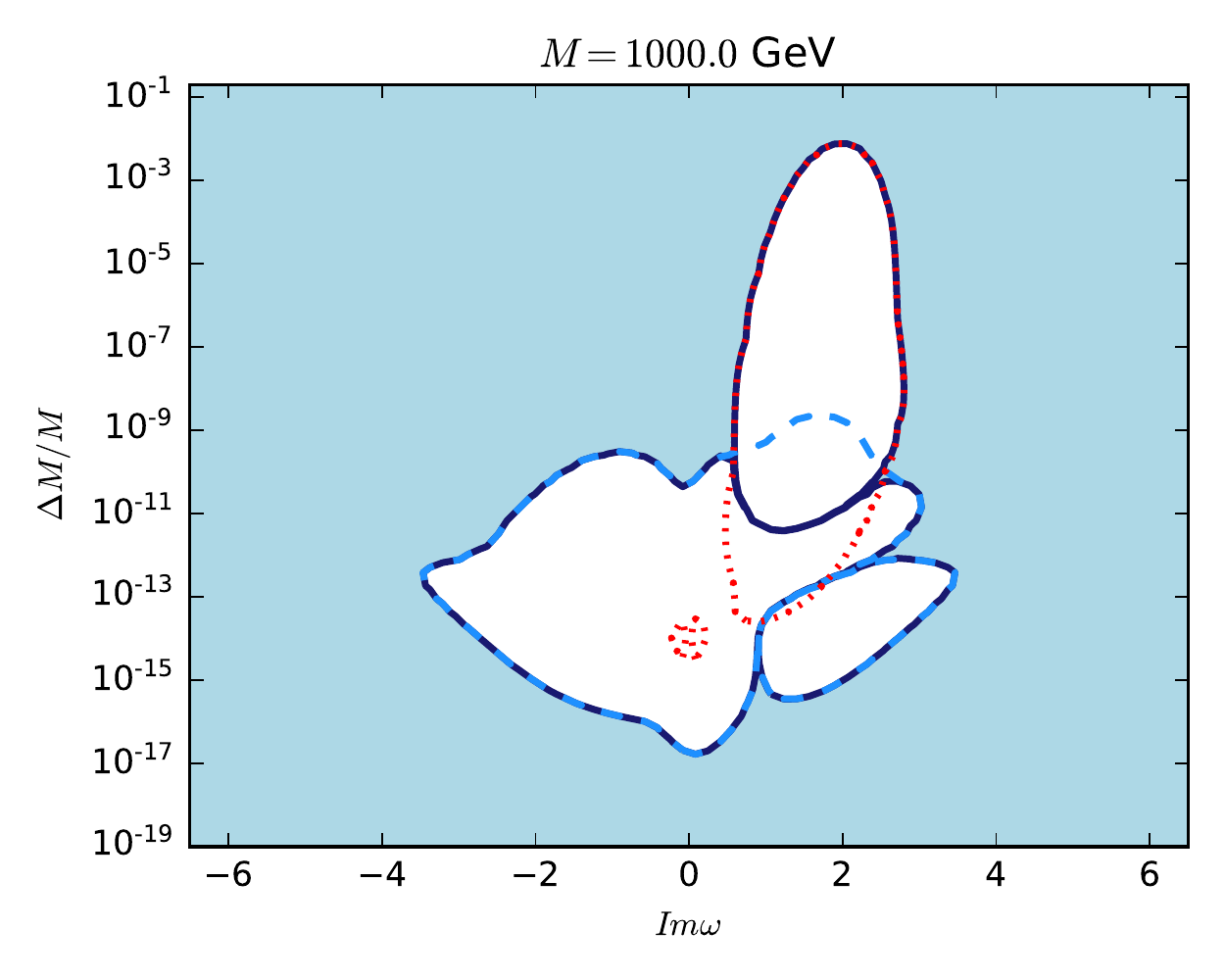}
	\caption{The allowed range of mass splittings and mixing angles for fixed phases.
		Left column: normal hierarchy, $\Re \omega = \pi/4$, \;$\delta = \pi$, \;$\eta = 3\pi/2$. Right column: inverted hierarchy, ${\Re \omega = \pi/4}$,\;$\delta = 0$,\; $\eta = \pi/2$.
		The different contours correspond to different initial conditions for the heavy neutrinos. The area inside the regions corresponds to a BAU greater than the observed asymmetry.
		The (dark blue, full) curve includes both the contributions from freeze-in and freeze-out.
		The (light blue, dashed) curve corresponds to freeze-out only and the (red, dotted) curve corresponds to baryogenesis via freeze-in.
		It is interesting that the largest mass splitting is realized exactly during freeze-in, as the HNL oscillations happen at high temperatures, and therefore before the HNLs begin to decay.
	}
	\label{fig:dMiOmega}
\end{figure}

The results of these parameter scans are shown in figure~\ref{fig:dMiOmega}. For each mass we show three lines, one corresponding to leptogenesis via freeze-in, one to leptogenesis via freeze-out, and one containing both contributions. This helps us better understand the transition between the two regimes.

\subsection{Main reasons why freeze-in and freeze-out leptogenesis parameter spaces are connected}

For heavier HNLs which start to decay before the sphaleron freeze-out, two opposite effects
take place. Their decays are driving the system towards equilibrium, whereas the expansion of the Universe driving it out of equilibrium.
For heavy enough HNLs, decay channels into $W$ and $Z$ bosons open up, which leads to an enhancement of the total heavy neutrino equilibration rate.
This increased washout can in principle erase all asymmetries produced during the freeze-in and oscillations of the heavy neutrinos~\cite{Blondel:2014bra}.
However, the lepton number washout does not remain large indefinitely, as it also depends on the equilibrium density of the heavy neutrinos.
This gives us the usual Boltzmann suppression factor~\cite{Buchmuller:2004nz} for the lepton number washout
\begin{align}
	\bar{\Gamma}_\nu = \int \frac{d^3 k}{(2\pi)^3} \Gamma f_N^\mathrm{eq} (1-f_N^\mathrm{eq}) \sim (F F^\dagger)_{\alpha \alpha} M \exp(-M / T)\quad \text{for}\; M \gtrsim T\,.
\end{align}
The lepton number washout rate therefore reaches its maximum when $T\sim M$, and the lepton number typically freezes out when $M/T \sim 10$~\cite{Buchmuller:2004nz}.
To estimate the strength of the lepton number washout, we compute the rate close to its maximum, at $T=M$, and compare it to the Hubble rate $\mathtt{H} = T^2/M_0$, $M_{0}=\sqrt{90 /\left(8 \pi^{3} g_{*}\right)} M_{\mathrm{Pl}}$.
This gives us the decay parameter~\cite{Buchmuller:2004nz}:
\begin{align}
	\label{washout:avg}
	\bar{K} = \frac{\Tr{\Gamma_N}}{2 \, \mathtt{H} } \Biggr|_{T=M} \approx
	\begin{cases}
		30 \cosh(2 \Im \omega)\, \text{for NH,}\\
		50 \cosh(2 \Im \omega)\, \text{for IH.}
	\end{cases}
\end{align}
which we average over the two (almost degenerate) heavy neutrinos $\bar{K}=(K_1+K_2)/2$.
The value of $\bar{K}$ indicates whether the washout is strong $\gg 1$ or weak $\ll 1$.
It is interesting that the washout rate exceeds the Hubble rate by at least a factor $\mathcal{O}(30)$,
which implies that the washout is already strong for small values of $\Im \omega$, and becomes even larger when $\Im \omega = \mathcal{O}(5)$.

At the same time, the expansion of the universe drives the heavy neutrinos out of equilibrium.
When the universe cools down to $T\sim M$, the heavy neutrinos are no longer relativistic, and they become over-abundant compared to their Boltzmann-suppressed equilibrium density.
This late-time deviation from equilibrium is exactly what leads to freeze-out leptogenesis.
It can typically be neglected in studies of leptogenesis with GeV-scale HNLs,
since it is suppressed by a factor $(M/T)^2$ for $M \lesssim T$, and only reaches its maximum when $M \approx T$.

The results from the parameter scan in Section~\ref{subsec:paramspace} imply that freeze-in and freeze-out leptogenesis parameter spaces have significant overlap:
freeze-in leptognesis remains viable for masses much higher than what one could have expected, and freeze-out leptogenesis is already possible for masses as light as $M\sim 5$~GeV.
We find that the main reason why freeze-in leptogenesis remains possible is the difference of the flavored washout compared to the average washout strength $\bar{K}$, which we further discuss in section~\ref{sss:wwffh}.
We show how freeze-out leptogenesis can be realized in a scenario with \si{GeV}-scale heavy neutrinos section~\ref{sss:fo-GeV}.
Finally, we discuss why large mixing angles do not necessarily imply a large lepton number washout in section~\ref{sss:cons_chg}.

\subsubsection{Freeze-in leptogenesis for TeV-scale HNLs}
\label{sss:wwffh}

For \si{\GeV}-scale heavy neutrinos the washout of the baryon number stops when the sphaleron processes freeze-out, i.e.\ at $T_\text{sph}\approx \SI{130}{\GeV}$.
The washout of the asymmetry in a particular lepton flavor does not only depend on the washout parameter $\bar{K}$, but also on how strongly a particular flavor couples to the heavy neutrinos.
In a realistic computation, we also need to take into account the so-called \textit{spectator} effects (see, e.g~\cite{Ghiglieri:2016xye}), which redistribute the asymmetries among the remaining SM species.
These effects are typically encoded in the susceptibility matrices $\omega_{\alpha \beta}$, as included in equation~\eqref{kin_eq_a}.

\begin{figure}
	\centering
	\includegraphics[width=0.75\textwidth]{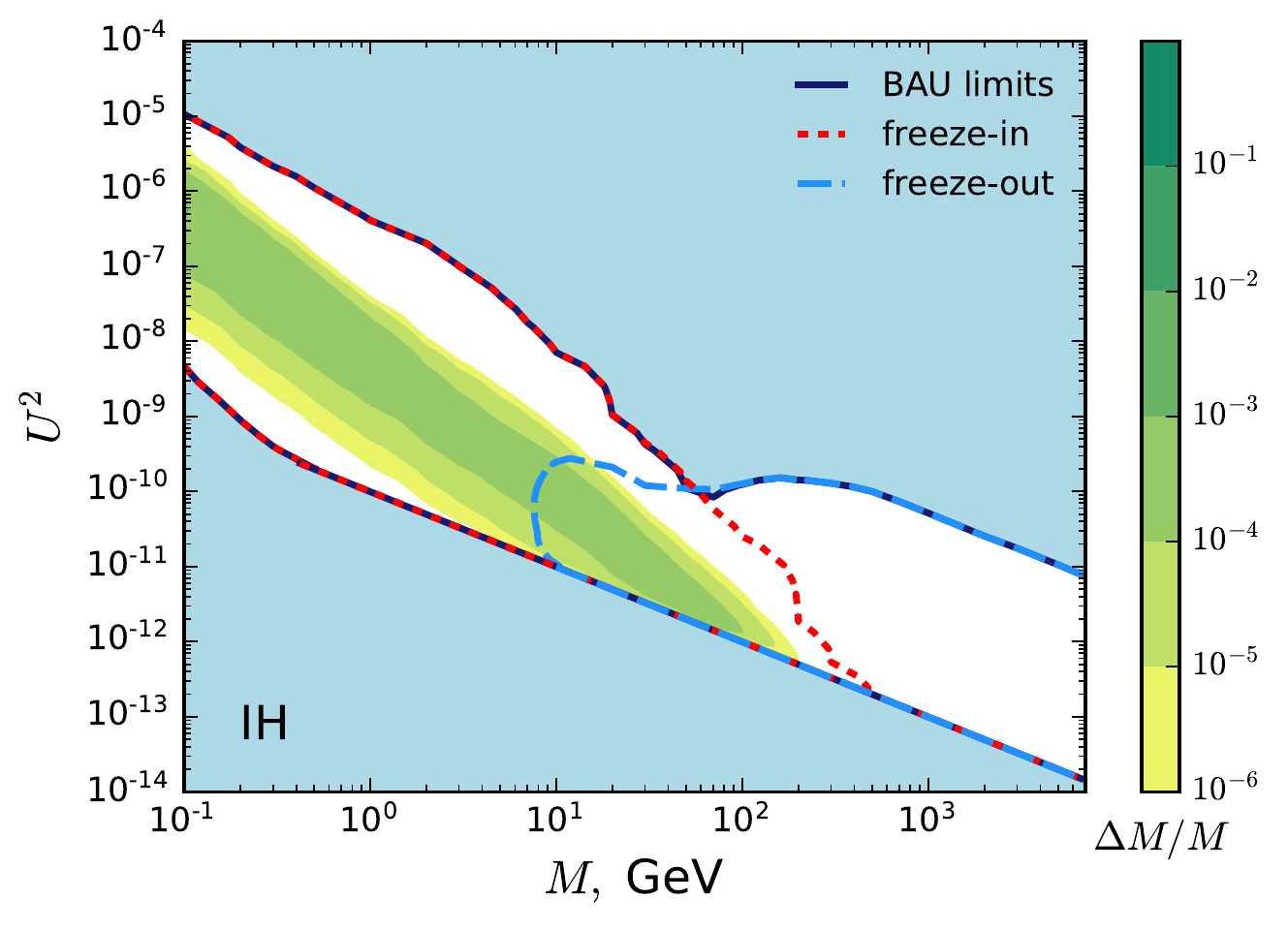}
	\caption{
		The parameter space of leptogenesis in the case of inverted hierarchy (IH) and phases $\delta=\pi$ and $\eta = 0$, which do not lead to a \emph{flavor hierarchical washout}.
		In contrast to Fig.~\ref{fig:MU2}, the freeze-in parameter space (red) stops at $M\sim 500$ GeV,
		where the flavored washout becomes too large for any prior asymmetry to survive.
		This choice of parameters corresponds to the \emph{strong washout} regime, where all of the BAU is generated in the HNL decays.
	Note that the phase $\Re \omega = \pi/4$ is the same as in figure~\ref{fig:MU2}, as it does not affect the washout of the lepton number.}
	\label{fig:flvDep}
\end{figure}

The SM lepton number washout is determined by the smallest flavored washout $W_\alpha$ from Eq.~\eqref{eqs:Boltzmann}.
To determine the size of the lepton number washout, we have to take the spectator effects into account via the susceptibility matrix $\omega_{\alpha \beta}$.
The susceptibility matrix is close to diagonal, and we can estimate the SM lepton number washout strength as
\begin{align}
	K_\mathrm{min} \approx \frac{\omega_{\alpha \alpha}T^2}{6} 2 \bar{K} \cdot \frac{\min_\alpha \sum_i |F_{\alpha i}|^2 }{ \sum_{\beta\,,j} |F_{\beta j}|^2}\,.
\end{align}
The key quantity governing the size of the lepton number washout compared to the heavy neutrino washout rate is therefore the minimal branching ratio
\begin{align}
	\frac{\min_\alpha \sum_i |F_{\alpha i}|^2 }{ \sum_{\beta\,,j} |F_{\beta j}|^2}
	\in
	\begin{cases}
		[5.2 \cdot 10^{-3}\,, 0.12] & \text{for NH}\,,\\
		[3.8 \cdot 10^{-4}\,, 1/3] & \text{for IH}\,.
	\end{cases}
	\label{def:br}
\end{align}
This means that the effective lepton number washout strength is given by
\begin{align}
	K_\mathrm{min} \in
	\begin{cases}
		[0.1\,, 3] \times \cosh(2 \Im \omega)\, \text{for NH,}\\
		[0.01\,, 10] \times \cosh(2 \Im \omega)\, \text{for IH.}
	\end{cases}
\end{align}
For normal ordering this implies that the washout can be weak to moderate when $\Im \omega \approx 0$,
whereas for inverted ordering it can also be strong - depending on the values of the $CP$-phases $\eta$ and $\delta$.
In figure~\ref{fig:flvDep} we show an example of the parameter space in the IH case with a choice of PMNS phases that leads to strong washout,
where the freeze-in parameter space completely closes around $500$ GeV, and the BAU for larger masses becomes independent of the initial conditions.

\subsubsection{Freeze-out leptogenesis for GeV-scale HNLs}
\label{sss:fo-GeV}
Another reason why the parameter space of resonant leptogenesis remains connected with leptogenesis through neutrino oscillations is that HNL freeze-out provides a sufficient deviation from equilibrium even if their masses are as low as a few GeV.
To obtain a qualitative estimate of the BAU, we can use the decay asymmetries~\eqref{effDecAsymm:LNV} from section~\ref{sec:boltzmannFromDnsMat}.
The deviation from equilibrium caused by the expansion of the Universe is suppressed by a factor $M^2/T^2$:
\begin{align}
	Y_{N_I} - Y_{N_I}^\mathrm{eq} \approx \frac{M^2 T \log 2}{4 \pi^2 s(T)}\,,
\end{align}
where $s(T)$ is the entropy density of the Universe.

If we neglect the lepton number washout, in equation~\eqref{eqs:Boltzmann} only the LNV part of the decay asymmetry survives, which gives us an estimate of the BAU
\begin{align}
	Y_B &\approx \sum_I \int d z\Gamma_I \epsilon_I^\mathrm{LNV} (Y_{N_I} - Y_{N_I}^\mathrm{eq})\,,\\\notag
	&\sim 10^{-10} \left( \frac{M}{3 \text{GeV}} \right)^5 \times \frac{\Im [F^\dagger F]_{23}^2}{[F^\dagger F]_{II}^2} \frac{[F^\dagger F]_{II}^2 v^2}{m_\nu M}\,,
\end{align}
where we used the mass splitting that saturates the resonant enhancement $\Delta M \sim \Gamma_I$, and only included the leading term in $\gamma_-$.

This approximation gives us an estimate similar to the results from~\cite{Hambye:2016sby,Granelli:2020ysj}, but the origin of this $M^4$ suppression is in reality quite different.
In~\cite{Hambye:2016sby}, the additional $M^2/T^2$ factor arises through the interplay of the thermal HNL masses, and the thermal decay rates,
while here it is a direct result of the helicity-dependent rates $\gamma_\pm$, where the FNV rate is suppressed by a factor $\gamma_- \sim M^2/T^2$.

We should however note that this approximation has very limited applicability,
and an accurate lower bound can only be obtained by using the full density matrix equations for the following reasons:
\begin{itemize}
	\item the mass splitting that is needed to saturate the decay asymmetry $\Delta M \sim \Gamma_I$ corresponds to an oscillation temperature very close to the sphaleron freeze-out temperature, which means that the approximation of fast oscillations does not hold\,,
	\item all the rates in the quantum Boltzman equations change significantly during the electroweak crossover,
		in particular the FNV rate $\gamma_-$ changes by two orders of magnitude between the temperature $T > 150$ GeV and $T \sim 130$ GeV (c.f. Figs.~\ref{fig:sigmapmApp} and~\ref{fig:sigmapmAppBroken})\,.
\end{itemize}

\subsection{Approximate scaling between the TeV scale and the Davidson-Ibarra Bound}
\label{sub:approximate_scaling_between_the_tev_scale_and_the_davidson_ibarra_bound}
For heavy neutrino masses above $\mathcal{O}(2)$ TeV, the freeze-out of the lepton asymmetry happens  before the sphaleron processes freeze out.
As a consequence of this, the only scales in the problem are the masses of the heavy neutrinos $M_N$ and the temperature $T$.

In this section we show how the solutions to the kinetic equations~\eqref{KE_1a} change for different values of the heavy neutrino masses $M_N$.

The coefficients $H_N$, and $\Gamma$ in Eqs.~\eqref{KE_1a} are all dimensionful quantities. When solving leptogenesis equations in an expanding universe it is convenient choose a dimensionless time variable $x=T_\mathrm{ref}/T$,
for $M_N \gtrsim 2$ TeV this is typically chosen as the average mass of the two heavy neutrinos $T_\mathrm{ref} = M$. Thre freeze-out of the lepton asymmetry then occurs for $x=\mathcal{O}(10)$.

If we change the heavy neutrino masses by a factor $M_N \rightarrow \zeta M_N$, the coefficients $H_0/M$ and $\Gamma/M$ then scale as
\begin{align}
	H_0/M = \frac{z^2 T_{com}^2}{E_N} \frac{\Delta M}{M^2} \rightarrow \zeta^{-1} H_0\,, && \Gamma/M = \gamma(x) \cdot Y / M \rightarrow \Gamma/M\,,
\end{align}
where $Y$ denotes either $Y_+$ ot $Y_-$ from Eq.~\eqref{YukawaMatrices} and we used $Y \sim F^2 \propto M$ for a fixed value of $m_\nu$, as can be seen from the parametrization~\eqref{parametrization:CI}.
We note that the equilibration rate remains the same, while the bare Hamiltonian term scales inversely with the mass of the heavy neutrinos.
Fortunately, the mass splitting $\Delta M$ in resonant leptogenesis is typically much smaller than the mass $M$.
This effectively allows us to scale $\Delta M$ independently of $M$. In order to cancel the $\zeta$ scaling of $H_0$ we replace $\Delta M \rightarrow \Delta M \zeta^2$.
This only induces a small correction to the Yukawa couplings $F \rightarrow F + \mathcal{O}(\zeta^2 \Delta M^2/M^2)$.
Finally, this means that we do not need to repeat the parameter scan for every mass above $M > 2$ TeV, as the results are related by an appropriate re-scaling of the mass splitting $\Delta M$.
This conclusion has several caveats:
\begin{itemize}
	\item For mass splittings of $\Delta M \sim M$, the scaling relation does not hold and the correction to $F$ becomes significant.
	\item The up quark only reaches equilibrium at temperatures below $10^6\,\mathrm{GeV}$, therefore if a significant portion of the lepton asymmetry is produced above these temperatures, the simple scaling solution will not be sufficient, as the susceptibility matrices change, and the distribution of the lepton asymmetries among the flavors may be modified. However, we note that this is a small change in the number of degrees of freedom entering the equations, and we expect it to lead to no more than an $\mathcal{O}(1)$ change in the baryon asymmetry.
	\item For temperatures above $10^9\,\mathrm{GeV}$, the ``flavored'' approximation breaks down, and we have to keep track of the lepton doublet correlation matrices.
	\item The typical energy scale of the interactions changes with temperature, which means that the running of the couplings has to be taken into account, which leads to a modification of the rates entering the kinetic equations.
	\item Finally, when solving the equations for the heavy neutrino oscillations we implicitly neglect the vertex diagrams which are important for thermal leptogenesis. However, the contribution of this diagram is subleading compared to the wave-function contribution below the Davidson-Ibarra bound.
\end{itemize}

We may conclude that the biggest differences occur when the majority of the BAU is produced at temperatures above $10^9\,\mathrm{GeV}$.

\section{Comparison with other studies of leptogenesis in the domain \texorpdfstring{$M\sim M_W$}{M~MW}}
\label{sec:comparison_with_other_works}

In this section we compare our results with several other works that discuss leptogenesis in the regime where resonant leptogenesis and leptogenesis via oscillations overlap.
We summarize the comparison with these works below:
\begin{itemize}
	\item \textbf{P. Hern\'{a}ndez, M. Kekic, J. L\'{o}pez-Pav\'{o}n, J. Racker and J. Salvado \cite{Hernandez:2016kel}}\\
		The authors consider the HNL masses between $0.1$ GeV and $100$ GeV.
		This work only considers the symmetric phase of the SM, and do not include the FNV processes
		(which have not yet been included in the quantum kinetic equations prior to~\cite{Eijima:2017anv,Ghiglieri:2017gjz}),
		nor the HNL decays to $W$ and $Z$ bosons.
		While this approach is valid in a large part of parameter space for GeV-scale HNLs,
		it leads to an underestimate of HNL equilibration rates when $M\sim 100$ GeV.
	\item \textbf{S. Antusch, E. Cazzato, M. Drewes, O. Fischer, B. Garbrecht, D. Gueter, and J. Klaric\ \cite{Antusch:2017pkq}}\\
		The authors consider the HNL masses between $0.1$ GeV and $50$ GeV. For heavier HNLs the  $M/T$ corrections become significant.
		Both FNV and FNC processes are included, albeit only in the symmetric phase.
		The enhancement of the FNV rate in the broken phase was not included, and the resulting FNV rate is underestimated for $T \leq 160$ GeV.
		Nonetheless, it is interesting to note that even these underestimated FNV rates exceed the Hubble rate when $M\sim 50$ GeV, without preventing successful baryogenesis.
	\item \textbf{T. Hambye and D. Teresi~\cite{Hambye:2016sby,Hambye:2017elz}}\\
		The authors study resonant leptogenesis in the GeV-regime.
		They apply the language of decay asymmetries to GeV-scale heavy neutrinos and indicate that
		leptogenesis in both freeze-in and freeze-out are possible for GeV-scale HNLs.
		One should note that although the decay asymmetry obtained in this work qualitatively differs from the expressions in~\eqref{effDecAsymm:LNV},
		the authors find the same parametric suppression of the LNV decay asymmetry of order $\mathcal{O}(M^2/T^2)$.
		In~\cite{Hambye:2017elz}, the authors used the density matrix approach, and found semi-quantitative agreement with the approach using Boltzmann equations.
		The authors also find that the LNV asymmetry scales with the product of the helicity-dependent rates $\gamma_+ \gamma_-$, similar to the LNV source term in Eq.~\eqref{effDecAsymm:LNV}.
		This study of parameter space is limited to $M < 10$ GeV.
		As, we summarize in section~\ref{sec:boltzmannFromDnsMat}, the Boltzmann equations can be understood as an approximate limit of the density matrix approach,
		however in a limited range of validity.
		A conclusive study of the leptogenesis parameter space in this regime requires a unified treatment that can reproduce
		both baryogenesis via oscillations and resonant leptogenesis in the appropriate limits (as we present in this work).
	\item \textbf{A. Granelli, K. Moffat and S. Petcov~\cite{Granelli:2020ysj}}\\
		The authors consider resonant and flavored leptogenesis in the GeV-regime.
		They find that the flavored decay asymmetry can lead to a BAU several orders of magnitude larger than what the LNV contribution alone would imply.
		Due to the reliance on the Boltzmann equations, this approach also has limited applicability (in the same sense as~\cite{Hambye:2016sby}).
		One should also note that the flavored decay asymmetry studied here can be understood as a limit of fast oscillations from~\eqref{effDecAsymm:LFV}.
		If interpreted this way, these results can be understood as an alternative approach to leptogenesis via neutrino oscillations using the language of Boltzmann equations.
\end{itemize}
One should note that~\cite{Hambye:2016sby,Hambye:2017elz,Granelli:2020ysj} often discuss two separate decay asymmetries---one coming from mixing,
and the other one coming from the oscillations~\cite{Dev:2014laa,Dev:2015wpa,Dev:2014wsa,Kartavtsev:2015vto} of the HNLs.
We do not make such a separation, as no separate sources of the asymmetry appear in derivation of the equations using either the
canonical Raffelt-Sigl formalism of neutrino oscillations~(see e.g.\ \cite{Akhmedov:1998qx,Asaka:2005pn,Ghiglieri:2017gjz,Bodeker:2019rvr}),
nor in the CTP approach in Wigner space~\cite{Garbrecht:2011aw,Drewes:2016gmt,Antusch:2017pkq}.
Because of this, we do not introduce any such additional terms by hand as this can lead to double-counting, and therefore violate the symmetries that are present in the equations in the relativistic limit
(for a detailed account of the conserved charges see section~\ref{sss:cons_chg}).

\section{Discussion and conclusions}
\label{sec:discussion_and_conclusions}
In this work we investigated the similarities and differences between resonant leptogenesis and baryogenesis via neutrino oscillations in the minimal extension of the standard model by two heavy neutrinos.
We found that the two mechanisms are closely related, and that the equations used to describe the two mechanisms are virtually the same.
Since the defining feature of resonant leptogenesis, namely the resonant production of the baryon asymmetry is also present in baryogenesis via neutrino oscillations, we focus on the major difference between the two mechanisms, namely the question whether the majority of the BAU is produced during the \textit{freeze-in}, or \textit{freeze-out} of the heavy neutrinos.

We found significant overlap between the two regimes, namely, freeze-in leptogenesis turns out to play a major role in generating the BAU even for \si{\TeV} and heavier Majorana neutrinos.
This regime mainly coincides with relatively large $\Delta M / M^2 \sim 10^{-7}$~GeV$^{-1}$ mass splitting,
compared to the one optimal for a resonant enhancement $\Delta M / M^2 \sim 10^{-13}$~GeV$^{-1}$.
Furthermore, this also implies a strong dependence on the initial condition which is absent in freeze-out leptogenesis.

On the other hand, we also find viable freeze-out leptogenesis with masses as low as $M = \SI{5}{\GeV}$.
This can qualitatively be understood via the relatively large decay asymmetries, which lead to a BAU only suppressed by $(M/T)^2$.

\acknowledgments
We thank Marco Drewes, Shintaro Eijima, Bj{\"o}rn Garbrecht, Jacopo Ghiglieri, Mikko Laine, Apostolos Pilaftsis and Daniele Teresi for helpful comments and discussions.
This work was supported by the
ERC-AdG-2015 grant 694896 and by the Swiss National Science Foundation Excellence grant 200020B\underline{ }182864.
\appendix

\section{Numerical implementation of the leptogenesis equations}
\label{App:Numerical}

The equations governing the evolution of the heavy neutrino densities, as well as the lepton asymmetries depend on multiple scales, and can in principle be stiff and therefore numerically expensive.
Since the parameter space spans many orders of magnitude, there is no single analytical approximation that can be applied everywhere in the parameter space.
Nonetheless, there are several simplifications of the equations that can lead to improvents in accuracy and speed of the numerical calculations.
In the following we summarize the main approximations and simplifications that we use.

\paragraph*{Momentum averaging.}
Equations~\eqref{KE_1a} are integro-differential equations. In principle one has to solve a differential equation for each of the HNL momentum modes, and integrate over them at each time step to accurately describe the evolution of the lepton asymmetries.
Such a calculation was performed in~\cite{Asaka:2011wq,Ghiglieri:2017csp,Ghiglieri:2018wbs}, however, the numerical intensity of such calculations makes it unfeasible for large scale parameter scans.
In this work we use a momentum-averaging procedure, whereby we only consider how the time evolution of the integrated HNL matrix of densities:
\begin{align}
	n_N \equiv \int \frac{d^3 k}{(2\pi)^3} \rho_N\,.
\end{align}
We then assume that the density matrices remain proportional to the equilibrium distribution, but differ by a total normalization factor:
\begin{align}
	\rho_N (k) \approx \frac{n_N}{n_N^\mathrm{eq}} f_N^\mathrm{eq}\,,
\end{align}
which allows us to replace the momentum-dependent matrices $\rho_N$ in Eq.~\eqref{KE_1a} by the HNL number $n_N$.
This also gives us a procedure to calculate the momentum-average of the HNL rate coefficients:
\begin{align}
	\langle X \rangle = \frac{1}{n_N^\mathrm{eq}} \int \frac{d^3 k}{(2\pi)^3} X(k) f_N^\mathrm{eq}\,,
\end{align}
for the coefficient in Eqs.~(\ref{H_N},\ref{damping},\ref{communication}): $\gamma_\pm$, $h_\pm$ and $1/E_N$.

This approximation gives rise to some artifacts, like the oscillatory behavior of the lepton asymmetries which are not present in the full system,
where the oscillations of the different momentum modes do not happen simultaneously.
However, in~\cite{Asaka:2011wq,Ghiglieri:2017csp,Ghiglieri:2018wbs} it was shown that the momentum-dependent treatment leads to $\mathcal{O}(1)$ corrections, which justifies the approach taken in this work.

\paragraph*{Transformation of the equations into vector form.}
The variables in the system of Eqs.~\eqref{KE_1a} are a combination of different objects---numerically, the lepton asymmetries form a real vector, while the HNL densities $\delta \rho_N$, and $\delta \bar{\rho}_N$ are hermitian matrices with complex elements.
In general, matrices with complex elements will have $n_s^2\times 2$ degrees of freedom, while a hermitian matrix only has $n_s^2$ degrees of freedom (where $n_s$ is the number of HNLs).
Any numerical ODE solver will treat these matrices as general complex matrices and introduce more degrees of freedom than are strictly necessary to solve the system.
To avoid this, we may choose parametrize all hermitian matrices as real four-component vectors:
\begin{align}
	\rho_N = \sum_{i=0}^3 \sigma_i \rho_N^i\,, &&
	\bar{\rho}_N = \sum_{i=0}^3 \sigma_i \bar{\rho}_N^i\,.
\end{align}
Once we transform all the equations accordingly, the full system~\eqref{KE_1a} can be written in matrix form:
\begin{align}
	\dot{q}_i = -A_{ij} q_j\,,
\end{align}
where all degrees of freedom can be encoded in an $11$-dimensional vector:
\begin{align}
	q = (\rho_N^i \,, \bar{\rho_N}^i\,, n_{\Delta_\alpha})^T\,.
\end{align}

\paragraph*{Approximate integration of fast modes.}
In general, the system of equations~\eqref{KE_1a} contains many different time scales.
The presence of significantly different time scales can makes the system stiff, which limits the speed of the numerical computations.

Several strategies of reducing the stiffness of the system have been laid out in~\cite{Drewes:2016gmt},
where the fast equilibration modes can be integrated-out in the case of large mixing angles and small mass splittings, and the equations can even be solved semi-analytically.

Let us briefly overview the main idea of this calculation. Let us assume that we can separate the fast and slow degrees of freedom as $q= (q_F\,, q_S)^T$, with
\begin{align}
	\dot{q_F} = - A_{FF} q_F - A_{FS} q_S\,,\\
	\dot{q_S} = - A_{SF} q_F - A_{SS} q_S\,,
\end{align}
where $A_{AB}$ are block matrices, and the eigenvalues of $A_{FF}$ are much bigger than the eigenvalues of $A_{SS}$.
Due to the large eigenvalues of $A_{FF}$, the fast modes reach the quasi-static equilibrium:
\begin{align}
	q_F \rightarrow q_F^\mathrm{QS} \approx - A_{FF}^{-1} A_{FS} q_S\,.
	\label{qFQS}
\end{align}
We can re-insert this solution into the equation for the slow modes to track the evolution of the rest of the system,
which gives us a differential equation for the slow modes:
\begin{align}
	\dot q_S \approx - (A_{SS} - A_{SF} A_{FF}^{-1} A_{FS}) q_S
	\label{dqS}
\end{align}

In practice, it is however impractical to keep track of the fast and slow modes, as these change with temperature (see, e.g.~\cite{Eijima:2020shs}).
What we can do instead is to replace the matrix $A$ with one that approaches the same quasi-static limit for the fast modes,
but leaves the slow modes intact.

One such ansatz is:
\begin{align}
	A \rightarrow A (A/\Lambda + 1)^{-1}\,,
\end{align}
where we choose the cutoff scale $\Lambda$ between the fast and the slow scales $A_{FF} \gg \Lambda \gg A_{SS}$.
This choice ensures that the fast modes approach the same quasi-static limit as in Eq.~\eqref{qFQS}, but at the time-scale $\Lambda$, i.e.\ for $A_{FF} \gg \Lambda$ we have:
\begin{align}
	\dot q_F \approx - \Lambda ( q_F - A_{FF}^{-1} A_{FS} q_S)\,,
\end{align}
whereas for the slow modes we recover Eq.~\eqref{dqS}.

What is perhaps most important is that we do not need to block-diagonalize the system into fast and slow modes at each point in time as the matrix $A$ evolves.
Instead, the fast and slow modes are automatically separated based on how they compare to the scale $\Lambda$.
Another useful feature of this approach is that it equally applies when the HNL oscillations become fast, since it does not depend on $A_{FF}$ being real.
This ensures that the quasi-static limit of the off-diagonal source term is the same as in~\eqref{rhoij:quasistatic},
but also prevents the system from becoming too stiff.

\section{Pseudo-Dirac limit}
\label{sub:pseudo_dirac_limit}

In this work we are interested in resonant leptogenesis and baryogenesis via oscillations which both require that the two HNLs have nearly equal masses.
In this limit there is an approxiamte $U(1)$ symmetry in the theory (two exactly degenerate Majorana particles form a Dirac one with the associated $U(1)$).
This symmetry allows us to assign a lepton number to different flavors of the heavy neutrinos, and to effectively combine the two Majorana neutrinos into a single Dirac neutrino.
\begin{align}
	P_R \Psi = \frac{\nu_{R_2} + i \nu_{R_3}}{\sqrt{2}}\,,
\end{align}
and the Lagrangian takes the form
\begin{align}
	\mathcal{L} &=  \mathcal{L}_{SM} +   \overline{\Psi} i \partial_\mu \gamma^\mu \Psi
	- M \overline{\Psi}\Psi +  \mathcal{L}_{int}, \\
	\mathcal{L}_{int} &=
	- \frac{\Delta M}{2} (\overline{\Psi}\Psi^c + \overline{\Psi^c}\Psi)
	- (h_{\alpha 2} \langle \Phi \rangle \overline{\nu_{L \alpha}} \Psi + h_{\alpha 3} \langle \Phi \rangle \overline{\nu_{L\alpha}} \Psi^c + h.c.).
	\label{Lagrangian}
\end{align}
This limit is also realized in the Yukawa couplings.
For large values of $|\Im \omega|$, they are much bigger than the naive estimate
\begin{align}
	|F_{\alpha I}|^2 \gg m_\nu M_M /v^2\,.
\end{align}
In such a scenario it is convenient to introduce the matrix of Yukawa couplings $h_{\alpha I}$ related to the matrix
$F_{\alpha I}$ defined in~\eqref{Lagr} as follows
\begin{align}
	F_{\alpha I} = h_{\alpha J} [U_N^\ast]_{JI}, \quad
	U_N = \frac{1}{\sqrt{2}}\begin{pmatrix} -i & 1 \\ i & 1 \end{pmatrix},
\end{align}
where the couplings $h_{\alpha I}$ are manifestly hierarchical with
\begin{align}
	\frac{\sum_\alpha |h_{\alpha 2}|^2}{\sum_\alpha |h_{\alpha 3}|^2} = e^{4 \Im \omega} + \mathcal{O}(M_3-M_2)\,.
\end{align}

\section{Conserved \textit{lepton} numbers}
\label{sss:cons_chg}

One of the main effects that lead to large values of the mixing angle even for $U^2 \gg m_\nu/M$ is the approximate conservation of lepton number.

There are several ways of assigning lepton number to heavy neutrinos so that it is approximately conserved. In the following we will discuss three that are important for leptogenesis.

\paragraph{Helicity as a lepton number}
The first, and most commonly discussed in low-scale leptogenesis mechanisms is the association of a lepton number with the helicity of the heavy neutrinos.
This regime effectively corresponds to heavy neutrinos being relativistic so that their masses can be neglected.
It was discussed already in the work~\citep{Akhmedov:1998qx}, and is the main reason why three heavy neutrinos were required for successful baryogenesis via oscillations. When this was further developed in~\cite{Asaka:2005pn}, it was noted that this conservation is not necessarily a barrier for a production of a total lepton asymmetry as a part of the total (conserved) lepton asymmetry is hidden in the heavy neutrino sector.
Furthermore, it is interesting that this lepton number remains conserved regardless of the number of the heavy neutrinos, as long as they are all relativistic.

This lepton number conservation is a direct consequence of one of the helicity-dependent rates dominating, in the case when $\gamma_+ \gg \gamma_+$,
we can verify that $\sum_\alpha \tilde{\Gamma}_\alpha \approx - \Gamma$.
Summing over the lepton asymmetries and the HNL helicity asymmetry, we find the conserved combination:
\begin{align}
	\frac{d}{d t} \left[ \sum_\alpha n_{\Delta_\alpha} - \int \frac{d^3 k}{(2 \pi)^3} \Tr [ \delta \rho_N - \delta \bar{\rho}_N ] \right] \approx 0\,,
	\label{consChgFNC}
\end{align}
which can be verified by inserting the expressions from~\eqref{KE_1a} into the derivatives of the components.
On the other hand, in the broken phase, the FNV rate $\gamma_-$ can be orders of magnitude larger than $\gamma_+$, giving us $\sum_\alpha \tilde{\Gamma}_\alpha \approx \Gamma$.
This gives us a different assignment of lepton number~\citep{Eijima:2017anv}:
\begin{align}
	\frac{d}{d t} \left[ \sum_\alpha n_{\Delta_\alpha} + \int \frac{d^3 k}{(2 \pi)^3} \Tr [ \delta \rho_N - \delta \bar{\rho}_N ] \right] \approx 0\,.
	\label{consChgFNV}
\end{align}

\paragraph{Approximate Dirac lepton number}

There is another type of lepton number that is typically considered in low-scale seesaw mechanisms. Instead of assigning a lepton number to all right-handed neutrinos, we may assign different lepton numbers to the superpositions of the heavy neutrino flavor eigenstates.
We effectively combine the two Majorana spinors into a single Dirac spinor. There are however terms that violate this symmetry, namely the mass splitting between the two heavy neutrinos, and particular structure of the Yukawa couplings.

The conserved combination is most evident in the non-relativistic limit (when $\gamma_+ = \gamma_-$, as well as $h_+ = h_-$).
In the pseudo-Dirac limit we can neglect the mass difference between the HNLs, as well as the terms suppressed by $e^{- 4 |\Im \omega|}$ in the Yukawa couplings.
This makes the equilibration matrix proportional to the identity ($\Gamma \sim 1_{2\times 2}$), and all of the $\tilde{\Gamma}_\alpha \sim \tilde{\Gamma}$ are proportional to each other.
We can introduce a matrix $\kappa \equiv \tilde{\Gamma}/\Gamma$, which selects the combination of HNL flavors that carries the lepton number.

We can then verify that the combination:
\begin{align}
	\frac{d}{d t} \left[ \sum_\alpha n_{\Delta_\alpha} + \int \frac{d^3 k}{(2 \pi)^3} \Tr [ \kappa (\delta \rho_N - \delta \bar{\rho}_N )] \right] \approx 0\,,
\end{align}
as long as the superpositions of the HNL flavors remain coherent.

This effect was taken into account in~\cite{Blanchet:2009kk,Deppisch:2010fr}, and leads to a suppression of the flavored washout parameter (which enter the washout term in the Boltzmann Equation~\eqref{eqs:Boltzmann}):
\begin{align*}
	K_\alpha^\mathrm{effective} =
	K_\alpha  \left[ 1 +4 \frac{\Re[F^\dagger_{2 \alpha} F_{\alpha 3}] \Re [(F^\dagger F)_{23}] - \Im[F^\dagger_{2 \alpha} F_{\alpha 3}] \Im [(F^\dagger F)_{23}]}{
	(F F^\dagger)_{\alpha \alpha} \Tr (F^\dagger F)} \right. \\
	\left. \times \frac{(\Gamma_{22} + \Gamma_{33})^2 }{(\Gamma_{22} + \Gamma_{33})^2 + 4 (M_2^2 - M_3^2)^2}\right]\,.
\end{align*}
It is important to note that this effect is already included in the density-matrix equations used here, which do not need to be modified further.

\section{Fine tuning}
\begin{figure}[h!]
	\centering
	\includegraphics[width=0.4\textwidth]{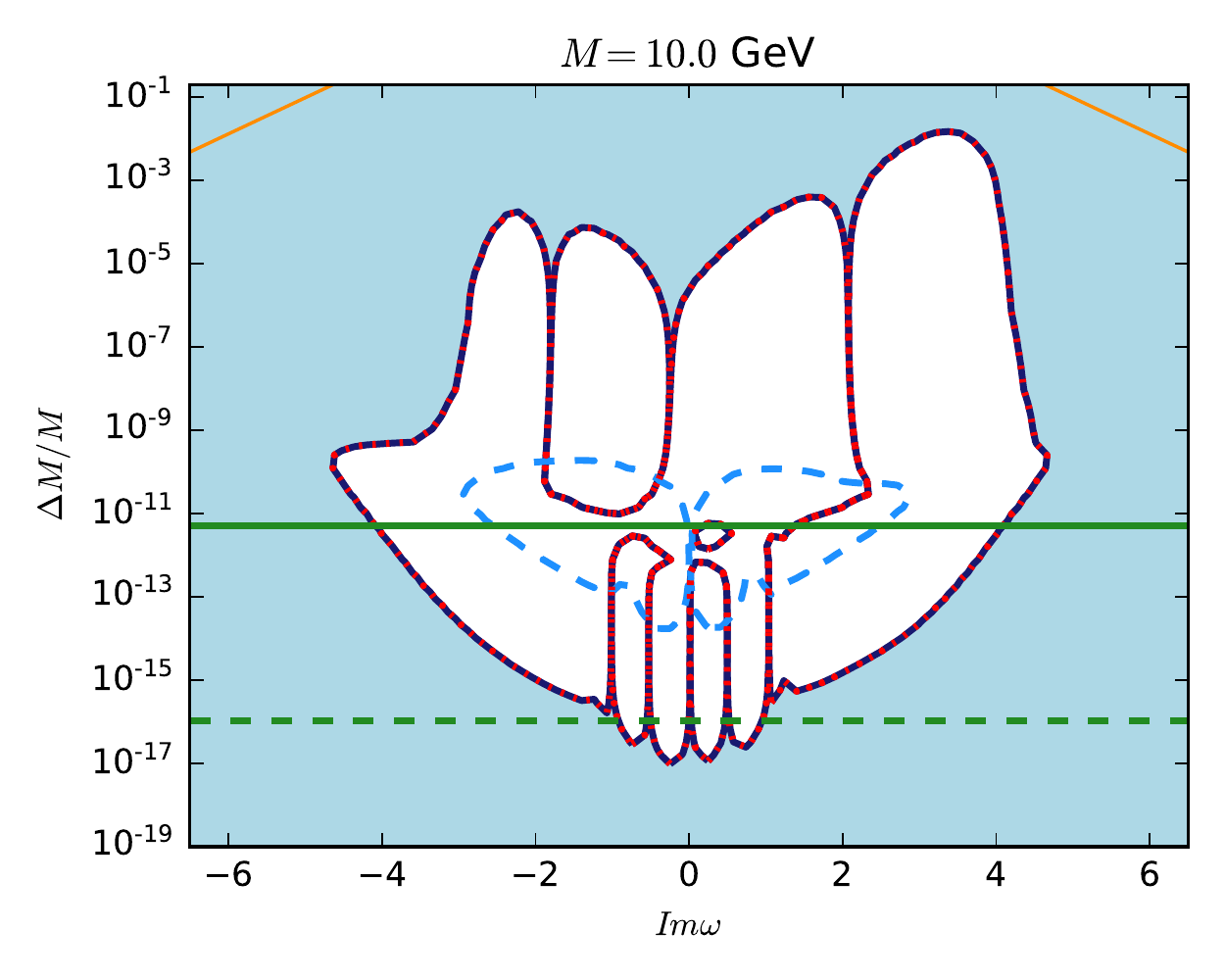} \includegraphics[width=0.4\textwidth]{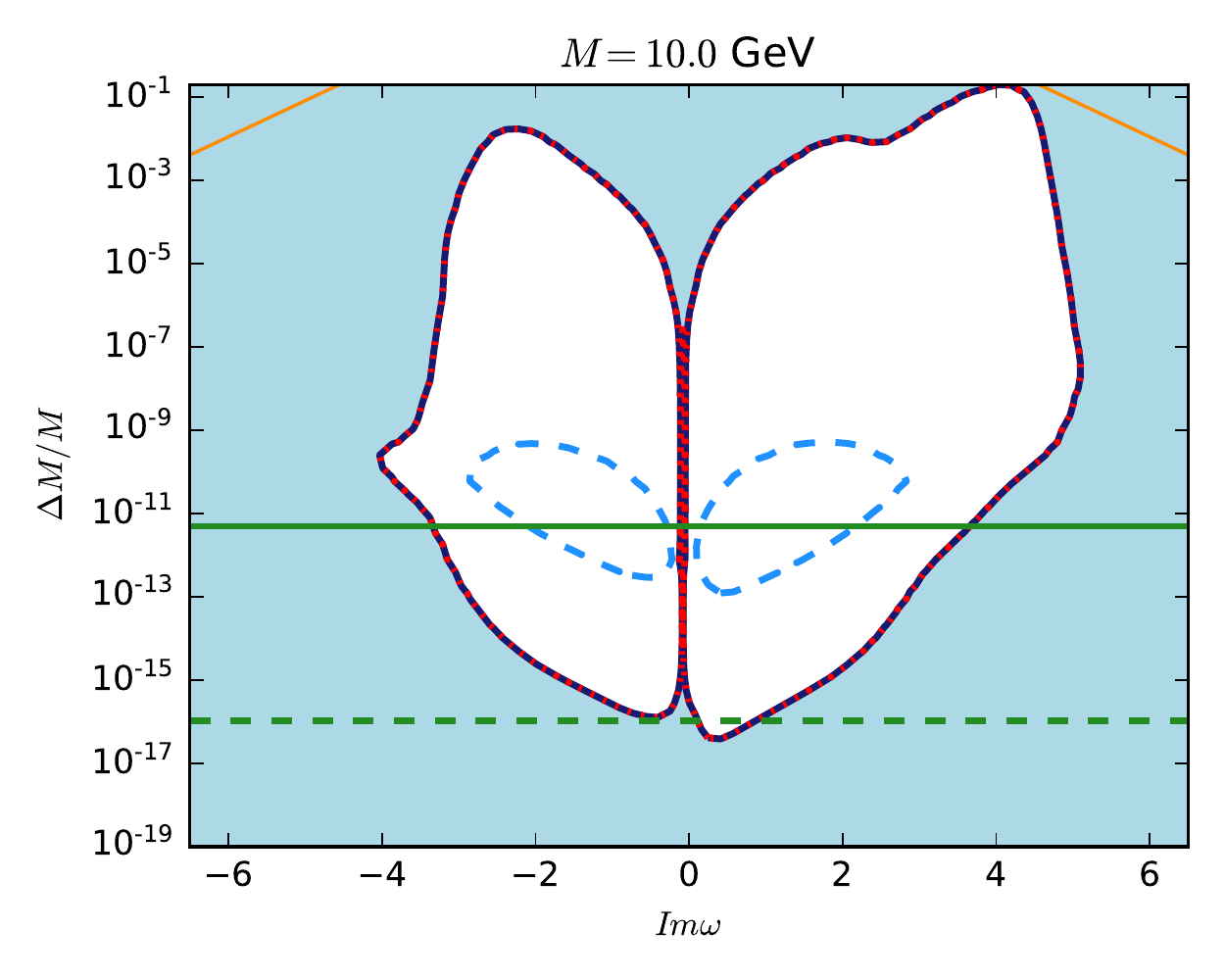}\\
	\includegraphics[width=0.4\textwidth]{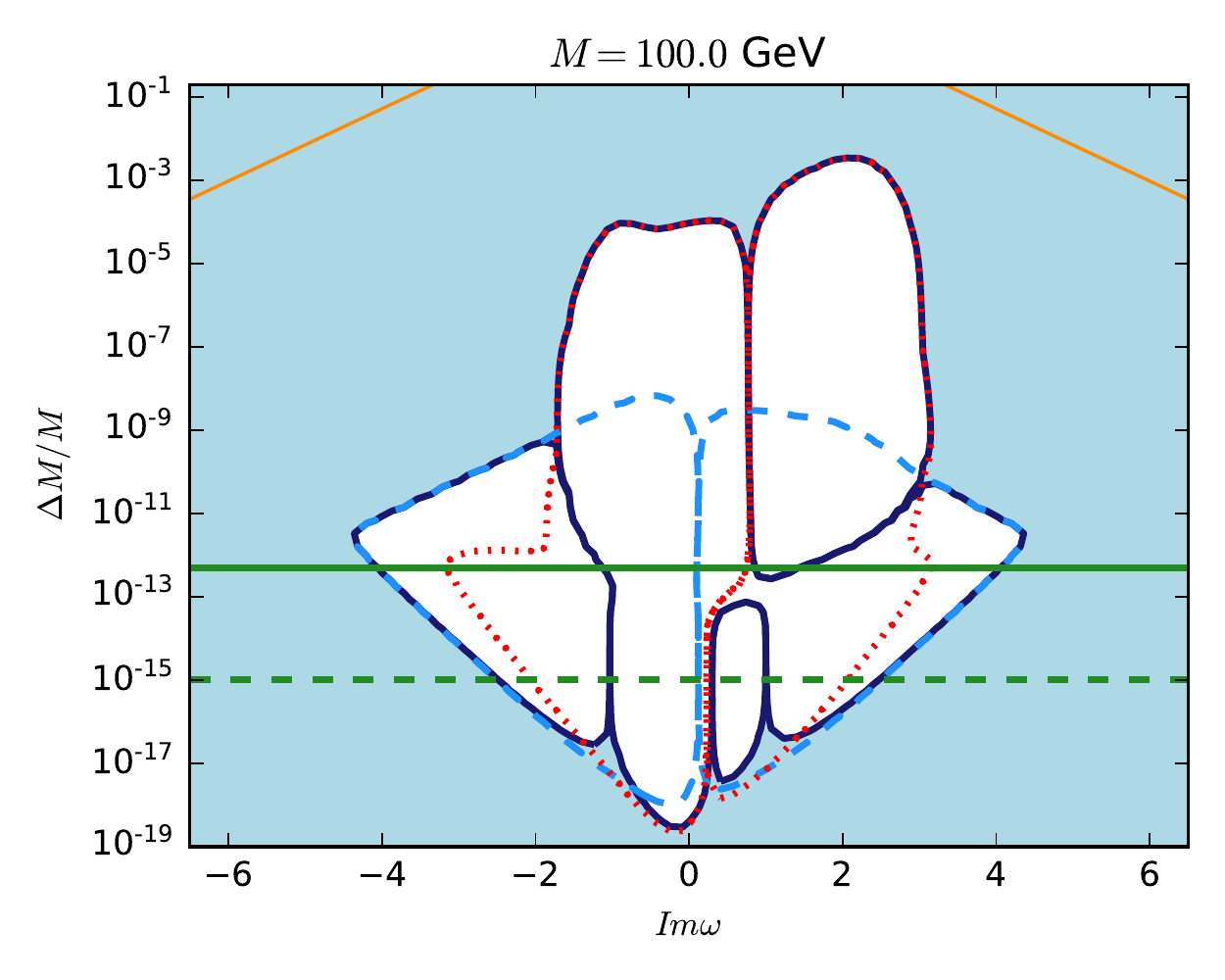} \includegraphics[width=0.4\textwidth]{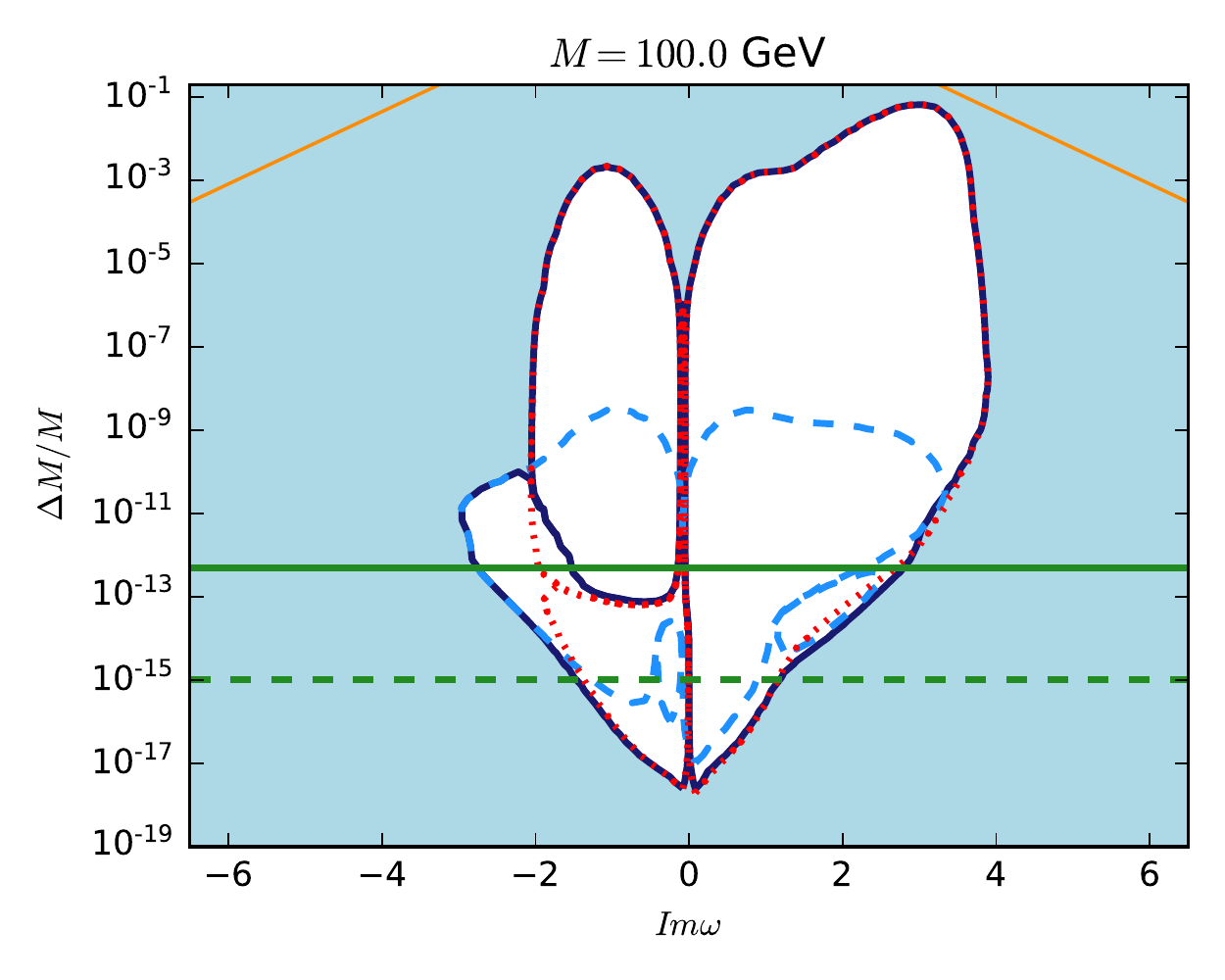}\\
	\includegraphics[width=0.4\textwidth]{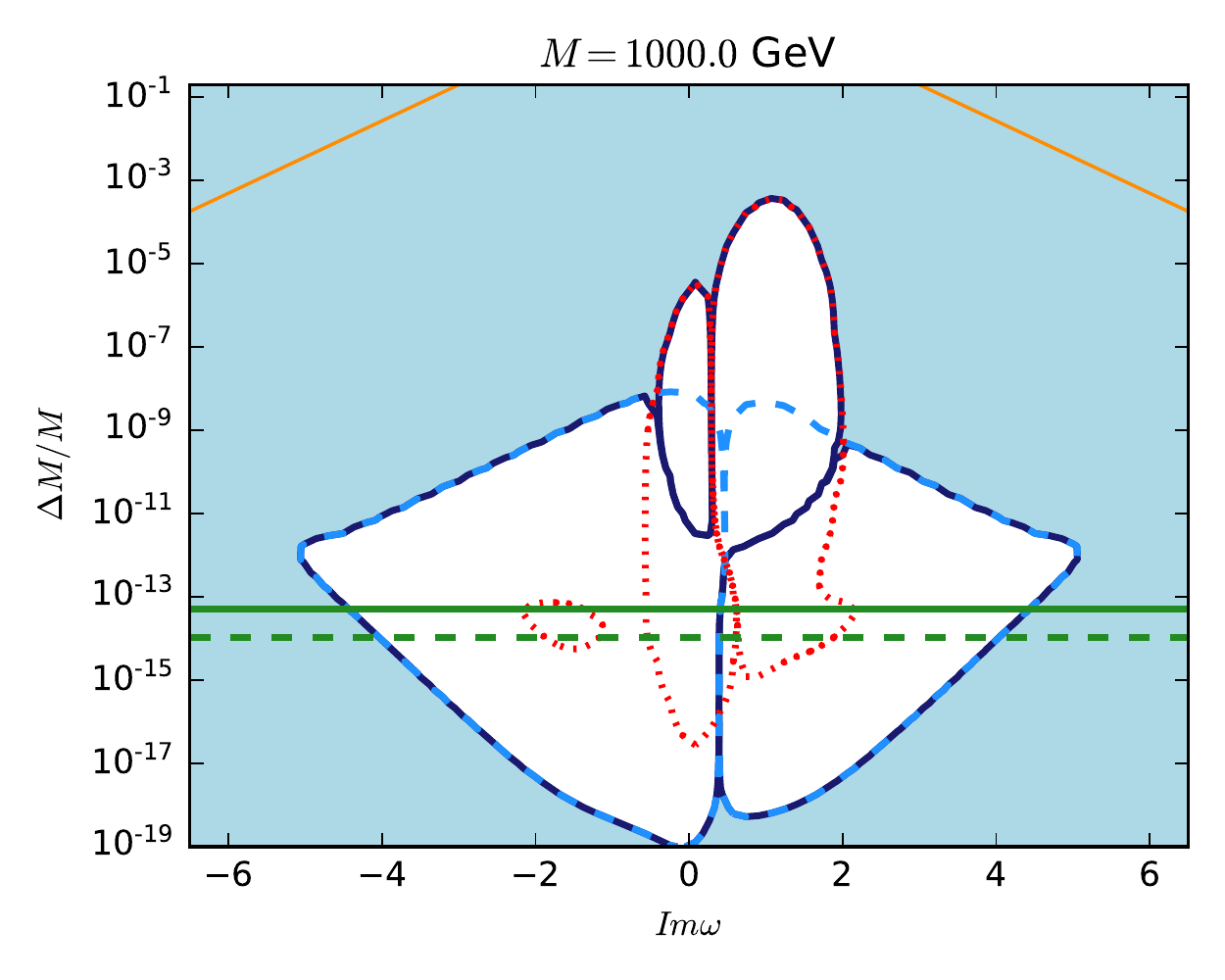} \includegraphics[width=0.4\textwidth]{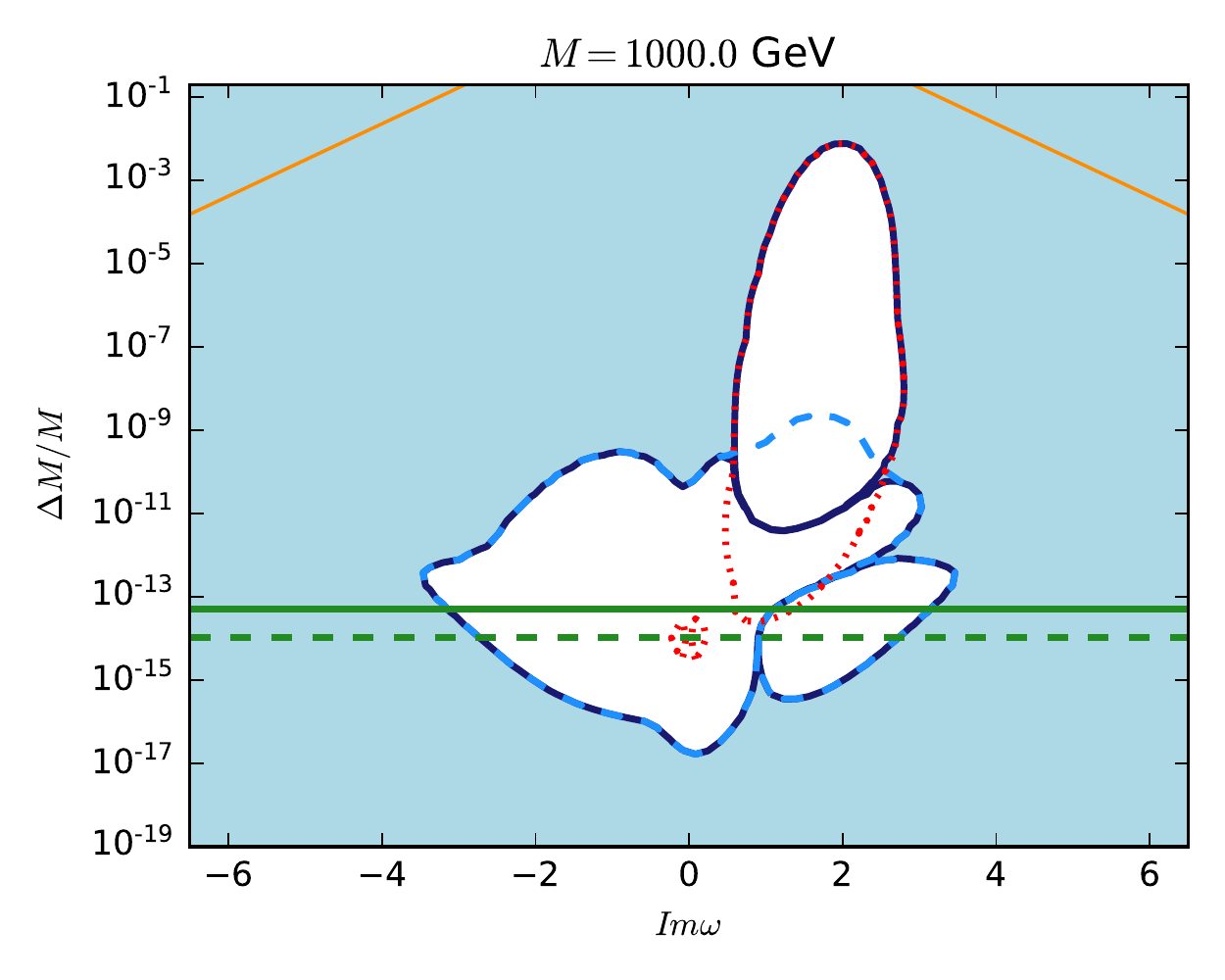}
	\caption{Same as Fig.~\ref{fig:dMiOmega}, but including the tuning limits on the mass splitting from Eqs.~\eqref{dMupper} and~\eqref{dMlower}.
		The allowed range of mass splittings and mixing angles for a benchmark point with normal hierarchy and parameters $\Re \omega = \pi/4$,$\delta = \pi$, $\eta = 3\pi/2$ for NH, and $\Re \omega = \pi/4$,$\delta = 0$, $\eta = \pi/2$ for IH.
		The (yellow, full) line shows the upper limit on the mass splitting~\eqref{dMupper}, and the (green, dashed) line shows the lower bound on the mass splitting~\eqref{dMlower}.
		For comparison we also show the magnitude of the mass splitting that is induced by the Higgs vev corrections to the HNL mass matrix.
		The Lagrangian mass splittings smaller than this value are completely obscured by this contribution, and because of this cannot be measured in any experiment.
	}
	\label{fig:dMiOmegaTune}
\end{figure}

It is often argued that the mass splittings required for low-scale leptogenesis are fine-tuned (see, e.g.~\cite{Shuve:2014zua,Abada:2018oly} for definitions of the fine-tuning).
As we see in figure~\ref{fig:dMiOmega}, the range of mass splittings can in fact be quite large, as leptogenesis is possible for $0 \leq \Delta M/M \lesssim 10^{-2}$.
Compared to thermal leptogenesis (where the HNL masses can be hierarchical), a mass splitting of $\Delta M/M \lesssim 10^{-2}$ can appear too small, and hard to justify.
However, from the technical naturalness standpoint the  model with two hierarchical HNLs and large values of Yukawas appears to be tuned.
If there are no cancellations between the tree-level and loop corrections to the light neutrino masses, the mass splitting may not exceed~\cite{Kersten:2007vk}
\begin{align}
	\label{dMupper}
	\frac{\Delta M}{M} \leq \mathrm{sech}( 2 \Im \omega) \frac{v^2}{M^2} \frac{(4\pi)^2}{4 l^\prime(M^2)}
\end{align}
On the other hand, the small mass differences between the heavy neutrinos receive radiative corrections (see e.g.~\cite{Antusch:2002rr,Lin:2009sq,Ibarra:2020eia}).
If we consider the correction from the Planck scale $M_{Pl}$, the lower bound on the mass splitting is given by~\cite{Roy:2010xq}:
\begin{align}
	\label{dMlower}
	\frac{\Delta M}{M} \gtrsim \frac{M \Delta m_\nu}{(4 \pi v)^2} \log \frac{M_{Pl}}{\mu_0}\,,
\end{align}
where a choice of $\mu_0 \sim M$ gives us a logarithm of $\mathcal{O}(50)$.
In Fig.~\ref{fig:dMiOmegaTune} we confront the range of mass splittings from leptogenesis with the tuning bounds discussed above.

\end{fmffile}
\newpage
\bibliography{refs}

\begin{thebibliography}{166}%
\makeatletter
\providecommand \@ifxundefined [1]{%
 \@ifx{#1\undefined}
}%
\providecommand \@ifnum [1]{%
 \ifnum #1\expandafter \@firstoftwo
 \else \expandafter \@secondoftwo
 \fi
}%
\providecommand \@ifx [1]{%
 \ifx #1\expandafter \@firstoftwo
 \else \expandafter \@secondoftwo
 \fi
}%
\providecommand \natexlab [1]{#1}%
\providecommand \enquote  [1]{``#1''}%
\providecommand \bibnamefont  [1]{#1}%
\providecommand \bibfnamefont [1]{#1}%
\providecommand \citenamefont [1]{#1}%
\providecommand \href@noop [0]{\@secondoftwo}%
\providecommand \href [0]{\begingroup \@sanitize@url \@href}%
\providecommand \@href[1]{\@@startlink{#1}\@@href}%
\providecommand \@@href[1]{\endgroup#1\@@endlink}%
\providecommand \@sanitize@url [0]{\catcode `\\12\catcode `\$12\catcode
  `\&12\catcode `\#12\catcode `\^12\catcode `\_12\catcode `\%12\relax}%
\providecommand \@@startlink[1]{}%
\providecommand \@@endlink[0]{}%
\providecommand \url  [0]{\begingroup\@sanitize@url \@url }%
\providecommand \@url [1]{\endgroup\@href {#1}{\urlprefix }}%
\providecommand \urlprefix  [0]{URL }%
\providecommand \Eprint [0]{\href }%
\providecommand \doibase [0]{http://dx.doi.org/}%
\providecommand \selectlanguage [0]{\@gobble}%
\providecommand \bibinfo  [0]{\@secondoftwo}%
\providecommand \bibfield  [0]{\@secondoftwo}%
\providecommand \translation [1]{[#1]}%
\providecommand \BibitemOpen [0]{}%
\providecommand \bibitemStop [0]{}%
\providecommand \bibitemNoStop [0]{.\EOS\space}%
\providecommand \EOS [0]{\spacefactor3000\relax}%
\providecommand \BibitemShut  [1]{\csname bibitem#1\endcsname}%
\let\auto@bib@innerbib\@empty
\bibitem [{\citenamefont {Minkowski}(1977)}]{Minkowski:1977sc}%
  \BibitemOpen
  \bibfield  {author} {\bibinfo {author} {\bibfnamefont {Peter}\ \bibnamefont
  {Minkowski}},\ }\bibfield  {title} {\enquote {\bibinfo {title} {{$\mu \to
  e\gamma$ at a Rate of One Out of $10^{9}$ Muon Decays?}}}\ }\href {\doibase
  10.1016/0370-2693(77)90435-X} {\bibfield  {journal} {\bibinfo  {journal}
  {Phys. Lett.}\ }\textbf {\bibinfo {volume} {67B}},\ \bibinfo {pages}
  {421--428} (\bibinfo {year} {1977})}\BibitemShut {NoStop}%
\bibitem [{\citenamefont {Gell-Mann}\ \emph {et~al.}(1979)\citenamefont
  {Gell-Mann}, \citenamefont {Ramond},\ and\ \citenamefont
  {Slansky}}]{GellMann:1980vs}%
  \BibitemOpen
  \bibfield  {author} {\bibinfo {author} {\bibfnamefont {Murray}\ \bibnamefont
  {Gell-Mann}}, \bibinfo {author} {\bibfnamefont {Pierre}\ \bibnamefont
  {Ramond}}, \ and\ \bibinfo {author} {\bibfnamefont {Richard}\ \bibnamefont
  {Slansky}},\ }\bibfield  {title} {\enquote {\bibinfo {title} {{Complex
  Spinors and Unified Theories}},}\ }\bibfield  {booktitle} {\emph {\bibinfo
  {booktitle} {{Supergravity Workshop Stony Brook, New York, September 27-28,
  1979}}},\ }\href@noop {} {\bibfield  {journal} {\bibinfo  {journal} {Conf.
  Proc.}\ }\textbf {\bibinfo {volume} {C790927}},\ \bibinfo {pages} {315--321}
  (\bibinfo {year} {1979})},\ \Eprint {http://arxiv.org/abs/1306.4669}
  {arXiv:1306.4669 [hep-th]} \BibitemShut {NoStop}%
\bibitem [{\citenamefont {Mohapatra}\ and\ \citenamefont
  {Senjanovic}(1980)}]{Mohapatra:1979ia}%
  \BibitemOpen
  \bibfield  {author} {\bibinfo {author} {\bibfnamefont {Rabindra~N.}\
  \bibnamefont {Mohapatra}}\ and\ \bibinfo {author} {\bibfnamefont {Goran}\
  \bibnamefont {Senjanovic}},\ }\bibfield  {title} {\enquote {\bibinfo {title}
  {{Neutrino Mass and Spontaneous Parity Nonconservation}},}\ }\href {\doibase
  10.1103/PhysRevLett.44.912} {\bibfield  {journal} {\bibinfo  {journal} {Phys.
  Rev. Lett.}\ }\textbf {\bibinfo {volume} {44}},\ \bibinfo {pages} {912}
  (\bibinfo {year} {1980})}\BibitemShut {NoStop}%
\bibitem [{\citenamefont {Yanagida}(1980)}]{Yanagida:1980xy}%
  \BibitemOpen
  \bibfield  {author} {\bibinfo {author} {\bibfnamefont {Tsutomu}\ \bibnamefont
  {Yanagida}},\ }\bibfield  {title} {\enquote {\bibinfo {title} {{Horizontal
  Symmetry and Masses of Neutrinos}},}\ }\href {\doibase 10.1143/PTP.64.1103}
  {\bibfield  {journal} {\bibinfo  {journal} {Prog. Theor. Phys.}\ }\textbf
  {\bibinfo {volume} {64}},\ \bibinfo {pages} {1103} (\bibinfo {year}
  {1980})}\BibitemShut {NoStop}%
\bibitem [{\citenamefont {Schechter}\ and\ \citenamefont
  {Valle}(1980)}]{Schechter:1980gr}%
  \BibitemOpen
  \bibfield  {author} {\bibinfo {author} {\bibfnamefont {J.}~\bibnamefont
  {Schechter}}\ and\ \bibinfo {author} {\bibfnamefont {J.~W.~F.}\ \bibnamefont
  {Valle}},\ }\bibfield  {title} {\enquote {\bibinfo {title} {{Neutrino Masses
  in SU(2) x U(1) Theories}},}\ }\href {\doibase 10.1103/PhysRevD.22.2227}
  {\bibfield  {journal} {\bibinfo  {journal} {Phys. Rev.}\ }\textbf {\bibinfo
  {volume} {D22}},\ \bibinfo {pages} {2227} (\bibinfo {year}
  {1980})}\BibitemShut {NoStop}%
\bibitem [{\citenamefont {Schechter}\ and\ \citenamefont
  {Valle}(1982)}]{Schechter:1981cv}%
  \BibitemOpen
  \bibfield  {author} {\bibinfo {author} {\bibfnamefont {J.}~\bibnamefont
  {Schechter}}\ and\ \bibinfo {author} {\bibfnamefont {J.~W.~F.}\ \bibnamefont
  {Valle}},\ }\bibfield  {title} {\enquote {\bibinfo {title} {{Neutrino Decay
  and Spontaneous Violation of Lepton Number}},}\ }\href {\doibase
  10.1103/PhysRevD.25.774} {\bibfield  {journal} {\bibinfo  {journal} {Phys.
  Rev.}\ }\textbf {\bibinfo {volume} {D25}},\ \bibinfo {pages} {774} (\bibinfo
  {year} {1982})}\BibitemShut {NoStop}%
\bibitem [{\citenamefont {Kuzmin}\ \emph {et~al.}(1985)\citenamefont {Kuzmin},
  \citenamefont {Rubakov},\ and\ \citenamefont {Shaposhnikov}}]{Kuzmin:1985mm}%
  \BibitemOpen
  \bibfield  {author} {\bibinfo {author} {\bibfnamefont {V.~A.}\ \bibnamefont
  {Kuzmin}}, \bibinfo {author} {\bibfnamefont {V.~A.}\ \bibnamefont {Rubakov}},
  \ and\ \bibinfo {author} {\bibfnamefont {M.~E.}\ \bibnamefont
  {Shaposhnikov}},\ }\bibfield  {title} {\enquote {\bibinfo {title} {{On the
  Anomalous Electroweak Baryon Number Nonconservation in the Early
  Universe}},}\ }\href {\doibase 10.1016/0370-2693(85)91028-7} {\bibfield
  {journal} {\bibinfo  {journal} {Phys. Lett.}\ }\textbf {\bibinfo {volume}
  {155B}},\ \bibinfo {pages} {36} (\bibinfo {year} {1985})}\BibitemShut
  {NoStop}%
\bibitem [{\citenamefont {Davidson}\ and\ \citenamefont
  {Ibarra}(2002)}]{Davidson:2002qv}%
  \BibitemOpen
  \bibfield  {author} {\bibinfo {author} {\bibfnamefont {Sacha}\ \bibnamefont
  {Davidson}}\ and\ \bibinfo {author} {\bibfnamefont {Alejandro}\ \bibnamefont
  {Ibarra}},\ }\bibfield  {title} {\enquote {\bibinfo {title} {{A Lower bound
  on the right-handed neutrino mass from leptogenesis}},}\ }\href {\doibase
  10.1016/S0370-2693(02)01735-5} {\bibfield  {journal} {\bibinfo  {journal}
  {Phys. Lett.}\ }\textbf {\bibinfo {volume} {B535}},\ \bibinfo {pages}
  {25--32} (\bibinfo {year} {2002})},\ \Eprint
  {http://arxiv.org/abs/hep-ph/0202239} {arXiv:hep-ph/0202239 [hep-ph]}
  \BibitemShut {NoStop}%
\bibitem [{\citenamefont {Canetti}\ and\ \citenamefont
  {Shaposhnikov}(2010)}]{Canetti:2010aw}%
  \BibitemOpen
  \bibfield  {author} {\bibinfo {author} {\bibfnamefont {Laurent}\ \bibnamefont
  {Canetti}}\ and\ \bibinfo {author} {\bibfnamefont {Mikhail}\ \bibnamefont
  {Shaposhnikov}},\ }\bibfield  {title} {\enquote {\bibinfo {title} {{Baryon
  Asymmetry of the Universe in the $\nu$MSM}},}\ }\href {\doibase
  10.1088/1475-7516/2010/09/001} {\bibfield  {journal} {\bibinfo  {journal}
  {JCAP}\ }\textbf {\bibinfo {volume} {1009}},\ \bibinfo {pages} {001}
  (\bibinfo {year} {2010})},\ \Eprint {http://arxiv.org/abs/1006.0133}
  {arXiv:1006.0133 [hep-ph]} \BibitemShut {NoStop}%
\bibitem [{\citenamefont {Liu}\ and\ \citenamefont {Segre}(1993)}]{Liu:1993tg}%
  \BibitemOpen
  \bibfield  {author} {\bibinfo {author} {\bibfnamefont {Jiang}\ \bibnamefont
  {Liu}}\ and\ \bibinfo {author} {\bibfnamefont {Gino}\ \bibnamefont {Segre}},\
  }\bibfield  {title} {\enquote {\bibinfo {title} {{Reexamination of generation
  of baryon and lepton number asymmetries by heavy particle decay}},}\ }\href
  {\doibase 10.1103/PhysRevD.48.4609} {\bibfield  {journal} {\bibinfo
  {journal} {Phys. Rev.}\ }\textbf {\bibinfo {volume} {D48}},\ \bibinfo {pages}
  {4609--4612} (\bibinfo {year} {1993})},\ \Eprint
  {http://arxiv.org/abs/hep-ph/9304241} {arXiv:hep-ph/9304241 [hep-ph]}
  \BibitemShut {NoStop}%
\bibitem [{\citenamefont {Flanz}\ \emph {et~al.}(1995)\citenamefont {Flanz},
  \citenamefont {Paschos},\ and\ \citenamefont {Sarkar}}]{Flanz:1994yx}%
  \BibitemOpen
  \bibfield  {author} {\bibinfo {author} {\bibfnamefont {Marion}\ \bibnamefont
  {Flanz}}, \bibinfo {author} {\bibfnamefont {Emmanuel~A.}\ \bibnamefont
  {Paschos}}, \ and\ \bibinfo {author} {\bibfnamefont {Utpal}\ \bibnamefont
  {Sarkar}},\ }\bibfield  {title} {\enquote {\bibinfo {title} {{Baryogenesis
  from a lepton asymmetric universe}},}\ }\href {\doibase
  10.1016/0370-2693(96)00866-0, 10.1016/0370-2693(96)00842-8,
  10.1016/0370-2693(94)01555-Q} {\bibfield  {journal} {\bibinfo  {journal}
  {Phys. Lett.}\ }\textbf {\bibinfo {volume} {B345}},\ \bibinfo {pages}
  {248--252} (\bibinfo {year} {1995})},\ \bibinfo {note} {[Erratum: Phys.
  Lett.B384,487(1996); Erratum: Phys. Lett.B382,447(1996)]},\ \Eprint
  {http://arxiv.org/abs/hep-ph/9411366} {arXiv:hep-ph/9411366 [hep-ph]}
  \BibitemShut {NoStop}%
\bibitem [{\citenamefont {Flanz}\ \emph {et~al.}(1996)\citenamefont {Flanz},
  \citenamefont {Paschos}, \citenamefont {Sarkar},\ and\ \citenamefont
  {Weiss}}]{Flanz:1996fb}%
  \BibitemOpen
  \bibfield  {author} {\bibinfo {author} {\bibfnamefont {Marion}\ \bibnamefont
  {Flanz}}, \bibinfo {author} {\bibfnamefont {Emmanuel~A.}\ \bibnamefont
  {Paschos}}, \bibinfo {author} {\bibfnamefont {Utpal}\ \bibnamefont {Sarkar}},
  \ and\ \bibinfo {author} {\bibfnamefont {Jan}\ \bibnamefont {Weiss}},\
  }\bibfield  {title} {\enquote {\bibinfo {title} {{Baryogenesis through mixing
  of heavy Majorana neutrinos}},}\ }\href {\doibase
  10.1016/S0370-2693(96)01337-8, 10.1016/S0370-2693(96)80011-6} {\bibfield
  {journal} {\bibinfo  {journal} {Phys. Lett.}\ }\textbf {\bibinfo {volume}
  {B389}},\ \bibinfo {pages} {693--699} (\bibinfo {year} {1996})},\ \Eprint
  {http://arxiv.org/abs/hep-ph/9607310} {arXiv:hep-ph/9607310 [hep-ph]}
  \BibitemShut {NoStop}%
\bibitem [{\citenamefont {Covi}\ \emph {et~al.}(1996)\citenamefont {Covi},
  \citenamefont {Roulet},\ and\ \citenamefont {Vissani}}]{Covi:1996wh}%
  \BibitemOpen
  \bibfield  {author} {\bibinfo {author} {\bibfnamefont {Laura}\ \bibnamefont
  {Covi}}, \bibinfo {author} {\bibfnamefont {Esteban}\ \bibnamefont {Roulet}},
  \ and\ \bibinfo {author} {\bibfnamefont {Francesco}\ \bibnamefont
  {Vissani}},\ }\bibfield  {title} {\enquote {\bibinfo {title} {{CP violating
  decays in leptogenesis scenarios}},}\ }\href {\doibase
  10.1016/0370-2693(96)00817-9} {\bibfield  {journal} {\bibinfo  {journal}
  {Phys. Lett.}\ }\textbf {\bibinfo {volume} {B384}},\ \bibinfo {pages}
  {169--174} (\bibinfo {year} {1996})},\ \Eprint
  {http://arxiv.org/abs/hep-ph/9605319} {arXiv:hep-ph/9605319 [hep-ph]}
  \BibitemShut {NoStop}%
\bibitem [{\citenamefont {Covi}\ and\ \citenamefont
  {Roulet}(1997)}]{Covi:1996fm}%
  \BibitemOpen
  \bibfield  {author} {\bibinfo {author} {\bibfnamefont {Laura}\ \bibnamefont
  {Covi}}\ and\ \bibinfo {author} {\bibfnamefont {Esteban}\ \bibnamefont
  {Roulet}},\ }\bibfield  {title} {\enquote {\bibinfo {title} {{Baryogenesis
  from mixed particle decays}},}\ }\href {\doibase
  10.1016/S0370-2693(97)00287-6} {\bibfield  {journal} {\bibinfo  {journal}
  {Phys. Lett.}\ }\textbf {\bibinfo {volume} {B399}},\ \bibinfo {pages}
  {113--118} (\bibinfo {year} {1997})},\ \Eprint
  {http://arxiv.org/abs/hep-ph/9611425} {arXiv:hep-ph/9611425 [hep-ph]}
  \BibitemShut {NoStop}%
\bibitem [{\citenamefont {Pilaftsis}(1997{\natexlab{a}})}]{Pilaftsis:1997jf}%
  \BibitemOpen
  \bibfield  {author} {\bibinfo {author} {\bibfnamefont {Apostolos}\
  \bibnamefont {Pilaftsis}},\ }\bibfield  {title} {\enquote {\bibinfo {title}
  {{CP violation and baryogenesis due to heavy Majorana neutrinos}},}\ }\href
  {\doibase 10.1103/PhysRevD.56.5431} {\bibfield  {journal} {\bibinfo
  {journal} {Phys. Rev.}\ }\textbf {\bibinfo {volume} {D56}},\ \bibinfo {pages}
  {5431--5451} (\bibinfo {year} {1997}{\natexlab{a}})},\ \Eprint
  {http://arxiv.org/abs/hep-ph/9707235} {arXiv:hep-ph/9707235 [hep-ph]}
  \BibitemShut {NoStop}%
\bibitem [{\citenamefont {Pilaftsis}(1997{\natexlab{b}})}]{Pilaftsis:1997dr}%
  \BibitemOpen
  \bibfield  {author} {\bibinfo {author} {\bibfnamefont {Apostolos}\
  \bibnamefont {Pilaftsis}},\ }\bibfield  {title} {\enquote {\bibinfo {title}
  {{Resonant CP violation induced by particle mixing in transition
  amplitudes}},}\ }\href {\doibase 10.1016/S0550-3213(97)00469-0} {\bibfield
  {journal} {\bibinfo  {journal} {Nucl. Phys.}\ }\textbf {\bibinfo {volume}
  {B504}},\ \bibinfo {pages} {61--107} (\bibinfo {year}
  {1997}{\natexlab{b}})},\ \Eprint {http://arxiv.org/abs/hep-ph/9702393}
  {arXiv:hep-ph/9702393 [hep-ph]} \BibitemShut {NoStop}%
\bibitem [{\citenamefont {Pilaftsis}(1999)}]{Pilaftsis:1998pd}%
  \BibitemOpen
  \bibfield  {author} {\bibinfo {author} {\bibfnamefont {Apostolos}\
  \bibnamefont {Pilaftsis}},\ }\bibfield  {title} {\enquote {\bibinfo {title}
  {{Heavy Majorana neutrinos and baryogenesis}},}\ }\href {\doibase
  10.1142/S0217751X99000932} {\bibfield  {journal} {\bibinfo  {journal} {Int.
  J. Mod. Phys.}\ }\textbf {\bibinfo {volume} {A14}},\ \bibinfo {pages}
  {1811--1858} (\bibinfo {year} {1999})},\ \Eprint
  {http://arxiv.org/abs/hep-ph/9812256} {arXiv:hep-ph/9812256 [hep-ph]}
  \BibitemShut {NoStop}%
\bibitem [{\citenamefont {Buchmuller}\ and\ \citenamefont
  {Plumacher}(1998)}]{Buchmuller:1997yu}%
  \BibitemOpen
  \bibfield  {author} {\bibinfo {author} {\bibfnamefont {W.}~\bibnamefont
  {Buchmuller}}\ and\ \bibinfo {author} {\bibfnamefont {M.}~\bibnamefont
  {Plumacher}},\ }\bibfield  {title} {\enquote {\bibinfo {title} {{CP asymmetry
  in Majorana neutrino decays}},}\ }\href {\doibase
  10.1016/S0370-2693(97)01548-7} {\bibfield  {journal} {\bibinfo  {journal}
  {Phys. Lett.}\ }\textbf {\bibinfo {volume} {B431}},\ \bibinfo {pages}
  {354--362} (\bibinfo {year} {1998})},\ \Eprint
  {http://arxiv.org/abs/hep-ph/9710460} {arXiv:hep-ph/9710460 [hep-ph]}
  \BibitemShut {NoStop}%
\bibitem [{\citenamefont {Pilaftsis}\ and\ \citenamefont
  {Underwood}(2004)}]{Pilaftsis:2003gt}%
  \BibitemOpen
  \bibfield  {author} {\bibinfo {author} {\bibfnamefont {Apostolos}\
  \bibnamefont {Pilaftsis}}\ and\ \bibinfo {author} {\bibfnamefont {Thomas
  E.~J.}\ \bibnamefont {Underwood}},\ }\bibfield  {title} {\enquote {\bibinfo
  {title} {{Resonant leptogenesis}},}\ }\href {\doibase
  10.1016/j.nuclphysb.2004.05.029} {\bibfield  {journal} {\bibinfo  {journal}
  {Nucl. Phys.}\ }\textbf {\bibinfo {volume} {B692}},\ \bibinfo {pages}
  {303--345} (\bibinfo {year} {2004})},\ \Eprint
  {http://arxiv.org/abs/hep-ph/0309342} {arXiv:hep-ph/0309342 [hep-ph]}
  \BibitemShut {NoStop}%
\bibitem [{\citenamefont {Akhmedov}\ \emph {et~al.}(1998)\citenamefont
  {Akhmedov}, \citenamefont {Rubakov},\ and\ \citenamefont
  {Smirnov}}]{Akhmedov:1998qx}%
  \BibitemOpen
  \bibfield  {author} {\bibinfo {author} {\bibfnamefont {Evgeny~K.}\
  \bibnamefont {Akhmedov}}, \bibinfo {author} {\bibfnamefont {V.~A.}\
  \bibnamefont {Rubakov}}, \ and\ \bibinfo {author} {\bibfnamefont {A.~{\relax
  Yu}.}\ \bibnamefont {Smirnov}},\ }\bibfield  {title} {\enquote {\bibinfo
  {title} {{Baryogenesis via neutrino oscillations}},}\ }\href {\doibase
  10.1103/PhysRevLett.81.1359} {\bibfield  {journal} {\bibinfo  {journal}
  {Phys. Rev. Lett.}\ }\textbf {\bibinfo {volume} {81}},\ \bibinfo {pages}
  {1359--1362} (\bibinfo {year} {1998})},\ \Eprint
  {http://arxiv.org/abs/hep-ph/9803255} {arXiv:hep-ph/9803255 [hep-ph]}
  \BibitemShut {NoStop}%
\bibitem [{\citenamefont {Asaka}\ and\ \citenamefont
  {Shaposhnikov}(2005)}]{Asaka:2005pn}%
  \BibitemOpen
  \bibfield  {author} {\bibinfo {author} {\bibfnamefont {Takehiko}\
  \bibnamefont {Asaka}}\ and\ \bibinfo {author} {\bibfnamefont {Mikhail}\
  \bibnamefont {Shaposhnikov}},\ }\bibfield  {title} {\enquote {\bibinfo
  {title} {{The $\nu$MSM, dark matter and baryon asymmetry of the universe}},}\
  }\href {\doibase 10.1016/j.physletb.2005.06.020} {\bibfield  {journal}
  {\bibinfo  {journal} {Phys. Lett.}\ }\textbf {\bibinfo {volume} {B620}},\
  \bibinfo {pages} {17--26} (\bibinfo {year} {2005})},\ \Eprint
  {http://arxiv.org/abs/hep-ph/0505013} {arXiv:hep-ph/0505013 [hep-ph]}
  \BibitemShut {NoStop}%
\bibitem [{\citenamefont {Shaposhnikov}(2007)}]{Shaposhnikov:2006nn}%
  \BibitemOpen
  \bibfield  {author} {\bibinfo {author} {\bibfnamefont {Mikhail}\ \bibnamefont
  {Shaposhnikov}},\ }\bibfield  {title} {\enquote {\bibinfo {title} {{A
  Possible symmetry of the $\nu$MSM}},}\ }\href {\doibase
  10.1016/j.nuclphysb.2006.11.003} {\bibfield  {journal} {\bibinfo  {journal}
  {Nucl. Phys.}\ }\textbf {\bibinfo {volume} {B763}},\ \bibinfo {pages}
  {49--59} (\bibinfo {year} {2007})},\ \Eprint
  {http://arxiv.org/abs/hep-ph/0605047} {arXiv:hep-ph/0605047 [hep-ph]}
  \BibitemShut {NoStop}%
\bibitem [{\citenamefont {Shaposhnikov}(2008)}]{Shaposhnikov:2008pf}%
  \BibitemOpen
  \bibfield  {author} {\bibinfo {author} {\bibfnamefont {Mikhail}\ \bibnamefont
  {Shaposhnikov}},\ }\bibfield  {title} {\enquote {\bibinfo {title} {{The
  $\nu$MSM, leptonic asymmetries, and properties of singlet fermions}},}\
  }\href {\doibase 10.1088/1126-6708/2008/08/008} {\bibfield  {journal}
  {\bibinfo  {journal} {JHEP}\ }\textbf {\bibinfo {volume} {08}},\ \bibinfo
  {pages} {008} (\bibinfo {year} {2008})},\ \Eprint
  {http://arxiv.org/abs/0804.4542} {arXiv:0804.4542 [hep-ph]} \BibitemShut
  {NoStop}%
\bibitem [{\citenamefont {Asaka}\ and\ \citenamefont
  {Ishida}(2010)}]{Asaka:2010kk}%
  \BibitemOpen
  \bibfield  {author} {\bibinfo {author} {\bibfnamefont {Takehiko}\
  \bibnamefont {Asaka}}\ and\ \bibinfo {author} {\bibfnamefont {Hiroyuki}\
  \bibnamefont {Ishida}},\ }\bibfield  {title} {\enquote {\bibinfo {title}
  {{Flavour Mixing of Neutrinos and Baryon Asymmetry of the Universe}},}\
  }\href {\doibase 10.1016/j.physletb.2010.07.016} {\bibfield  {journal}
  {\bibinfo  {journal} {Phys. Lett.}\ }\textbf {\bibinfo {volume} {B692}},\
  \bibinfo {pages} {105--113} (\bibinfo {year} {2010})},\ \Eprint
  {http://arxiv.org/abs/1004.5491} {arXiv:1004.5491 [hep-ph]} \BibitemShut
  {NoStop}%
\bibitem [{\citenamefont {Anisimov}\ \emph
  {et~al.}(2011{\natexlab{a}})\citenamefont {Anisimov}, \citenamefont {Besak},\
  and\ \citenamefont {Bodeker}}]{Anisimov:2010gy}%
  \BibitemOpen
  \bibfield  {author} {\bibinfo {author} {\bibfnamefont {Alexey}\ \bibnamefont
  {Anisimov}}, \bibinfo {author} {\bibfnamefont {Denis}\ \bibnamefont {Besak}},
  \ and\ \bibinfo {author} {\bibfnamefont {Dietrich}\ \bibnamefont {Bodeker}},\
  }\bibfield  {title} {\enquote {\bibinfo {title} {{Thermal production of
  relativistic Majorana neutrinos: Strong enhancement by multiple soft
  scattering}},}\ }\href {\doibase 10.1088/1475-7516/2011/03/042} {\bibfield
  {journal} {\bibinfo  {journal} {JCAP}\ }\textbf {\bibinfo {volume} {1103}},\
  \bibinfo {pages} {042} (\bibinfo {year} {2011}{\natexlab{a}})},\ \Eprint
  {http://arxiv.org/abs/1012.3784} {arXiv:1012.3784 [hep-ph]} \BibitemShut
  {NoStop}%
\bibitem [{\citenamefont {Asaka}\ \emph {et~al.}(2012)\citenamefont {Asaka},
  \citenamefont {Eijima},\ and\ \citenamefont {Ishida}}]{Asaka:2011wq}%
  \BibitemOpen
  \bibfield  {author} {\bibinfo {author} {\bibfnamefont {Takehiko}\
  \bibnamefont {Asaka}}, \bibinfo {author} {\bibfnamefont {Shintaro}\
  \bibnamefont {Eijima}}, \ and\ \bibinfo {author} {\bibfnamefont {Hiroyuki}\
  \bibnamefont {Ishida}},\ }\bibfield  {title} {\enquote {\bibinfo {title}
  {{Kinetic Equations for Baryogenesis via Sterile Neutrino Oscillation}},}\
  }\href {\doibase 10.1088/1475-7516/2012/02/021} {\bibfield  {journal}
  {\bibinfo  {journal} {JCAP}\ }\textbf {\bibinfo {volume} {1202}},\ \bibinfo
  {pages} {021} (\bibinfo {year} {2012})},\ \Eprint
  {http://arxiv.org/abs/1112.5565} {arXiv:1112.5565 [hep-ph]} \BibitemShut
  {NoStop}%
\bibitem [{\citenamefont {Besak}\ and\ \citenamefont
  {Bodeker}(2012)}]{Besak:2012qm}%
  \BibitemOpen
  \bibfield  {author} {\bibinfo {author} {\bibfnamefont {Denis}\ \bibnamefont
  {Besak}}\ and\ \bibinfo {author} {\bibfnamefont {Dietrich}\ \bibnamefont
  {Bodeker}},\ }\bibfield  {title} {\enquote {\bibinfo {title} {{Thermal
  production of ultrarelativistic right-handed neutrinos: Complete
  leading-order results}},}\ }\href {\doibase 10.1088/1475-7516/2012/03/029}
  {\bibfield  {journal} {\bibinfo  {journal} {JCAP}\ }\textbf {\bibinfo
  {volume} {1203}},\ \bibinfo {pages} {029} (\bibinfo {year} {2012})},\ \Eprint
  {http://arxiv.org/abs/1202.1288} {arXiv:1202.1288 [hep-ph]} \BibitemShut
  {NoStop}%
\bibitem [{\citenamefont {Canetti}\ \emph
  {et~al.}(2013{\natexlab{a}})\citenamefont {Canetti}, \citenamefont {Drewes},\
  and\ \citenamefont {Shaposhnikov}}]{Canetti:2012vf}%
  \BibitemOpen
  \bibfield  {author} {\bibinfo {author} {\bibfnamefont {Laurent}\ \bibnamefont
  {Canetti}}, \bibinfo {author} {\bibfnamefont {Marco}\ \bibnamefont {Drewes}},
  \ and\ \bibinfo {author} {\bibfnamefont {Mikhail}\ \bibnamefont
  {Shaposhnikov}},\ }\bibfield  {title} {\enquote {\bibinfo {title} {{Sterile
  Neutrinos as the Origin of Dark and Baryonic Matter}},}\ }\href {\doibase
  10.1103/PhysRevLett.110.061801} {\bibfield  {journal} {\bibinfo  {journal}
  {Phys. Rev. Lett.}\ }\textbf {\bibinfo {volume} {110}},\ \bibinfo {pages}
  {061801} (\bibinfo {year} {2013}{\natexlab{a}})},\ \Eprint
  {http://arxiv.org/abs/1204.3902} {arXiv:1204.3902 [hep-ph]} \BibitemShut
  {NoStop}%
\bibitem [{\citenamefont {Drewes}\ and\ \citenamefont
  {Garbrecht}(2013)}]{Drewes:2012ma}%
  \BibitemOpen
  \bibfield  {author} {\bibinfo {author} {\bibfnamefont {Marco}\ \bibnamefont
  {Drewes}}\ and\ \bibinfo {author} {\bibfnamefont {Björn}\ \bibnamefont
  {Garbrecht}},\ }\bibfield  {title} {\enquote {\bibinfo {title} {{Leptogenesis
  from a GeV Seesaw without Mass Degeneracy}},}\ }\href {\doibase
  10.1007/JHEP03(2013)096} {\bibfield  {journal} {\bibinfo  {journal} {JHEP}\
  }\textbf {\bibinfo {volume} {03}},\ \bibinfo {pages} {096} (\bibinfo {year}
  {2013})},\ \Eprint {http://arxiv.org/abs/1206.5537} {arXiv:1206.5537
  [hep-ph]} \BibitemShut {NoStop}%
\bibitem [{\citenamefont {Canetti}\ \emph
  {et~al.}(2013{\natexlab{b}})\citenamefont {Canetti}, \citenamefont {Drewes},
  \citenamefont {Frossard},\ and\ \citenamefont
  {Shaposhnikov}}]{Canetti:2012kh}%
  \BibitemOpen
  \bibfield  {author} {\bibinfo {author} {\bibfnamefont {Laurent}\ \bibnamefont
  {Canetti}}, \bibinfo {author} {\bibfnamefont {Marco}\ \bibnamefont {Drewes}},
  \bibinfo {author} {\bibfnamefont {Tibor}\ \bibnamefont {Frossard}}, \ and\
  \bibinfo {author} {\bibfnamefont {Mikhail}\ \bibnamefont {Shaposhnikov}},\
  }\bibfield  {title} {\enquote {\bibinfo {title} {{Dark Matter, Baryogenesis
  and Neutrino Oscillations from Right Handed Neutrinos}},}\ }\href {\doibase
  10.1103/PhysRevD.87.093006} {\bibfield  {journal} {\bibinfo  {journal} {Phys.
  Rev.}\ }\textbf {\bibinfo {volume} {D87}},\ \bibinfo {pages} {093006}
  (\bibinfo {year} {2013}{\natexlab{b}})},\ \Eprint
  {http://arxiv.org/abs/1208.4607} {arXiv:1208.4607 [hep-ph]} \BibitemShut
  {NoStop}%
\bibitem [{\citenamefont {Shuve}\ and\ \citenamefont
  {Yavin}(2014)}]{Shuve:2014zua}%
  \BibitemOpen
  \bibfield  {author} {\bibinfo {author} {\bibfnamefont {Brian}\ \bibnamefont
  {Shuve}}\ and\ \bibinfo {author} {\bibfnamefont {Itay}\ \bibnamefont
  {Yavin}},\ }\bibfield  {title} {\enquote {\bibinfo {title} {{Baryogenesis
  through Neutrino Oscillations: A Unified Perspective}},}\ }\href {\doibase
  10.1103/PhysRevD.89.075014} {\bibfield  {journal} {\bibinfo  {journal} {Phys.
  Rev.}\ }\textbf {\bibinfo {volume} {D89}},\ \bibinfo {pages} {075014}
  (\bibinfo {year} {2014})},\ \Eprint {http://arxiv.org/abs/1401.2459}
  {arXiv:1401.2459 [hep-ph]} \BibitemShut {NoStop}%
\bibitem [{\citenamefont {Bodeker}\ and\ \citenamefont
  {Laine}(2014)}]{Bodeker:2014hqa}%
  \BibitemOpen
  \bibfield  {author} {\bibinfo {author} {\bibfnamefont {D.}~\bibnamefont
  {Bodeker}}\ and\ \bibinfo {author} {\bibfnamefont {M.}~\bibnamefont
  {Laine}},\ }\bibfield  {title} {\enquote {\bibinfo {title} {{Kubo relations
  and radiative corrections for lepton number washout}},}\ }\href {\doibase
  10.1088/1475-7516/2014/05/041} {\bibfield  {journal} {\bibinfo  {journal}
  {JCAP}\ }\textbf {\bibinfo {volume} {1405}},\ \bibinfo {pages} {041}
  (\bibinfo {year} {2014})},\ \Eprint {http://arxiv.org/abs/1403.2755}
  {arXiv:1403.2755 [hep-ph]} \BibitemShut {NoStop}%
\bibitem [{\citenamefont {Abada}\ \emph {et~al.}(2015)\citenamefont {Abada},
  \citenamefont {Arcadi}, \citenamefont {Domcke},\ and\ \citenamefont
  {Lucente}}]{Abada:2015rta}%
  \BibitemOpen
  \bibfield  {author} {\bibinfo {author} {\bibfnamefont {Asmaa}\ \bibnamefont
  {Abada}}, \bibinfo {author} {\bibfnamefont {Giorgio}\ \bibnamefont {Arcadi}},
  \bibinfo {author} {\bibfnamefont {Valerie}\ \bibnamefont {Domcke}}, \ and\
  \bibinfo {author} {\bibfnamefont {Michele}\ \bibnamefont {Lucente}},\
  }\bibfield  {title} {\enquote {\bibinfo {title} {{Lepton number violation as
  a key to low-scale leptogenesis}},}\ }\href {\doibase
  10.1088/1475-7516/2015/11/041} {\bibfield  {journal} {\bibinfo  {journal}
  {JCAP}\ }\textbf {\bibinfo {volume} {1511}},\ \bibinfo {pages} {041}
  (\bibinfo {year} {2015})},\ \Eprint {http://arxiv.org/abs/1507.06215}
  {arXiv:1507.06215 [hep-ph]} \BibitemShut {NoStop}%
\bibitem [{\citenamefont {Hernández}\ \emph {et~al.}(2015)\citenamefont
  {Hernández}, \citenamefont {Kekic}, \citenamefont {López-Pavón},
  \citenamefont {Racker},\ and\ \citenamefont {Rius}}]{Hernandez:2015wna}%
  \BibitemOpen
  \bibfield  {author} {\bibinfo {author} {\bibfnamefont {P.}~\bibnamefont
  {Hernández}}, \bibinfo {author} {\bibfnamefont {M.}~\bibnamefont {Kekic}},
  \bibinfo {author} {\bibfnamefont {J.}~\bibnamefont {López-Pavón}}, \bibinfo
  {author} {\bibfnamefont {J.}~\bibnamefont {Racker}}, \ and\ \bibinfo {author}
  {\bibfnamefont {N.}~\bibnamefont {Rius}},\ }\bibfield  {title} {\enquote
  {\bibinfo {title} {{Leptogenesis in GeV scale seesaw models}},}\ }\href
  {\doibase 10.1007/JHEP10(2015)067} {\bibfield  {journal} {\bibinfo  {journal}
  {JHEP}\ }\textbf {\bibinfo {volume} {10}},\ \bibinfo {pages} {067} (\bibinfo
  {year} {2015})},\ \Eprint {http://arxiv.org/abs/1508.03676} {arXiv:1508.03676
  [hep-ph]} \BibitemShut {NoStop}%
\bibitem [{\citenamefont {Ghiglieri}\ and\ \citenamefont
  {Laine}(2016)}]{Ghiglieri:2016xye}%
  \BibitemOpen
  \bibfield  {author} {\bibinfo {author} {\bibfnamefont {J.}~\bibnamefont
  {Ghiglieri}}\ and\ \bibinfo {author} {\bibfnamefont {M.}~\bibnamefont
  {Laine}},\ }\bibfield  {title} {\enquote {\bibinfo {title} {{Neutrino
  dynamics below the electroweak crossover}},}\ }\href {\doibase
  10.1088/1475-7516/2016/07/015} {\bibfield  {journal} {\bibinfo  {journal}
  {JCAP}\ }\textbf {\bibinfo {volume} {1607}},\ \bibinfo {pages} {015}
  (\bibinfo {year} {2016})},\ \Eprint {http://arxiv.org/abs/1605.07720}
  {arXiv:1605.07720 [hep-ph]} \BibitemShut {NoStop}%
\bibitem [{\citenamefont {Hambye}\ and\ \citenamefont
  {Teresi}(2016)}]{Hambye:2016sby}%
  \BibitemOpen
  \bibfield  {author} {\bibinfo {author} {\bibfnamefont {Thomas}\ \bibnamefont
  {Hambye}}\ and\ \bibinfo {author} {\bibfnamefont {Daniele}\ \bibnamefont
  {Teresi}},\ }\bibfield  {title} {\enquote {\bibinfo {title} {{Higgs doublet
  decay as the origin of the baryon asymmetry}},}\ }\href {\doibase
  10.1103/PhysRevLett.117.091801} {\bibfield  {journal} {\bibinfo  {journal}
  {Phys. Rev. Lett.}\ }\textbf {\bibinfo {volume} {117}},\ \bibinfo {pages}
  {091801} (\bibinfo {year} {2016})},\ \Eprint
  {http://arxiv.org/abs/1606.00017} {arXiv:1606.00017 [hep-ph]} \BibitemShut
  {NoStop}%
\bibitem [{\citenamefont {Hambye}\ and\ \citenamefont
  {Teresi}(2017)}]{Hambye:2017elz}%
  \BibitemOpen
  \bibfield  {author} {\bibinfo {author} {\bibfnamefont {Thomas}\ \bibnamefont
  {Hambye}}\ and\ \bibinfo {author} {\bibfnamefont {Daniele}\ \bibnamefont
  {Teresi}},\ }\bibfield  {title} {\enquote {\bibinfo {title} {{Baryogenesis
  from L-Violating Higgs-Doublet Decay in the Density-Matrix Formalism}},}\
  }\href {\doibase 10.1103/PhysRevD.96.015031} {\bibfield  {journal} {\bibinfo
  {journal} {Phys. Rev. D}\ }\textbf {\bibinfo {volume} {96}},\ \bibinfo
  {pages} {015031} (\bibinfo {year} {2017})},\ \Eprint
  {http://arxiv.org/abs/1705.00016} {arXiv:1705.00016 [hep-ph]} \BibitemShut
  {NoStop}%
\bibitem [{\citenamefont {Drewes}\ and\ \citenamefont
  {Eijima}(2016)}]{Drewes:2016lqo}%
  \BibitemOpen
  \bibfield  {author} {\bibinfo {author} {\bibfnamefont {Marco}\ \bibnamefont
  {Drewes}}\ and\ \bibinfo {author} {\bibfnamefont {Shintaro}\ \bibnamefont
  {Eijima}},\ }\bibfield  {title} {\enquote {\bibinfo {title} {{Neutrinoless
  double $\beta$ decay and low scale leptogenesis}},}\ }\href {\doibase
  10.1016/j.physletb.2016.09.054} {\bibfield  {journal} {\bibinfo  {journal}
  {Phys. Lett.}\ }\textbf {\bibinfo {volume} {B763}},\ \bibinfo {pages}
  {72--79} (\bibinfo {year} {2016})},\ \Eprint
  {http://arxiv.org/abs/1606.06221} {arXiv:1606.06221 [hep-ph]} \BibitemShut
  {NoStop}%
\bibitem [{\citenamefont {Asaka}\ \emph {et~al.}(2016)\citenamefont {Asaka},
  \citenamefont {Eijima},\ and\ \citenamefont {Ishida}}]{Asaka:2016zib}%
  \BibitemOpen
  \bibfield  {author} {\bibinfo {author} {\bibfnamefont {Takehiko}\
  \bibnamefont {Asaka}}, \bibinfo {author} {\bibfnamefont {Shintaro}\
  \bibnamefont {Eijima}}, \ and\ \bibinfo {author} {\bibfnamefont {Hiroyuki}\
  \bibnamefont {Ishida}},\ }\bibfield  {title} {\enquote {\bibinfo {title} {{On
  neutrinoless double beta decay in the $\nu$MSM}},}\ }\href {\doibase
  10.1016/j.physletb.2016.09.044} {\bibfield  {journal} {\bibinfo  {journal}
  {Phys. Lett.}\ }\textbf {\bibinfo {volume} {B762}},\ \bibinfo {pages}
  {371--375} (\bibinfo {year} {2016})},\ \Eprint
  {http://arxiv.org/abs/1606.06686} {arXiv:1606.06686 [hep-ph]} \BibitemShut
  {NoStop}%
\bibitem [{\citenamefont {Drewes}\ \emph {et~al.}(2016)\citenamefont {Drewes},
  \citenamefont {Garbrecht}, \citenamefont {Gueter},\ and\ \citenamefont
  {Klaric}}]{Drewes:2016gmt}%
  \BibitemOpen
  \bibfield  {author} {\bibinfo {author} {\bibfnamefont {Marco}\ \bibnamefont
  {Drewes}}, \bibinfo {author} {\bibfnamefont {Bjorn}\ \bibnamefont
  {Garbrecht}}, \bibinfo {author} {\bibfnamefont {Dario}\ \bibnamefont
  {Gueter}}, \ and\ \bibinfo {author} {\bibfnamefont {Juraj}\ \bibnamefont
  {Klaric}},\ }\bibfield  {title} {\enquote {\bibinfo {title} {{Leptogenesis
  from Oscillations of Heavy Neutrinos with Large Mixing Angles}},}\ }\href
  {\doibase 10.1007/JHEP12(2016)150} {\bibfield  {journal} {\bibinfo  {journal}
  {JHEP}\ }\textbf {\bibinfo {volume} {12}},\ \bibinfo {pages} {150} (\bibinfo
  {year} {2016})},\ \Eprint {http://arxiv.org/abs/1606.06690} {arXiv:1606.06690
  [hep-ph]} \BibitemShut {NoStop}%
\bibitem [{\citenamefont {Hernández}\ \emph {et~al.}(2016)\citenamefont
  {Hernández}, \citenamefont {Kekic}, \citenamefont {López-Pavón},
  \citenamefont {Racker},\ and\ \citenamefont {Salvado}}]{Hernandez:2016kel}%
  \BibitemOpen
  \bibfield  {author} {\bibinfo {author} {\bibfnamefont {P.}~\bibnamefont
  {Hernández}}, \bibinfo {author} {\bibfnamefont {M.}~\bibnamefont {Kekic}},
  \bibinfo {author} {\bibfnamefont {J.}~\bibnamefont {López-Pavón}}, \bibinfo
  {author} {\bibfnamefont {J.}~\bibnamefont {Racker}}, \ and\ \bibinfo {author}
  {\bibfnamefont {J.}~\bibnamefont {Salvado}},\ }\bibfield  {title} {\enquote
  {\bibinfo {title} {{Testable Baryogenesis in Seesaw Models}},}\ }\href
  {\doibase 10.1007/JHEP08(2016)157} {\bibfield  {journal} {\bibinfo  {journal}
  {JHEP}\ }\textbf {\bibinfo {volume} {08}},\ \bibinfo {pages} {157} (\bibinfo
  {year} {2016})},\ \Eprint {http://arxiv.org/abs/1606.06719} {arXiv:1606.06719
  [hep-ph]} \BibitemShut {NoStop}%
\bibitem [{\citenamefont {Drewes}\ \emph {et~al.}(2017)\citenamefont {Drewes},
  \citenamefont {Garbrecht}, \citenamefont {Gueter},\ and\ \citenamefont
  {Klaric}}]{Drewes:2016jae}%
  \BibitemOpen
  \bibfield  {author} {\bibinfo {author} {\bibfnamefont {Marco}\ \bibnamefont
  {Drewes}}, \bibinfo {author} {\bibfnamefont {Bjorn}\ \bibnamefont
  {Garbrecht}}, \bibinfo {author} {\bibfnamefont {Dario}\ \bibnamefont
  {Gueter}}, \ and\ \bibinfo {author} {\bibfnamefont {Juraj}\ \bibnamefont
  {Klaric}},\ }\bibfield  {title} {\enquote {\bibinfo {title} {{Testing the low
  scale seesaw and leptogenesis}},}\ }\href {\doibase 10.1007/JHEP08(2017)018}
  {\bibfield  {journal} {\bibinfo  {journal} {JHEP}\ }\textbf {\bibinfo
  {volume} {08}},\ \bibinfo {pages} {018} (\bibinfo {year} {2017})},\ \Eprint
  {http://arxiv.org/abs/1609.09069} {arXiv:1609.09069 [hep-ph]} \BibitemShut
  {NoStop}%
\bibitem [{\citenamefont {Asaka}\ \emph {et~al.}(2017)\citenamefont {Asaka},
  \citenamefont {Eijima}, \citenamefont {Ishida}, \citenamefont {Minogawa},\
  and\ \citenamefont {Yoshii}}]{Asaka:2017rdj}%
  \BibitemOpen
  \bibfield  {author} {\bibinfo {author} {\bibfnamefont {Takehiko}\
  \bibnamefont {Asaka}}, \bibinfo {author} {\bibfnamefont {Shintaro}\
  \bibnamefont {Eijima}}, \bibinfo {author} {\bibfnamefont {Hiroyuki}\
  \bibnamefont {Ishida}}, \bibinfo {author} {\bibfnamefont {Kosuke}\
  \bibnamefont {Minogawa}}, \ and\ \bibinfo {author} {\bibfnamefont {Tomoya}\
  \bibnamefont {Yoshii}},\ }\bibfield  {title} {\enquote {\bibinfo {title}
  {{Initial condition for baryogenesis via neutrino oscillation}},}\ }\href
  {\doibase 10.1103/PhysRevD.96.083010} {\bibfield  {journal} {\bibinfo
  {journal} {Phys. Rev.}\ }\textbf {\bibinfo {volume} {D96}},\ \bibinfo {pages}
  {083010} (\bibinfo {year} {2017})},\ \Eprint
  {http://arxiv.org/abs/1704.02692} {arXiv:1704.02692 [hep-ph]} \BibitemShut
  {NoStop}%
\bibitem [{\citenamefont {Eijima}\ and\ \citenamefont
  {Shaposhnikov}(2017)}]{Eijima:2017anv}%
  \BibitemOpen
  \bibfield  {author} {\bibinfo {author} {\bibfnamefont {Shintaro}\
  \bibnamefont {Eijima}}\ and\ \bibinfo {author} {\bibfnamefont {Mikhail}\
  \bibnamefont {Shaposhnikov}},\ }\bibfield  {title} {\enquote {\bibinfo
  {title} {{Fermion number violating effects in low scale leptogenesis}},}\
  }\href {\doibase 10.1016/j.physletb.2017.05.068} {\bibfield  {journal}
  {\bibinfo  {journal} {Phys. Lett.}\ }\textbf {\bibinfo {volume} {B771}},\
  \bibinfo {pages} {288--296} (\bibinfo {year} {2017})},\ \Eprint
  {http://arxiv.org/abs/1703.06085} {arXiv:1703.06085 [hep-ph]} \BibitemShut
  {NoStop}%
\bibitem [{\citenamefont {Ghiglieri}\ and\ \citenamefont
  {Laine}(2017)}]{Ghiglieri:2017gjz}%
  \BibitemOpen
  \bibfield  {author} {\bibinfo {author} {\bibfnamefont {J.}~\bibnamefont
  {Ghiglieri}}\ and\ \bibinfo {author} {\bibfnamefont {M.}~\bibnamefont
  {Laine}},\ }\bibfield  {title} {\enquote {\bibinfo {title} {{GeV-scale hot
  sterile neutrino oscillations: a derivation of evolution equations}},}\
  }\href {\doibase 10.1007/JHEP05(2017)132} {\bibfield  {journal} {\bibinfo
  {journal} {JHEP}\ }\textbf {\bibinfo {volume} {05}},\ \bibinfo {pages} {132}
  (\bibinfo {year} {2017})},\ \Eprint {http://arxiv.org/abs/1703.06087}
  {arXiv:1703.06087 [hep-ph]} \BibitemShut {NoStop}%
\bibitem [{\citenamefont {Eijima}\ \emph {et~al.}(2017)\citenamefont {Eijima},
  \citenamefont {Shaposhnikov},\ and\ \citenamefont
  {Timiryasov}}]{Eijima:2017cxr}%
  \BibitemOpen
  \bibfield  {author} {\bibinfo {author} {\bibfnamefont {S.}~\bibnamefont
  {Eijima}}, \bibinfo {author} {\bibfnamefont {M.}~\bibnamefont
  {Shaposhnikov}}, \ and\ \bibinfo {author} {\bibfnamefont {I.}~\bibnamefont
  {Timiryasov}},\ }\bibfield  {title} {\enquote {\bibinfo {title} {{Freeze-out
  of baryon number in low-scale leptogenesis}},}\ }\href {\doibase
  10.1088/1475-7516/2017/11/030} {\bibfield  {journal} {\bibinfo  {journal}
  {JCAP}\ }\textbf {\bibinfo {volume} {1711}},\ \bibinfo {pages} {030}
  (\bibinfo {year} {2017})},\ \Eprint {http://arxiv.org/abs/1709.07834}
  {arXiv:1709.07834 [hep-ph]} \BibitemShut {NoStop}%
\bibitem [{\citenamefont {Antusch}\ \emph {et~al.}(2018)\citenamefont
  {Antusch}, \citenamefont {Cazzato}, \citenamefont {Drewes}, \citenamefont
  {Fischer}, \citenamefont {Garbrecht}, \citenamefont {Gueter},\ and\
  \citenamefont {Klaric}}]{Antusch:2017pkq}%
  \BibitemOpen
  \bibfield  {author} {\bibinfo {author} {\bibfnamefont {Stefan}\ \bibnamefont
  {Antusch}}, \bibinfo {author} {\bibfnamefont {Eros}\ \bibnamefont {Cazzato}},
  \bibinfo {author} {\bibfnamefont {Marco}\ \bibnamefont {Drewes}}, \bibinfo
  {author} {\bibfnamefont {Oliver}\ \bibnamefont {Fischer}}, \bibinfo {author}
  {\bibfnamefont {Bjorn}\ \bibnamefont {Garbrecht}}, \bibinfo {author}
  {\bibfnamefont {Dario}\ \bibnamefont {Gueter}}, \ and\ \bibinfo {author}
  {\bibfnamefont {Juraj}\ \bibnamefont {Klaric}},\ }\bibfield  {title}
  {\enquote {\bibinfo {title} {{Probing Leptogenesis at Future Colliders}},}\
  }\href {\doibase 10.1007/JHEP09(2018)124} {\bibfield  {journal} {\bibinfo
  {journal} {JHEP}\ }\textbf {\bibinfo {volume} {09}},\ \bibinfo {pages} {124}
  (\bibinfo {year} {2018})},\ \Eprint {http://arxiv.org/abs/1710.03744}
  {arXiv:1710.03744 [hep-ph]} \BibitemShut {NoStop}%
\bibitem [{\citenamefont {Ghiglieri}\ and\ \citenamefont
  {Laine}(2018)}]{Ghiglieri:2017csp}%
  \BibitemOpen
  \bibfield  {author} {\bibinfo {author} {\bibfnamefont {J.}~\bibnamefont
  {Ghiglieri}}\ and\ \bibinfo {author} {\bibfnamefont {M.}~\bibnamefont
  {Laine}},\ }\bibfield  {title} {\enquote {\bibinfo {title} {{GeV-scale hot
  sterile neutrino oscillations: a numerical solution}},}\ }\href {\doibase
  10.1007/JHEP02(2018)078} {\bibfield  {journal} {\bibinfo  {journal} {JHEP}\
  }\textbf {\bibinfo {volume} {02}},\ \bibinfo {pages} {078} (\bibinfo {year}
  {2018})},\ \Eprint {http://arxiv.org/abs/1711.08469} {arXiv:1711.08469
  [hep-ph]} \BibitemShut {NoStop}%
\bibitem [{\citenamefont {Eijima}\ \emph {et~al.}(2019)\citenamefont {Eijima},
  \citenamefont {Shaposhnikov},\ and\ \citenamefont
  {Timiryasov}}]{Eijima:2018qke}%
  \BibitemOpen
  \bibfield  {author} {\bibinfo {author} {\bibfnamefont {S.}~\bibnamefont
  {Eijima}}, \bibinfo {author} {\bibfnamefont {M.}~\bibnamefont
  {Shaposhnikov}}, \ and\ \bibinfo {author} {\bibfnamefont {I.}~\bibnamefont
  {Timiryasov}},\ }\bibfield  {title} {\enquote {\bibinfo {title} {{Parameter
  space of baryogenesis in the $\nu$MSM}},}\ }\href {\doibase
  10.1007/JHEP07(2019)077} {\bibfield  {journal} {\bibinfo  {journal} {JHEP}\
  }\textbf {\bibinfo {volume} {07}},\ \bibinfo {pages} {077} (\bibinfo {year}
  {2019})},\ \Eprint {http://arxiv.org/abs/1808.10833} {arXiv:1808.10833
  [hep-ph]} \BibitemShut {NoStop}%
\bibitem [{\citenamefont {Ghiglieri}\ and\ \citenamefont
  {Laine}(2019{\natexlab{a}})}]{Ghiglieri:2018wbs}%
  \BibitemOpen
  \bibfield  {author} {\bibinfo {author} {\bibfnamefont {J.}~\bibnamefont
  {Ghiglieri}}\ and\ \bibinfo {author} {\bibfnamefont {M.}~\bibnamefont
  {Laine}},\ }\bibfield  {title} {\enquote {\bibinfo {title} {{Precision study
  of GeV-scale resonant leptogenesis}},}\ }\href {\doibase
  10.1007/JHEP02(2019)014} {\bibfield  {journal} {\bibinfo  {journal} {JHEP}\
  }\textbf {\bibinfo {volume} {02}},\ \bibinfo {pages} {014} (\bibinfo {year}
  {2019}{\natexlab{a}})},\ \Eprint {http://arxiv.org/abs/1811.01971}
  {arXiv:1811.01971 [hep-ph]} \BibitemShut {NoStop}%
\bibitem [{\citenamefont {Ghiglieri}\ and\ \citenamefont
  {Laine}(2019{\natexlab{b}})}]{Ghiglieri:2019kbw}%
  \BibitemOpen
  \bibfield  {author} {\bibinfo {author} {\bibfnamefont {J.}~\bibnamefont
  {Ghiglieri}}\ and\ \bibinfo {author} {\bibfnamefont {M.}~\bibnamefont
  {Laine}},\ }\bibfield  {title} {\enquote {\bibinfo {title} {{Sterile Neutrino
  Dark Matter via Gev-Scale Leptogenesis?}}}\ }\href {\doibase
  10.1007/JHEP07(2019)078} {\bibfield  {journal} {\bibinfo  {journal} {JHEP}\
  }\textbf {\bibinfo {volume} {07}},\ \bibinfo {pages} {078} (\bibinfo {year}
  {2019}{\natexlab{b}})},\ \Eprint {http://arxiv.org/abs/1905.08814}
  {arXiv:1905.08814 [hep-ph]} \BibitemShut {NoStop}%
\bibitem [{\citenamefont {Bödeker}\ and\ \citenamefont
  {Schröder}(2020)}]{Bodeker:2019rvr}%
  \BibitemOpen
  \bibfield  {author} {\bibinfo {author} {\bibfnamefont {Dietrich}\
  \bibnamefont {Bödeker}}\ and\ \bibinfo {author} {\bibfnamefont {Dennis}\
  \bibnamefont {Schröder}},\ }\bibfield  {title} {\enquote {\bibinfo {title}
  {{Kinetic Equations for Sterile Neutrinos from Thermal Fluctuations}},}\
  }\href {\doibase 10.1088/1475-7516/2020/02/033} {\bibfield  {journal}
  {\bibinfo  {journal} {JCAP}\ }\textbf {\bibinfo {volume} {02}},\ \bibinfo
  {pages} {033} (\bibinfo {year} {2020})},\ \Eprint
  {http://arxiv.org/abs/1911.05092} {arXiv:1911.05092 [hep-ph]} \BibitemShut
  {NoStop}%
\bibitem [{\citenamefont {Ghiglieri}\ and\ \citenamefont
  {Laine}(2020)}]{Ghiglieri:2020ulj}%
  \BibitemOpen
  \bibfield  {author} {\bibinfo {author} {\bibfnamefont {J.}~\bibnamefont
  {Ghiglieri}}\ and\ \bibinfo {author} {\bibfnamefont {M.}~\bibnamefont
  {Laine}},\ }\bibfield  {title} {\enquote {\bibinfo {title} {{Sterile Neutrino
  Dark Matter via Coinciding Resonances}},}\ }\href {\doibase
  10.1088/1475-7516/2020/07/012} {\bibfield  {journal} {\bibinfo  {journal}
  {JCAP}\ }\textbf {\bibinfo {volume} {07}},\ \bibinfo {pages} {012} (\bibinfo
  {year} {2020})},\ \Eprint {http://arxiv.org/abs/2004.10766} {arXiv:2004.10766
  [hep-ph]} \BibitemShut {NoStop}%
\bibitem [{\citenamefont {Klari\'c}\ \emph {et~al.}(2020)\citenamefont
  {Klari\'c}, \citenamefont {Shaposhnikov},\ and\ \citenamefont
  {Timiryasov}}]{Klaric:2020lov}%
  \BibitemOpen
  \bibfield  {author} {\bibinfo {author} {\bibfnamefont {Juraj}\ \bibnamefont
  {Klari\'c}}, \bibinfo {author} {\bibfnamefont {Mikhail}\ \bibnamefont
  {Shaposhnikov}}, \ and\ \bibinfo {author} {\bibfnamefont {Inar}\ \bibnamefont
  {Timiryasov}},\ }\bibfield  {title} {\enquote {\bibinfo {title} {{Uniting
  low-scale leptogeneses}},}\ }\href@noop {} {\  (\bibinfo {year} {2020})},\
  \Eprint {http://arxiv.org/abs/2008.13771} {arXiv:2008.13771 [hep-ph]}
  \BibitemShut {NoStop}%
\bibitem [{\citenamefont {Domcke}\ \emph {et~al.}(2020)\citenamefont {Domcke},
  \citenamefont {Drewes}, \citenamefont {Hufnagel},\ and\ \citenamefont
  {Lucente}}]{Domcke:2020ety}%
  \BibitemOpen
  \bibfield  {author} {\bibinfo {author} {\bibfnamefont {Valerie}\ \bibnamefont
  {Domcke}}, \bibinfo {author} {\bibfnamefont {Marco}\ \bibnamefont {Drewes}},
  \bibinfo {author} {\bibfnamefont {Marco}\ \bibnamefont {Hufnagel}}, \ and\
  \bibinfo {author} {\bibfnamefont {Michele}\ \bibnamefont {Lucente}},\
  }\bibfield  {title} {\enquote {\bibinfo {title} {{Mev-Scale Seesaw and
  Leptogenesis}},}\ }\href@noop {} {\  (\bibinfo {year} {2020})},\ \Eprint
  {http://arxiv.org/abs/2009.11678} {arXiv:2009.11678 [hep-ph]} \BibitemShut
  {NoStop}%
\bibitem [{\citenamefont {Eijima}\ \emph {et~al.}(2020)\citenamefont {Eijima},
  \citenamefont {Shaposhnikov},\ and\ \citenamefont
  {Timiryasov}}]{Eijima:2020shs}%
  \BibitemOpen
  \bibfield  {author} {\bibinfo {author} {\bibfnamefont {Shintaro}\
  \bibnamefont {Eijima}}, \bibinfo {author} {\bibfnamefont {Mikhail}\
  \bibnamefont {Shaposhnikov}}, \ and\ \bibinfo {author} {\bibfnamefont {Inar}\
  \bibnamefont {Timiryasov}},\ }\bibfield  {title} {\enquote {\bibinfo {title}
  {{Freeze-in generation of lepton asymmetries after baryogenesis in the
  $\nu$MSM}},}\ }\href@noop {} {\  (\bibinfo {year} {2020})},\ \Eprint
  {http://arxiv.org/abs/2011.12637} {arXiv:2011.12637 [hep-ph]} \BibitemShut
  {NoStop}%
\bibitem [{\citenamefont {De~Simone}\ and\ \citenamefont
  {Riotto}(2007{\natexlab{a}})}]{DeSimone:2007edo}%
  \BibitemOpen
  \bibfield  {author} {\bibinfo {author} {\bibfnamefont {Andrea}\ \bibnamefont
  {De~Simone}}\ and\ \bibinfo {author} {\bibfnamefont {Antonio}\ \bibnamefont
  {Riotto}},\ }\bibfield  {title} {\enquote {\bibinfo {title} {{On Resonant
  Leptogenesis}},}\ }\href {\doibase 10.1088/1475-7516/2007/08/013} {\bibfield
  {journal} {\bibinfo  {journal} {JCAP}\ }\textbf {\bibinfo {volume} {0708}},\
  \bibinfo {pages} {013} (\bibinfo {year} {2007}{\natexlab{a}})},\ \Eprint
  {http://arxiv.org/abs/0705.2183} {arXiv:0705.2183 [hep-ph]} \BibitemShut
  {NoStop}%
\bibitem [{\citenamefont {De~Simone}\ and\ \citenamefont
  {Riotto}(2007{\natexlab{b}})}]{DeSimone:2007gkc}%
  \BibitemOpen
  \bibfield  {author} {\bibinfo {author} {\bibfnamefont {Andrea}\ \bibnamefont
  {De~Simone}}\ and\ \bibinfo {author} {\bibfnamefont {Antonio}\ \bibnamefont
  {Riotto}},\ }\bibfield  {title} {\enquote {\bibinfo {title} {{Quantum
  Boltzmann Equations and Leptogenesis}},}\ }\href {\doibase
  10.1088/1475-7516/2007/08/002} {\bibfield  {journal} {\bibinfo  {journal}
  {JCAP}\ }\textbf {\bibinfo {volume} {0708}},\ \bibinfo {pages} {002}
  (\bibinfo {year} {2007}{\natexlab{b}})},\ \Eprint
  {http://arxiv.org/abs/hep-ph/0703175} {arXiv:hep-ph/0703175 [hep-ph]}
  \BibitemShut {NoStop}%
\bibitem [{\citenamefont {Garny}\ \emph {et~al.}(2010)\citenamefont {Garny},
  \citenamefont {Hohenegger}, \citenamefont {Kartavtsev},\ and\ \citenamefont
  {Lindner}}]{Garny:2009qn}%
  \BibitemOpen
  \bibfield  {author} {\bibinfo {author} {\bibfnamefont {M.}~\bibnamefont
  {Garny}}, \bibinfo {author} {\bibfnamefont {A.}~\bibnamefont {Hohenegger}},
  \bibinfo {author} {\bibfnamefont {A.}~\bibnamefont {Kartavtsev}}, \ and\
  \bibinfo {author} {\bibfnamefont {M.}~\bibnamefont {Lindner}},\ }\bibfield
  {title} {\enquote {\bibinfo {title} {{Systematic approach to leptogenesis in
  nonequilibrium QFT: Self-energy contribution to the CP-violating
  parameter}},}\ }\href {\doibase 10.1103/PhysRevD.81.085027} {\bibfield
  {journal} {\bibinfo  {journal} {Phys. Rev.}\ }\textbf {\bibinfo {volume}
  {D81}},\ \bibinfo {pages} {085027} (\bibinfo {year} {2010})},\ \Eprint
  {http://arxiv.org/abs/0911.4122} {arXiv:0911.4122 [hep-ph]} \BibitemShut
  {NoStop}%
\bibitem [{\citenamefont {Garny}\ \emph {et~al.}(2013)\citenamefont {Garny},
  \citenamefont {Kartavtsev},\ and\ \citenamefont {Hohenegger}}]{Garny:2011hg}%
  \BibitemOpen
  \bibfield  {author} {\bibinfo {author} {\bibfnamefont {Mathias}\ \bibnamefont
  {Garny}}, \bibinfo {author} {\bibfnamefont {Alexander}\ \bibnamefont
  {Kartavtsev}}, \ and\ \bibinfo {author} {\bibfnamefont {Andreas}\
  \bibnamefont {Hohenegger}},\ }\bibfield  {title} {\enquote {\bibinfo {title}
  {{Leptogenesis from first principles in the resonant regime}},}\ }\href
  {\doibase 10.1016/j.aop.2012.10.007} {\bibfield  {journal} {\bibinfo
  {journal} {Annals Phys.}\ }\textbf {\bibinfo {volume} {328}},\ \bibinfo
  {pages} {26--63} (\bibinfo {year} {2013})},\ \Eprint
  {http://arxiv.org/abs/1112.6428} {arXiv:1112.6428 [hep-ph]} \BibitemShut
  {NoStop}%
\bibitem [{\citenamefont {Iso}\ \emph {et~al.}(2014)\citenamefont {Iso},
  \citenamefont {Shimada},\ and\ \citenamefont {Yamanaka}}]{Iso:2013lba}%
  \BibitemOpen
  \bibfield  {author} {\bibinfo {author} {\bibfnamefont {Satoshi}\ \bibnamefont
  {Iso}}, \bibinfo {author} {\bibfnamefont {Kengo}\ \bibnamefont {Shimada}}, \
  and\ \bibinfo {author} {\bibfnamefont {Masato}\ \bibnamefont {Yamanaka}},\
  }\bibfield  {title} {\enquote {\bibinfo {title} {{Kadanoff-Baym approach to
  the thermal resonant leptogenesis}},}\ }\href {\doibase
  10.1007/JHEP04(2014)062} {\bibfield  {journal} {\bibinfo  {journal} {JHEP}\
  }\textbf {\bibinfo {volume} {04}},\ \bibinfo {pages} {062} (\bibinfo {year}
  {2014})},\ \Eprint {http://arxiv.org/abs/1312.7680} {arXiv:1312.7680
  [hep-ph]} \BibitemShut {NoStop}%
\bibitem [{\citenamefont {Garbrecht}\ and\ \citenamefont
  {Herranen}(2012)}]{Garbrecht:2011aw}%
  \BibitemOpen
  \bibfield  {author} {\bibinfo {author} {\bibfnamefont {Bjorn}\ \bibnamefont
  {Garbrecht}}\ and\ \bibinfo {author} {\bibfnamefont {Matti}\ \bibnamefont
  {Herranen}},\ }\bibfield  {title} {\enquote {\bibinfo {title} {{Effective
  Theory of Resonant Leptogenesis in the Closed-Time-Path Approach}},}\ }\href
  {\doibase 10.1016/j.nuclphysb.2012.03.009} {\bibfield  {journal} {\bibinfo
  {journal} {Nucl. Phys.}\ }\textbf {\bibinfo {volume} {B861}},\ \bibinfo
  {pages} {17--52} (\bibinfo {year} {2012})},\ \Eprint
  {http://arxiv.org/abs/1112.5954} {arXiv:1112.5954 [hep-ph]} \BibitemShut
  {NoStop}%
\bibitem [{\citenamefont {Bhupal~Dev}\ \emph {et~al.}(2014)\citenamefont
  {Bhupal~Dev}, \citenamefont {Millington}, \citenamefont {Pilaftsis},\ and\
  \citenamefont {Teresi}}]{Dev:2014laa}%
  \BibitemOpen
  \bibfield  {author} {\bibinfo {author} {\bibfnamefont {P.~S.}\ \bibnamefont
  {Bhupal~Dev}}, \bibinfo {author} {\bibfnamefont {Peter}\ \bibnamefont
  {Millington}}, \bibinfo {author} {\bibfnamefont {Apostolos}\ \bibnamefont
  {Pilaftsis}}, \ and\ \bibinfo {author} {\bibfnamefont {Daniele}\ \bibnamefont
  {Teresi}},\ }\bibfield  {title} {\enquote {\bibinfo {title} {{Flavour
  Covariant Transport Equations: an Application to Resonant Leptogenesis}},}\
  }\href {\doibase 10.1016/j.nuclphysb.2014.06.020} {\bibfield  {journal}
  {\bibinfo  {journal} {Nucl. Phys.}\ }\textbf {\bibinfo {volume} {B886}},\
  \bibinfo {pages} {569--664} (\bibinfo {year} {2014})},\ \Eprint
  {http://arxiv.org/abs/1404.1003} {arXiv:1404.1003 [hep-ph]} \BibitemShut
  {NoStop}%
\bibitem [{\citenamefont {Bhupal~Dev}\ \emph {et~al.}(2016)\citenamefont
  {Bhupal~Dev}, \citenamefont {Millington}, \citenamefont {Pilaftsis},\ and\
  \citenamefont {Teresi}}]{Dev:2014tpa}%
  \BibitemOpen
  \bibfield  {author} {\bibinfo {author} {\bibfnamefont {P.~S.}\ \bibnamefont
  {Bhupal~Dev}}, \bibinfo {author} {\bibfnamefont {Peter}\ \bibnamefont
  {Millington}}, \bibinfo {author} {\bibfnamefont {Apostolos}\ \bibnamefont
  {Pilaftsis}}, \ and\ \bibinfo {author} {\bibfnamefont {Daniele}\ \bibnamefont
  {Teresi}},\ }\bibfield  {title} {\enquote {\bibinfo {title} {{Flavour
  Covariant Formalism for Resonant Leptogenesis}},}\ }\bibfield  {booktitle}
  {\emph {\bibinfo {booktitle} {{Proceedings, 37th International Conference on
  High Energy Physics (ICHEP 2014): Valencia, Spain, July 2-9, 2014}}},\ }\href
  {\doibase 10.1016/j.nuclphysbps.2015.09.037} {\bibfield  {journal} {\bibinfo
  {journal} {Nucl. Part. Phys. Proc.}\ }\textbf {\bibinfo {volume} {273-275}},\
  \bibinfo {pages} {268--274} (\bibinfo {year} {2016})},\ \Eprint
  {http://arxiv.org/abs/1409.8263} {arXiv:1409.8263 [hep-ph]} \BibitemShut
  {NoStop}%
\bibitem [{\citenamefont {Garbrecht}\ \emph {et~al.}(2014)\citenamefont
  {Garbrecht}, \citenamefont {Gautier},\ and\ \citenamefont
  {Klaric}}]{Garbrecht:2014aga}%
  \BibitemOpen
  \bibfield  {author} {\bibinfo {author} {\bibfnamefont {Bjorn}\ \bibnamefont
  {Garbrecht}}, \bibinfo {author} {\bibfnamefont {Florian}\ \bibnamefont
  {Gautier}}, \ and\ \bibinfo {author} {\bibfnamefont {Juraj}\ \bibnamefont
  {Klaric}},\ }\bibfield  {title} {\enquote {\bibinfo {title} {{Strong Washout
  Approximation to Resonant Leptogenesis}},}\ }\href {\doibase
  10.1088/1475-7516/2014/09/033} {\bibfield  {journal} {\bibinfo  {journal}
  {JCAP}\ }\textbf {\bibinfo {volume} {1409}},\ \bibinfo {pages} {033}
  (\bibinfo {year} {2014})},\ \Eprint {http://arxiv.org/abs/1406.4190}
  {arXiv:1406.4190 [hep-ph]} \BibitemShut {NoStop}%
\bibitem [{\citenamefont {Dev}\ \emph {et~al.}(2015)\citenamefont {Dev},
  \citenamefont {Millington}, \citenamefont {Pilaftsis},\ and\ \citenamefont
  {Teresi}}]{Dev:2015wpa}%
  \BibitemOpen
  \bibfield  {author} {\bibinfo {author} {\bibfnamefont {P.~S.~Bhupal}\
  \bibnamefont {Dev}}, \bibinfo {author} {\bibfnamefont {Peter}\ \bibnamefont
  {Millington}}, \bibinfo {author} {\bibfnamefont {Apostolos}\ \bibnamefont
  {Pilaftsis}}, \ and\ \bibinfo {author} {\bibfnamefont {Daniele}\ \bibnamefont
  {Teresi}},\ }\bibfield  {title} {\enquote {\bibinfo {title} {{Corrigendum to
  "Flavour Covariant Transport Equations: an Application to Resonant
  Leptogenesis"}},}\ }\href {\doibase 10.1016/j.nuclphysb.2015.06.015}
  {\bibfield  {journal} {\bibinfo  {journal} {Nucl. Phys.}\ }\textbf {\bibinfo
  {volume} {B897}},\ \bibinfo {pages} {749--756} (\bibinfo {year} {2015})},\
  \Eprint {http://arxiv.org/abs/1504.07640} {arXiv:1504.07640 [hep-ph]}
  \BibitemShut {NoStop}%
\bibitem [{\citenamefont {Jiang}\ \emph {et~al.}(2020)\citenamefont {Jiang},
  \citenamefont {Tang}, \citenamefont {Yu},\ and\ \citenamefont
  {Zhang}}]{Jiang:2020kbt}%
  \BibitemOpen
  \bibfield  {author} {\bibinfo {author} {\bibfnamefont {Xue-Min}\ \bibnamefont
  {Jiang}}, \bibinfo {author} {\bibfnamefont {Yi-Lei}\ \bibnamefont {Tang}},
  \bibinfo {author} {\bibfnamefont {Zhao-Huan}\ \bibnamefont {Yu}}, \ and\
  \bibinfo {author} {\bibfnamefont {Hong-Hao}\ \bibnamefont {Zhang}},\
  }\bibfield  {title} {\enquote {\bibinfo {title} {{$1 \leftrightarrow 2$
  Processes of a Sterile Neutrino Around Electroweak Scale in the Thermal
  Plasma}},}\ }\href@noop {} {\  (\bibinfo {year} {2020})},\ \Eprint
  {http://arxiv.org/abs/2008.00642} {arXiv:2008.00642 [hep-ph]} \BibitemShut
  {NoStop}%
\bibitem [{\citenamefont {Liventsev}\ \emph {et~al.}(2013)\citenamefont
  {Liventsev} \emph {et~al.}}]{Liventsev:2013zz}%
  \BibitemOpen
  \bibfield  {author} {\bibinfo {author} {\bibfnamefont {D.}~\bibnamefont
  {Liventsev}} \emph {et~al.} (\bibinfo {collaboration} {Belle}),\ }\bibfield
  {title} {\enquote {\bibinfo {title} {{Search for heavy neutrinos at
  Belle}},}\ }\href {\doibase 10.1103/PhysRevD.95.099903,
  10.1103/PhysRevD.87.071102} {\bibfield  {journal} {\bibinfo  {journal} {Phys.
  Rev.}\ }\textbf {\bibinfo {volume} {D87}},\ \bibinfo {pages} {071102}
  (\bibinfo {year} {2013})},\ \bibinfo {note} {[Erratum: Phys.
  Rev.D95,no.9,099903(2017)]},\ \Eprint {http://arxiv.org/abs/1301.1105}
  {arXiv:1301.1105 [hep-ex]} \BibitemShut {NoStop}%
\bibitem [{\citenamefont {Aaij}\ \emph {et~al.}(2014)\citenamefont {Aaij} \emph
  {et~al.}}]{Aaij:2014aba}%
  \BibitemOpen
  \bibfield  {author} {\bibinfo {author} {\bibfnamefont {Roel}\ \bibnamefont
  {Aaij}} \emph {et~al.} (\bibinfo {collaboration} {LHCb}),\ }\bibfield
  {title} {\enquote {\bibinfo {title} {{Search for Majorana neutrinos in $B^-
  \to \pi^+\mu^-\mu^-$ decays}},}\ }\href {\doibase
  10.1103/PhysRevLett.112.131802} {\bibfield  {journal} {\bibinfo  {journal}
  {Phys. Rev. Lett.}\ }\textbf {\bibinfo {volume} {112}},\ \bibinfo {pages}
  {131802} (\bibinfo {year} {2014})},\ \Eprint {http://arxiv.org/abs/1401.5361}
  {arXiv:1401.5361 [hep-ex]} \BibitemShut {NoStop}%
\bibitem [{\citenamefont {Artamonov}\ \emph {et~al.}(2015)\citenamefont
  {Artamonov} \emph {et~al.}}]{Artamonov:2014urb}%
  \BibitemOpen
  \bibfield  {author} {\bibinfo {author} {\bibfnamefont {A.~V.}\ \bibnamefont
  {Artamonov}} \emph {et~al.} (\bibinfo {collaboration} {E949}),\ }\bibfield
  {title} {\enquote {\bibinfo {title} {{Search for heavy neutrinos in
  $K^+\to\mu^+\nu_H$ decays}},}\ }\href {\doibase 10.1103/PhysRevD.91.059903,
  10.1103/PhysRevD.91.052001} {\bibfield  {journal} {\bibinfo  {journal} {Phys.
  Rev.}\ }\textbf {\bibinfo {volume} {D91}},\ \bibinfo {pages} {052001}
  (\bibinfo {year} {2015})},\ \bibinfo {note} {[Erratum: Phys.
  Rev.D91,no.5,059903(2015)]},\ \Eprint {http://arxiv.org/abs/1411.3963}
  {arXiv:1411.3963 [hep-ex]} \BibitemShut {NoStop}%
\bibitem [{\citenamefont {Aad}\ \emph {et~al.}(2015)\citenamefont {Aad} \emph
  {et~al.}}]{Aad:2015xaa}%
  \BibitemOpen
  \bibfield  {author} {\bibinfo {author} {\bibfnamefont {Georges}\ \bibnamefont
  {Aad}} \emph {et~al.} (\bibinfo {collaboration} {ATLAS}),\ }\bibfield
  {title} {\enquote {\bibinfo {title} {{Search for heavy Majorana neutrinos
  with the ATLAS detector in pp collisions at $ \sqrt{s}=8 $ TeV}},}\ }\href
  {\doibase 10.1007/JHEP07(2015)162} {\bibfield  {journal} {\bibinfo  {journal}
  {JHEP}\ }\textbf {\bibinfo {volume} {07}},\ \bibinfo {pages} {162} (\bibinfo
  {year} {2015})},\ \Eprint {http://arxiv.org/abs/1506.06020} {arXiv:1506.06020
  [hep-ex]} \BibitemShut {NoStop}%
\bibitem [{\citenamefont {Khachatryan}\ \emph {et~al.}(2015)\citenamefont
  {Khachatryan} \emph {et~al.}}]{Khachatryan:2015gha}%
  \BibitemOpen
  \bibfield  {author} {\bibinfo {author} {\bibfnamefont {Vardan}\ \bibnamefont
  {Khachatryan}} \emph {et~al.} (\bibinfo {collaboration} {CMS}),\ }\bibfield
  {title} {\enquote {\bibinfo {title} {{Search for heavy Majorana neutrinos in
  $\mu^\pm \mu^\pm+$ jets events in proton-proton collisions at $\sqrt{s}$ = 8
  TeV}},}\ }\href {\doibase 10.1016/j.physletb.2015.06.070} {\bibfield
  {journal} {\bibinfo  {journal} {Phys. Lett.}\ }\textbf {\bibinfo {volume}
  {B748}},\ \bibinfo {pages} {144--166} (\bibinfo {year} {2015})},\ \Eprint
  {http://arxiv.org/abs/1501.05566} {arXiv:1501.05566 [hep-ex]} \BibitemShut
  {NoStop}%
\bibitem [{\citenamefont {Antusch}\ \emph {et~al.}(2017)\citenamefont
  {Antusch}, \citenamefont {Cazzato},\ and\ \citenamefont
  {Fischer}}]{Antusch:2017hhu}%
  \BibitemOpen
  \bibfield  {author} {\bibinfo {author} {\bibfnamefont {Stefan}\ \bibnamefont
  {Antusch}}, \bibinfo {author} {\bibfnamefont {Eros}\ \bibnamefont {Cazzato}},
  \ and\ \bibinfo {author} {\bibfnamefont {Oliver}\ \bibnamefont {Fischer}},\
  }\bibfield  {title} {\enquote {\bibinfo {title} {{Sterile Neutrino Searches
  via Displaced Vertices at Lhcb}},}\ }\href {\doibase
  10.1016/j.physletb.2017.09.057} {\bibfield  {journal} {\bibinfo  {journal}
  {Phys. Lett.}\ }\textbf {\bibinfo {volume} {B774}},\ \bibinfo {pages}
  {114--118} (\bibinfo {year} {2017})},\ \Eprint
  {http://arxiv.org/abs/1706.05990} {arXiv:1706.05990 [hep-ph]} \BibitemShut
  {NoStop}%
\bibitem [{\citenamefont {Cortina~Gil}\ \emph {et~al.}(2018)\citenamefont
  {Cortina~Gil} \emph {et~al.}}]{CortinaGil:2017mqf}%
  \BibitemOpen
  \bibfield  {author} {\bibinfo {author} {\bibfnamefont {Eduardo}\ \bibnamefont
  {Cortina~Gil}} \emph {et~al.} (\bibinfo {collaboration} {NA62}),\ }\bibfield
  {title} {\enquote {\bibinfo {title} {{Search for heavy neutral lepton
  production in $K^+$ decays}},}\ }\href {\doibase
  10.1016/j.physletb.2018.01.031} {\bibfield  {journal} {\bibinfo  {journal}
  {Phys. Lett.}\ }\textbf {\bibinfo {volume} {B778}},\ \bibinfo {pages}
  {137--145} (\bibinfo {year} {2018})},\ \Eprint
  {http://arxiv.org/abs/1712.00297} {arXiv:1712.00297 [hep-ex]} \BibitemShut
  {NoStop}%
\bibitem [{\citenamefont {Izmaylov}\ and\ \citenamefont
  {Suvorov}(2017)}]{Izmaylov:2017lkv}%
  \BibitemOpen
  \bibfield  {author} {\bibinfo {author} {\bibfnamefont {A.}~\bibnamefont
  {Izmaylov}}\ and\ \bibinfo {author} {\bibfnamefont {S.}~\bibnamefont
  {Suvorov}},\ }\bibfield  {title} {\enquote {\bibinfo {title} {{Search for
  Heavy Neutrinos in the Nd280 Near Detector of the T2K Experiment}},}\
  }\bibfield  {booktitle} {\emph {\bibinfo {booktitle} {{Proceedings,
  International Session-Conference of the Snp of Psd Ras "Physics of
  Fundamental Interactions": Dubna, Russia, April 12-15, 2016}}},\ }\href
  {\doibase 10.1134/S1063779617060223} {\bibfield  {journal} {\bibinfo
  {journal} {Phys. Part. Nucl.}\ }\textbf {\bibinfo {volume} {48}},\ \bibinfo
  {pages} {984--986} (\bibinfo {year} {2017})}\BibitemShut {NoStop}%
\bibitem [{\citenamefont {Mermod}(2017)}]{Mermod:2017ceo}%
  \BibitemOpen
  \bibfield  {author} {\bibinfo {author} {\bibfnamefont {Philippe}\
  \bibnamefont {Mermod}} (\bibinfo {collaboration} {SHiP}),\ }\bibfield
  {title} {\enquote {\bibinfo {title} {{Prospects of the SHiP and Na62
  Experiments at Cern for Hidden Sector Searches}},}\ }\bibfield  {booktitle}
  {\emph {\bibinfo {booktitle} {{Proceedings, 2017 International Workshop on
  Neutrinos from Accelerators (Nufact17): Uppsala University Main Building,
  Uppsala, Sweden, September 25-30, 2017}}},\ }\href {\doibase
  10.22323/1.295.0139} {\bibfield  {journal} {\bibinfo  {journal} {PoS}\
  }\textbf {\bibinfo {volume} {NuFact2017}},\ \bibinfo {pages} {139} (\bibinfo
  {year} {2017})},\ \Eprint {http://arxiv.org/abs/1712.01768} {arXiv:1712.01768
  [hep-ex]} \BibitemShut {NoStop}%
\bibitem [{\citenamefont {Drewes}\ \emph {et~al.}(2018)\citenamefont {Drewes},
  \citenamefont {Hajer}, \citenamefont {Klaric},\ and\ \citenamefont
  {Lanfranchi}}]{Drewes:2018gkc}%
  \BibitemOpen
  \bibfield  {author} {\bibinfo {author} {\bibfnamefont {Marco}\ \bibnamefont
  {Drewes}}, \bibinfo {author} {\bibfnamefont {Jan}\ \bibnamefont {Hajer}},
  \bibinfo {author} {\bibfnamefont {Juraj}\ \bibnamefont {Klaric}}, \ and\
  \bibinfo {author} {\bibfnamefont {Gaia}\ \bibnamefont {Lanfranchi}},\
  }\bibfield  {title} {\enquote {\bibinfo {title} {{Na62 Sensitivity to Heavy
  Neutral Leptons in the Low Scale Seesaw Model}},}\ }\href {\doibase
  10.1007/JHEP07(2018)105} {\bibfield  {journal} {\bibinfo  {journal} {JHEP}\
  }\textbf {\bibinfo {volume} {07}},\ \bibinfo {pages} {105} (\bibinfo {year}
  {2018})},\ \Eprint {http://arxiv.org/abs/1801.04207} {arXiv:1801.04207
  [hep-ph]} \BibitemShut {NoStop}%
\bibitem [{\citenamefont {Ballett}\ \emph {et~al.}(2020)\citenamefont
  {Ballett}, \citenamefont {Boschi},\ and\ \citenamefont
  {Pascoli}}]{Ballett:2019bgd}%
  \BibitemOpen
  \bibfield  {author} {\bibinfo {author} {\bibfnamefont {Peter}\ \bibnamefont
  {Ballett}}, \bibinfo {author} {\bibfnamefont {Tommaso}\ \bibnamefont
  {Boschi}}, \ and\ \bibinfo {author} {\bibfnamefont {Silvia}\ \bibnamefont
  {Pascoli}},\ }\bibfield  {title} {\enquote {\bibinfo {title} {{Heavy Neutral
  Leptons from Low-Scale Seesaws at the Dune Near Detector}},}\ }\href
  {\doibase 10.1007/JHEP03(2020)111} {\bibfield  {journal} {\bibinfo  {journal}
  {JHEP}\ }\textbf {\bibinfo {volume} {03}},\ \bibinfo {pages} {111} (\bibinfo
  {year} {2020})},\ \Eprint {http://arxiv.org/abs/1905.00284} {arXiv:1905.00284
  [hep-ph]} \BibitemShut {NoStop}%
\bibitem [{\citenamefont {Sirunyan}\ \emph {et~al.}(2018)\citenamefont
  {Sirunyan} \emph {et~al.}}]{Sirunyan:2018mtv}%
  \BibitemOpen
  \bibfield  {author} {\bibinfo {author} {\bibfnamefont {Albert~M}\
  \bibnamefont {Sirunyan}} \emph {et~al.} (\bibinfo {collaboration} {CMS}),\
  }\bibfield  {title} {\enquote {\bibinfo {title} {{Search for heavy neutral
  leptons in events with three charged leptons in proton-proton collisions at
  $\sqrt{s} =$ 13 TeV}},}\ }\href {\doibase 10.1103/PhysRevLett.120.221801}
  {\bibfield  {journal} {\bibinfo  {journal} {Phys. Rev. Lett.}\ }\textbf
  {\bibinfo {volume} {120}},\ \bibinfo {pages} {221801} (\bibinfo {year}
  {2018})},\ \Eprint {http://arxiv.org/abs/1802.02965} {arXiv:1802.02965
  [hep-ex]} \BibitemShut {NoStop}%
\bibitem [{\citenamefont {Ahdida}\ \emph {et~al.}(2019)\citenamefont {Ahdida}
  \emph {et~al.}}]{SHiP:2018xqw}%
  \BibitemOpen
  \bibfield  {author} {\bibinfo {author} {\bibfnamefont {C.}~\bibnamefont
  {Ahdida}} \emph {et~al.} (\bibinfo {collaboration} {SHiP}),\ }\bibfield
  {title} {\enquote {\bibinfo {title} {{Sensitivity of the Ship Experiment to
  Heavy Neutral Leptons}},}\ }\href {\doibase 10.1007/JHEP04(2019)077}
  {\bibfield  {journal} {\bibinfo  {journal} {JHEP}\ }\textbf {\bibinfo
  {volume} {04}},\ \bibinfo {pages} {077} (\bibinfo {year} {2019})},\ \Eprint
  {http://arxiv.org/abs/1811.00930} {arXiv:1811.00930 [hep-ph]} \BibitemShut
  {NoStop}%
\bibitem [{\citenamefont {Boiarska}\ \emph {et~al.}(2019)\citenamefont
  {Boiarska}, \citenamefont {Bondarenko}, \citenamefont {Boyarsky},
  \citenamefont {Eijima}, \citenamefont {Ovchynnikov}, \citenamefont
  {Ruchayskiy},\ and\ \citenamefont {Timiryasov}}]{Boiarska:2019jcw}%
  \BibitemOpen
  \bibfield  {author} {\bibinfo {author} {\bibfnamefont {Iryna}\ \bibnamefont
  {Boiarska}}, \bibinfo {author} {\bibfnamefont {Kyrylo}\ \bibnamefont
  {Bondarenko}}, \bibinfo {author} {\bibfnamefont {Alexey}\ \bibnamefont
  {Boyarsky}}, \bibinfo {author} {\bibfnamefont {Shintaro}\ \bibnamefont
  {Eijima}}, \bibinfo {author} {\bibfnamefont {Maksym}\ \bibnamefont
  {Ovchynnikov}}, \bibinfo {author} {\bibfnamefont {Oleg}\ \bibnamefont
  {Ruchayskiy}}, \ and\ \bibinfo {author} {\bibfnamefont {Inar}\ \bibnamefont
  {Timiryasov}},\ }\bibfield  {title} {\enquote {\bibinfo {title} {{Probing
  Baryon Asymmetry of the Universe at Lhc and Ship}},}\ }\href@noop {} {\
  (\bibinfo {year} {2019})},\ \Eprint {http://arxiv.org/abs/1902.04535}
  {arXiv:1902.04535 [hep-ph]} \BibitemShut {NoStop}%
\bibitem [{\citenamefont {Bolton}\ \emph {et~al.}(2020)\citenamefont {Bolton},
  \citenamefont {Deppisch},\ and\ \citenamefont {Bhupal~Dev}}]{Bolton:2019pcu}%
  \BibitemOpen
  \bibfield  {author} {\bibinfo {author} {\bibfnamefont {Patrick~D.}\
  \bibnamefont {Bolton}}, \bibinfo {author} {\bibfnamefont {Frank~F.}\
  \bibnamefont {Deppisch}}, \ and\ \bibinfo {author} {\bibfnamefont {P.~S.}\
  \bibnamefont {Bhupal~Dev}},\ }\bibfield  {title} {\enquote {\bibinfo {title}
  {{Neutrinoless Double Beta Decay Versus Other Probes of Heavy Sterile
  Neutrinos}},}\ }\href {\doibase 10.1007/JHEP03(2020)170} {\bibfield
  {journal} {\bibinfo  {journal} {JHEP}\ }\textbf {\bibinfo {volume} {03}},\
  \bibinfo {pages} {170} (\bibinfo {year} {2020})},\ \Eprint
  {http://arxiv.org/abs/1912.03058} {arXiv:1912.03058 [hep-ph]} \BibitemShut
  {NoStop}%
\bibitem [{\citenamefont {Cortina~Gil}\ \emph {et~al.}(2020)\citenamefont
  {Cortina~Gil} \emph {et~al.}}]{NA62:2020mcv}%
  \BibitemOpen
  \bibfield  {author} {\bibinfo {author} {\bibfnamefont {Eduardo}\ \bibnamefont
  {Cortina~Gil}} \emph {et~al.} (\bibinfo {collaboration} {NA62}),\ }\bibfield
  {title} {\enquote {\bibinfo {title} {{Search for heavy neutral lepton
  production in $K^+$ decays to positrons}},}\ }\href {\doibase
  10.1016/j.physletb.2020.135599} {\bibfield  {journal} {\bibinfo  {journal}
  {Phys. Lett. B}\ }\textbf {\bibinfo {volume} {807}},\ \bibinfo {pages}
  {135599} (\bibinfo {year} {2020})},\ \Eprint
  {http://arxiv.org/abs/2005.09575} {arXiv:2005.09575 [hep-ex]} \BibitemShut
  {NoStop}%
\bibitem [{\citenamefont {Tastet}\ \emph {et~al.}(2020)\citenamefont {Tastet},
  \citenamefont {Goudzovski}, \citenamefont {Timiryasov},\ and\ \citenamefont
  {Ruchayskiy}}]{Tastet:2020tzh}%
  \BibitemOpen
  \bibfield  {author} {\bibinfo {author} {\bibfnamefont {Jean-Loup}\
  \bibnamefont {Tastet}}, \bibinfo {author} {\bibfnamefont {Evgueni}\
  \bibnamefont {Goudzovski}}, \bibinfo {author} {\bibfnamefont {Inar}\
  \bibnamefont {Timiryasov}}, \ and\ \bibinfo {author} {\bibfnamefont {Oleg}\
  \bibnamefont {Ruchayskiy}},\ }\bibfield  {title} {\enquote {\bibinfo {title}
  {{Projected NA62 sensitivity to heavy neutral lepton production in $K^+ \to
  \pi^0 e^+ N$ decays}},}\ }\href@noop {} {\  (\bibinfo {year} {2020})},\
  \Eprint {http://arxiv.org/abs/2008.11654} {arXiv:2008.11654 [hep-ph]}
  \BibitemShut {NoStop}%
\bibitem [{\citenamefont {Bondarenko}\ \emph {et~al.}(2021)\citenamefont
  {Bondarenko}, \citenamefont {Boyarsky}, \citenamefont {Klaric}, \citenamefont
  {Mikulenko}, \citenamefont {Ruchayskiy}, \citenamefont {Syvolap},\ and\
  \citenamefont {Timiryasov}}]{Bondarenko:2021cpc}%
  \BibitemOpen
  \bibfield  {author} {\bibinfo {author} {\bibfnamefont {Kyrylo}\ \bibnamefont
  {Bondarenko}}, \bibinfo {author} {\bibfnamefont {Alexey}\ \bibnamefont
  {Boyarsky}}, \bibinfo {author} {\bibfnamefont {Juraj}\ \bibnamefont
  {Klaric}}, \bibinfo {author} {\bibfnamefont {Oleksii}\ \bibnamefont
  {Mikulenko}}, \bibinfo {author} {\bibfnamefont {Oleg}\ \bibnamefont
  {Ruchayskiy}}, \bibinfo {author} {\bibfnamefont {Vsevolod}\ \bibnamefont
  {Syvolap}}, \ and\ \bibinfo {author} {\bibfnamefont {Inar}\ \bibnamefont
  {Timiryasov}},\ }\bibfield  {title} {\enquote {\bibinfo {title} {{An Allowed
  Window for Heavy Neutral Leptons Below the Kaon Mass}},}\ }\href@noop {} {\
  (\bibinfo {year} {2021})},\ \Eprint {http://arxiv.org/abs/2101.09255}
  {arXiv:2101.09255 [hep-ph]} \BibitemShut {NoStop}%
\bibitem [{\citenamefont {Cortina~Gil}\ \emph {et~al.}(2021)\citenamefont
  {Cortina~Gil} \emph {et~al.}}]{CortinaGil:2021gga}%
  \BibitemOpen
  \bibfield  {author} {\bibinfo {author} {\bibfnamefont {Eduardo}\ \bibnamefont
  {Cortina~Gil}} \emph {et~al.} (\bibinfo {collaboration} {NA62}),\ }\bibfield
  {title} {\enquote {\bibinfo {title} {{Search for $K^+$ decays to a muon and
  invisible particles}},}\ }\href@noop {} {\  (\bibinfo {year} {2021})},\
  \Eprint {http://arxiv.org/abs/2101.12304} {arXiv:2101.12304 [hep-ex]}
  \BibitemShut {NoStop}%
\bibitem [{\citenamefont {Aad}\ \emph {et~al.}(2019)\citenamefont {Aad} \emph
  {et~al.}}]{Aad:2019kiz}%
  \BibitemOpen
  \bibfield  {author} {\bibinfo {author} {\bibfnamefont {Georges}\ \bibnamefont
  {Aad}} \emph {et~al.} (\bibinfo {collaboration} {ATLAS}),\ }\bibfield
  {title} {\enquote {\bibinfo {title} {{Search for heavy neutral leptons in
  decays of $W$ bosons produced in 13 TeV $pp$ collisions using prompt and
  displaced signatures with the ATLAS detector}},}\ }\href {\doibase
  10.1007/JHEP10(2019)265} {\bibfield  {journal} {\bibinfo  {journal} {JHEP}\
  }\textbf {\bibinfo {volume} {10}},\ \bibinfo {pages} {265} (\bibinfo {year}
  {2019})},\ \Eprint {http://arxiv.org/abs/1905.09787} {arXiv:1905.09787
  [hep-ex]} \BibitemShut {NoStop}%
\bibitem [{\citenamefont {Wulz}(2019)}]{Wulz:2019lsz}%
  \BibitemOpen
  \bibfield  {author} {\bibinfo {author} {\bibfnamefont {Claudia-Elisabeth}\
  \bibnamefont {Wulz}} (\bibinfo {collaboration} {ATLAS, CMS}),\ }\bibfield
  {title} {\enquote {\bibinfo {title} {{Techniques and Results of Neutral
  Long-Lived Particle Searches in Atlas and Cms in Lhc Run 2}},}\ }in\
  \href@noop {} {\emph {\bibinfo {booktitle} {{54Th Rencontres De Moriond on
  Electroweak Interactions and Unified Theories}}}}\ (\bibinfo {year} {2019})\
  pp.\ \bibinfo {pages} {77--84},\ \Eprint {http://arxiv.org/abs/1907.13588}
  {arXiv:1907.13588 [hep-ex]} \BibitemShut {NoStop}%
\bibitem [{\citenamefont {Alekhin}\ \emph {et~al.}(2016)\citenamefont {Alekhin}
  \emph {et~al.}}]{Alekhin:2015byh}%
  \BibitemOpen
  \bibfield  {author} {\bibinfo {author} {\bibfnamefont {Sergey}\ \bibnamefont
  {Alekhin}} \emph {et~al.},\ }\bibfield  {title} {\enquote {\bibinfo {title}
  {{A Facility to Search for Hidden Particles at the Cern Sps: the Ship Physics
  Case}},}\ }\href {\doibase 10.1088/0034-4885/79/12/124201} {\bibfield
  {journal} {\bibinfo  {journal} {Rept. Prog. Phys.}\ }\textbf {\bibinfo
  {volume} {79}},\ \bibinfo {pages} {124201} (\bibinfo {year} {2016})},\
  \Eprint {http://arxiv.org/abs/1504.04855} {arXiv:1504.04855 [hep-ph]}
  \BibitemShut {NoStop}%
\bibitem [{\citenamefont {Curtin}\ \emph {et~al.}(2019)\citenamefont {Curtin}
  \emph {et~al.}}]{Curtin:2018mvb}%
  \BibitemOpen
  \bibfield  {author} {\bibinfo {author} {\bibfnamefont {David}\ \bibnamefont
  {Curtin}} \emph {et~al.},\ }\bibfield  {title} {\enquote {\bibinfo {title}
  {{Long-Lived Particles at the Energy Frontier: the Mathusla Physics Case}},}\
  }\href {\doibase 10.1088/1361-6633/ab28d6} {\bibfield  {journal} {\bibinfo
  {journal} {Rept. Prog. Phys.}\ }\textbf {\bibinfo {volume} {82}},\ \bibinfo
  {pages} {116201} (\bibinfo {year} {2019})},\ \Eprint
  {http://arxiv.org/abs/1806.07396} {arXiv:1806.07396 [hep-ph]} \BibitemShut
  {NoStop}%
\bibitem [{\citenamefont {Gligorov}\ \emph {et~al.}(2018)\citenamefont
  {Gligorov}, \citenamefont {Knapen}, \citenamefont {Papucci},\ and\
  \citenamefont {Robinson}}]{Gligorov:2017nwh}%
  \BibitemOpen
  \bibfield  {author} {\bibinfo {author} {\bibfnamefont {Vladimir~V.}\
  \bibnamefont {Gligorov}}, \bibinfo {author} {\bibfnamefont {Simon}\
  \bibnamefont {Knapen}}, \bibinfo {author} {\bibfnamefont {Michele}\
  \bibnamefont {Papucci}}, \ and\ \bibinfo {author} {\bibfnamefont {Dean~J.}\
  \bibnamefont {Robinson}},\ }\bibfield  {title} {\enquote {\bibinfo {title}
  {{Searching for Long-Lived Particles: a Compact Detector for Exotics at
  LHCb}},}\ }\href {\doibase 10.1103/PhysRevD.97.015023} {\bibfield  {journal}
  {\bibinfo  {journal} {Phys. Rev.}\ }\textbf {\bibinfo {volume} {D97}},\
  \bibinfo {pages} {015023} (\bibinfo {year} {2018})},\ \Eprint
  {http://arxiv.org/abs/1708.09395} {arXiv:1708.09395 [hep-ph]} \BibitemShut
  {NoStop}%
\bibitem [{\citenamefont {Feng}\ \emph {et~al.}(2018)\citenamefont {Feng},
  \citenamefont {Galon}, \citenamefont {Kling},\ and\ \citenamefont
  {Trojanowski}}]{Feng:2017uoz}%
  \BibitemOpen
  \bibfield  {author} {\bibinfo {author} {\bibfnamefont {Jonathan~L.}\
  \bibnamefont {Feng}}, \bibinfo {author} {\bibfnamefont {Iftah}\ \bibnamefont
  {Galon}}, \bibinfo {author} {\bibfnamefont {Felix}\ \bibnamefont {Kling}}, \
  and\ \bibinfo {author} {\bibfnamefont {Sebastian}\ \bibnamefont
  {Trojanowski}},\ }\bibfield  {title} {\enquote {\bibinfo {title} {{Forward
  Search Experiment at the Lhc}},}\ }\href {\doibase
  10.1103/PhysRevD.97.035001} {\bibfield  {journal} {\bibinfo  {journal} {Phys.
  Rev.}\ }\textbf {\bibinfo {volume} {D97}},\ \bibinfo {pages} {035001}
  (\bibinfo {year} {2018})},\ \Eprint {http://arxiv.org/abs/1708.09389}
  {arXiv:1708.09389 [hep-ph]} \BibitemShut {NoStop}%
\bibitem [{\citenamefont {Kling}\ and\ \citenamefont
  {Trojanowski}(2018)}]{Kling:2018wct}%
  \BibitemOpen
  \bibfield  {author} {\bibinfo {author} {\bibfnamefont {Felix}\ \bibnamefont
  {Kling}}\ and\ \bibinfo {author} {\bibfnamefont {Sebastian}\ \bibnamefont
  {Trojanowski}},\ }\bibfield  {title} {\enquote {\bibinfo {title} {{Heavy
  Neutral Leptons at Faser}},}\ }\href {\doibase 10.1103/PhysRevD.97.095016}
  {\bibfield  {journal} {\bibinfo  {journal} {Phys. Rev.}\ }\textbf {\bibinfo
  {volume} {D97}},\ \bibinfo {pages} {095016} (\bibinfo {year} {2018})},\
  \Eprint {http://arxiv.org/abs/1801.08947} {arXiv:1801.08947 [hep-ph]}
  \BibitemShut {NoStop}%
\bibitem [{\citenamefont {Hirsch}\ and\ \citenamefont
  {Wang}(2020)}]{Hirsch:2020klk}%
  \BibitemOpen
  \bibfield  {author} {\bibinfo {author} {\bibfnamefont {Martin}\ \bibnamefont
  {Hirsch}}\ and\ \bibinfo {author} {\bibfnamefont {Zeren~Simon}\ \bibnamefont
  {Wang}},\ }\bibfield  {title} {\enquote {\bibinfo {title} {{Heavy Neutral
  Leptons at Anubis}},}\ }\href {\doibase 10.1103/PhysRevD.101.055034}
  {\bibfield  {journal} {\bibinfo  {journal} {Phys. Rev. D}\ }\textbf {\bibinfo
  {volume} {101}},\ \bibinfo {pages} {055034} (\bibinfo {year} {2020})},\
  \Eprint {http://arxiv.org/abs/2001.04750} {arXiv:2001.04750 [hep-ph]}
  \BibitemShut {NoStop}%
\bibitem [{\citenamefont {Beacham}\ \emph {et~al.}(2020)\citenamefont {Beacham}
  \emph {et~al.}}]{Beacham:2019nyx}%
  \BibitemOpen
  \bibfield  {author} {\bibinfo {author} {\bibfnamefont {J.}~\bibnamefont
  {Beacham}} \emph {et~al.},\ }\bibfield  {title} {\enquote {\bibinfo {title}
  {{Physics Beyond Colliders at Cern: Beyond the Standard Model Working Group
  Report}},}\ }\href {\doibase 10.1088/1361-6471/ab4cd2} {\bibfield  {journal}
  {\bibinfo  {journal} {J. Phys. G}\ }\textbf {\bibinfo {volume} {47}},\
  \bibinfo {pages} {010501} (\bibinfo {year} {2020})},\ \Eprint
  {http://arxiv.org/abs/1901.09966} {arXiv:1901.09966 [hep-ex]} \BibitemShut
  {NoStop}%
\bibitem [{\citenamefont {Lanfranchi}\ \emph {et~al.}(2020)\citenamefont
  {Lanfranchi}, \citenamefont {Pospelov},\ and\ \citenamefont
  {Schuster}}]{Lanfranchi:2020crw}%
  \BibitemOpen
  \bibfield  {author} {\bibinfo {author} {\bibfnamefont {Gaia}\ \bibnamefont
  {Lanfranchi}}, \bibinfo {author} {\bibfnamefont {Maxim}\ \bibnamefont
  {Pospelov}}, \ and\ \bibinfo {author} {\bibfnamefont {Philip}\ \bibnamefont
  {Schuster}},\ }\bibfield  {title} {\enquote {\bibinfo {title} {{The Search
  for Feebly-Interacting Particles}},}\ }\href {\doibase
  10.1146/annurev-nucl-102419-055056} {\  (\bibinfo {year} {2020}),\
  10.1146/annurev-nucl-102419-055056},\ \Eprint
  {http://arxiv.org/abs/2011.02157} {arXiv:2011.02157 [hep-ph]} \BibitemShut
  {NoStop}%
\bibitem [{\citenamefont {Sakharov}(1991)}]{Sakharov:1967dj}%
  \BibitemOpen
  \bibfield  {author} {\bibinfo {author} {\bibfnamefont {A.D.}\ \bibnamefont
  {Sakharov}},\ }\bibfield  {title} {\enquote {\bibinfo {title} {{Violation of
  CP Invariance, C Asymmetry, and Baryon Asymmetry of the Universe}},}\ }\href
  {\doibase 10.1070/PU1991v034n05ABEH002497} {\bibfield  {journal} {\bibinfo
  {journal} {Sov. Phys. Usp.}\ }\textbf {\bibinfo {volume} {34}},\ \bibinfo
  {pages} {392--393} (\bibinfo {year} {1991})}\BibitemShut {NoStop}%
\bibitem [{\citenamefont {Buchmuller}\ \emph {et~al.}(2005)\citenamefont
  {Buchmuller}, \citenamefont {Di~Bari},\ and\ \citenamefont
  {Plumacher}}]{Buchmuller:2004nz}%
  \BibitemOpen
  \bibfield  {author} {\bibinfo {author} {\bibfnamefont {W.}~\bibnamefont
  {Buchmuller}}, \bibinfo {author} {\bibfnamefont {P.}~\bibnamefont {Di~Bari}},
  \ and\ \bibinfo {author} {\bibfnamefont {M.}~\bibnamefont {Plumacher}},\
  }\bibfield  {title} {\enquote {\bibinfo {title} {{Leptogenesis for
  pedestrians}},}\ }\href {\doibase 10.1016/j.aop.2004.02.003} {\bibfield
  {journal} {\bibinfo  {journal} {Annals Phys.}\ }\textbf {\bibinfo {volume}
  {315}},\ \bibinfo {pages} {305--351} (\bibinfo {year} {2005})},\ \Eprint
  {http://arxiv.org/abs/hep-ph/0401240} {arXiv:hep-ph/0401240 [hep-ph]}
  \BibitemShut {NoStop}%
\bibitem [{\citenamefont {Blondel}\ \emph {et~al.}(2016)\citenamefont
  {Blondel}, \citenamefont {Graverini}, \citenamefont {Serra},\ and\
  \citenamefont {Shaposhnikov}}]{Blondel:2014bra}%
  \BibitemOpen
  \bibfield  {author} {\bibinfo {author} {\bibfnamefont {Alain}\ \bibnamefont
  {Blondel}}, \bibinfo {author} {\bibfnamefont {E.}~\bibnamefont {Graverini}},
  \bibinfo {author} {\bibfnamefont {N.}~\bibnamefont {Serra}}, \ and\ \bibinfo
  {author} {\bibfnamefont {M.}~\bibnamefont {Shaposhnikov}} (\bibinfo
  {collaboration} {FCC-ee study Team}),\ }\bibfield  {title} {\enquote
  {\bibinfo {title} {{Search for Heavy Right Handed Neutrinos at the
  FCC-ee}},}\ }\bibfield  {booktitle} {\emph {\bibinfo {booktitle}
  {{Proceedings, 37th International Conference on High Energy Physics (ICHEP
  2014): Valencia, Spain, July 2-9, 2014}}},\ }\href {\doibase
  10.1016/j.nuclphysbps.2015.09.304} {\bibfield  {journal} {\bibinfo  {journal}
  {Nucl. Part. Phys. Proc.}\ }\textbf {\bibinfo {volume} {273-275}},\ \bibinfo
  {pages} {1883--1890} (\bibinfo {year} {2016})},\ \Eprint
  {http://arxiv.org/abs/1411.5230} {arXiv:1411.5230 [hep-ex]} \BibitemShut
  {NoStop}%
\bibitem [{\citenamefont {Garbrecht}(2014)}]{Garbrecht:2014bfa}%
  \BibitemOpen
  \bibfield  {author} {\bibinfo {author} {\bibfnamefont {Bjorn}\ \bibnamefont
  {Garbrecht}},\ }\bibfield  {title} {\enquote {\bibinfo {title} {{More Viable
  Parameter Space for Leptogenesis}},}\ }\href {\doibase
  10.1103/PhysRevD.90.063522} {\bibfield  {journal} {\bibinfo  {journal} {Phys.
  Rev.}\ }\textbf {\bibinfo {volume} {D90}},\ \bibinfo {pages} {063522}
  (\bibinfo {year} {2014})},\ \Eprint {http://arxiv.org/abs/1401.3278}
  {arXiv:1401.3278 [hep-ph]} \BibitemShut {NoStop}%
\bibitem [{\citenamefont {Garbrecht}\ \emph {et~al.}(2020)\citenamefont
  {Garbrecht}, \citenamefont {Klose},\ and\ \citenamefont
  {Tamarit}}]{Garbrecht:2019zaa}%
  \BibitemOpen
  \bibfield  {author} {\bibinfo {author} {\bibfnamefont {Björn}\ \bibnamefont
  {Garbrecht}}, \bibinfo {author} {\bibfnamefont {Philipp}\ \bibnamefont
  {Klose}}, \ and\ \bibinfo {author} {\bibfnamefont {Carlos}\ \bibnamefont
  {Tamarit}},\ }\bibfield  {title} {\enquote {\bibinfo {title} {{Relativistic
  and spectator effects in leptogenesis with heavy sterile neutrinos}},}\
  }\href {\doibase 10.1007/JHEP02(2020)117} {\bibfield  {journal} {\bibinfo
  {journal} {JHEP}\ }\textbf {\bibinfo {volume} {02}},\ \bibinfo {pages} {117}
  (\bibinfo {year} {2020})},\ \Eprint {http://arxiv.org/abs/1904.09956}
  {arXiv:1904.09956 [hep-ph]} \BibitemShut {NoStop}%
\bibitem [{\citenamefont {Blanchet}\ \emph {et~al.}(2010)\citenamefont
  {Blanchet}, \citenamefont {Hambye},\ and\ \citenamefont
  {Josse-Michaux}}]{Blanchet:2009kk}%
  \BibitemOpen
  \bibfield  {author} {\bibinfo {author} {\bibfnamefont {Steve}\ \bibnamefont
  {Blanchet}}, \bibinfo {author} {\bibfnamefont {Thomas}\ \bibnamefont
  {Hambye}}, \ and\ \bibinfo {author} {\bibfnamefont {Francois-Xavier}\
  \bibnamefont {Josse-Michaux}},\ }\bibfield  {title} {\enquote {\bibinfo
  {title} {{Reconciling leptogenesis with observable $\mu \to e \gamma$
  rates}},}\ }\href {\doibase 10.1007/JHEP04(2010)023} {\bibfield  {journal}
  {\bibinfo  {journal} {JHEP}\ }\textbf {\bibinfo {volume} {04}},\ \bibinfo
  {pages} {023} (\bibinfo {year} {2010})},\ \Eprint
  {http://arxiv.org/abs/0912.3153} {arXiv:0912.3153 [hep-ph]} \BibitemShut
  {NoStop}%
\bibitem [{\citenamefont {Deppisch}\ and\ \citenamefont
  {Pilaftsis}(2011)}]{Deppisch:2010fr}%
  \BibitemOpen
  \bibfield  {author} {\bibinfo {author} {\bibfnamefont {Frank~F.}\
  \bibnamefont {Deppisch}}\ and\ \bibinfo {author} {\bibfnamefont {Apostolos}\
  \bibnamefont {Pilaftsis}},\ }\bibfield  {title} {\enquote {\bibinfo {title}
  {{Lepton Flavour Violation and theta(13) in Minimal Resonant
  Leptogenesis}},}\ }\href {\doibase 10.1103/PhysRevD.83.076007} {\bibfield
  {journal} {\bibinfo  {journal} {Phys. Rev.}\ }\textbf {\bibinfo {volume}
  {D83}},\ \bibinfo {pages} {076007} (\bibinfo {year} {2011})},\ \Eprint
  {http://arxiv.org/abs/1012.1834} {arXiv:1012.1834 [hep-ph]} \BibitemShut
  {NoStop}%
\bibitem [{\citenamefont {Kuzmin}(1970)}]{Kuzmin:1970nx}%
  \BibitemOpen
  \bibfield  {author} {\bibinfo {author} {\bibfnamefont {V.~A.}\ \bibnamefont
  {Kuzmin}},\ }\bibfield  {title} {\enquote {\bibinfo {title} {{CP Violation
  and Baryon Asymmetry of the Universe}},}\ }\href@noop {} {\bibfield
  {journal} {\bibinfo  {journal} {Pisma Zh. Eksp. Teor. Fiz.}\ }\textbf
  {\bibinfo {volume} {12}},\ \bibinfo {pages} {335--337} (\bibinfo {year}
  {1970})}\BibitemShut {NoStop}%
\bibitem [{\citenamefont {Pilaftsis}\ and\ \citenamefont
  {Underwood}(2005)}]{Pilaftsis:2005rv}%
  \BibitemOpen
  \bibfield  {author} {\bibinfo {author} {\bibfnamefont {Apostolos}\
  \bibnamefont {Pilaftsis}}\ and\ \bibinfo {author} {\bibfnamefont {Thomas
  E.~J.}\ \bibnamefont {Underwood}},\ }\bibfield  {title} {\enquote {\bibinfo
  {title} {{Electroweak-scale resonant leptogenesis}},}\ }\href {\doibase
  10.1103/PhysRevD.72.113001} {\bibfield  {journal} {\bibinfo  {journal} {Phys.
  Rev.}\ }\textbf {\bibinfo {volume} {D72}},\ \bibinfo {pages} {113001}
  (\bibinfo {year} {2005})},\ \Eprint {http://arxiv.org/abs/hep-ph/0506107}
  {arXiv:hep-ph/0506107 [hep-ph]} \BibitemShut {NoStop}%
\bibitem [{\citenamefont {Barbieri}\ \emph {et~al.}(2000)\citenamefont
  {Barbieri}, \citenamefont {Creminelli}, \citenamefont {Strumia},\ and\
  \citenamefont {Tetradis}}]{Barbieri:1999ma}%
  \BibitemOpen
  \bibfield  {author} {\bibinfo {author} {\bibfnamefont {Riccardo}\
  \bibnamefont {Barbieri}}, \bibinfo {author} {\bibfnamefont {Paolo}\
  \bibnamefont {Creminelli}}, \bibinfo {author} {\bibfnamefont {Alessandro}\
  \bibnamefont {Strumia}}, \ and\ \bibinfo {author} {\bibfnamefont {Nikolaos}\
  \bibnamefont {Tetradis}},\ }\bibfield  {title} {\enquote {\bibinfo {title}
  {{Baryogenesis through leptogenesis}},}\ }\href {\doibase
  10.1016/S0550-3213(00)00011-0} {\bibfield  {journal} {\bibinfo  {journal}
  {Nucl. Phys.}\ }\textbf {\bibinfo {volume} {B575}},\ \bibinfo {pages}
  {61--77} (\bibinfo {year} {2000})},\ \Eprint
  {http://arxiv.org/abs/hep-ph/9911315} {arXiv:hep-ph/9911315 [hep-ph]}
  \BibitemShut {NoStop}%
\bibitem [{\citenamefont {Buchmuller}\ and\ \citenamefont
  {Plumacher}(2001)}]{Buchmuller:2001sr}%
  \BibitemOpen
  \bibfield  {author} {\bibinfo {author} {\bibfnamefont {W.}~\bibnamefont
  {Buchmuller}}\ and\ \bibinfo {author} {\bibfnamefont {M.}~\bibnamefont
  {Plumacher}},\ }\bibfield  {title} {\enquote {\bibinfo {title} {{Spectator
  processes and baryogenesis}},}\ }\href {\doibase
  10.1016/S0370-2693(01)00614-1} {\bibfield  {journal} {\bibinfo  {journal}
  {Phys. Lett.}\ }\textbf {\bibinfo {volume} {B511}},\ \bibinfo {pages}
  {74--76} (\bibinfo {year} {2001})},\ \Eprint
  {http://arxiv.org/abs/hep-ph/0104189} {arXiv:hep-ph/0104189 [hep-ph]}
  \BibitemShut {NoStop}%
\bibitem [{\citenamefont {Davidson}\ \emph {et~al.}(2008)\citenamefont
  {Davidson}, \citenamefont {Nardi},\ and\ \citenamefont
  {Nir}}]{Davidson:2008bu}%
  \BibitemOpen
  \bibfield  {author} {\bibinfo {author} {\bibfnamefont {Sacha}\ \bibnamefont
  {Davidson}}, \bibinfo {author} {\bibfnamefont {Enrico}\ \bibnamefont
  {Nardi}}, \ and\ \bibinfo {author} {\bibfnamefont {Yosef}\ \bibnamefont
  {Nir}},\ }\bibfield  {title} {\enquote {\bibinfo {title} {{Leptogenesis}},}\
  }\href {\doibase 10.1016/j.physrep.2008.06.002} {\bibfield  {journal}
  {\bibinfo  {journal} {Phys. Rept.}\ }\textbf {\bibinfo {volume} {466}},\
  \bibinfo {pages} {105--177} (\bibinfo {year} {2008})},\ \Eprint
  {http://arxiv.org/abs/0802.2962} {arXiv:0802.2962 [hep-ph]} \BibitemShut
  {NoStop}%
\bibitem [{\citenamefont {Roulet}\ \emph {et~al.}(1998)\citenamefont {Roulet},
  \citenamefont {Covi},\ and\ \citenamefont {Vissani}}]{Roulet:1997xa}%
  \BibitemOpen
  \bibfield  {author} {\bibinfo {author} {\bibfnamefont {Esteban}\ \bibnamefont
  {Roulet}}, \bibinfo {author} {\bibfnamefont {Laura}\ \bibnamefont {Covi}}, \
  and\ \bibinfo {author} {\bibfnamefont {Francesco}\ \bibnamefont {Vissani}},\
  }\bibfield  {title} {\enquote {\bibinfo {title} {{On the CP asymmetries in
  Majorana neutrino decays}},}\ }\href {\doibase 10.1016/S0370-2693(98)00135-X}
  {\bibfield  {journal} {\bibinfo  {journal} {Phys. Lett.}\ }\textbf {\bibinfo
  {volume} {B424}},\ \bibinfo {pages} {101--105} (\bibinfo {year} {1998})},\
  \Eprint {http://arxiv.org/abs/hep-ph/9712468} {arXiv:hep-ph/9712468 [hep-ph]}
  \BibitemShut {NoStop}%
\bibitem [{\citenamefont {Covi}\ \emph {et~al.}(1998)\citenamefont {Covi},
  \citenamefont {Rius}, \citenamefont {Roulet},\ and\ \citenamefont
  {Vissani}}]{Covi:1997dr}%
  \BibitemOpen
  \bibfield  {author} {\bibinfo {author} {\bibfnamefont {Laura}\ \bibnamefont
  {Covi}}, \bibinfo {author} {\bibfnamefont {Nuria}\ \bibnamefont {Rius}},
  \bibinfo {author} {\bibfnamefont {Esteban}\ \bibnamefont {Roulet}}, \ and\
  \bibinfo {author} {\bibfnamefont {Francesco}\ \bibnamefont {Vissani}},\
  }\bibfield  {title} {\enquote {\bibinfo {title} {{Finite temperature effects
  on CP violating asymmetries}},}\ }\href {\doibase 10.1103/PhysRevD.57.93}
  {\bibfield  {journal} {\bibinfo  {journal} {Phys. Rev.}\ }\textbf {\bibinfo
  {volume} {D57}},\ \bibinfo {pages} {93--99} (\bibinfo {year} {1998})},\
  \Eprint {http://arxiv.org/abs/hep-ph/9704366} {arXiv:hep-ph/9704366 [hep-ph]}
  \BibitemShut {NoStop}%
\bibitem [{\citenamefont {Schwinger}(1961)}]{Schwinger:1960qe}%
  \BibitemOpen
  \bibfield  {author} {\bibinfo {author} {\bibfnamefont {Julian~S.}\
  \bibnamefont {Schwinger}},\ }\bibfield  {title} {\enquote {\bibinfo {title}
  {{Brownian motion of a quantum oscillator}},}\ }\href {\doibase
  10.1063/1.1703727} {\bibfield  {journal} {\bibinfo  {journal} {J. Math.
  Phys.}\ }\textbf {\bibinfo {volume} {2}},\ \bibinfo {pages} {407--432}
  (\bibinfo {year} {1961})}\BibitemShut {NoStop}%
\bibitem [{\citenamefont {Keldysh}(1964)}]{Keldysh:1964ud}%
  \BibitemOpen
  \bibfield  {author} {\bibinfo {author} {\bibfnamefont {L.~V.}\ \bibnamefont
  {Keldysh}},\ }\bibfield  {title} {\enquote {\bibinfo {title} {{Diagram
  technique for nonequilibrium processes}},}\ }\href@noop {} {\bibfield
  {journal} {\bibinfo  {journal} {Zh. Eksp. Teor. Fiz.}\ }\textbf {\bibinfo
  {volume} {47}},\ \bibinfo {pages} {1515--1527} (\bibinfo {year} {1964})},\
  \bibinfo {note} {[Sov. Phys. JETP20,1018(1965)]}\BibitemShut {NoStop}%
\bibitem [{\citenamefont {Baym}\ and\ \citenamefont
  {Kadanoff}(1961)}]{Baym:1961zz}%
  \BibitemOpen
  \bibfield  {author} {\bibinfo {author} {\bibfnamefont {Gordon}\ \bibnamefont
  {Baym}}\ and\ \bibinfo {author} {\bibfnamefont {Leo~P.}\ \bibnamefont
  {Kadanoff}},\ }\bibfield  {title} {\enquote {\bibinfo {title} {{Conservation
  Laws and Correlation Functions}},}\ }\href {\doibase 10.1103/PhysRev.124.287}
  {\bibfield  {journal} {\bibinfo  {journal} {Phys. Rev.}\ }\textbf {\bibinfo
  {volume} {124}},\ \bibinfo {pages} {287--299} (\bibinfo {year}
  {1961})}\BibitemShut {NoStop}%
\bibitem [{\citenamefont {Danielewicz}(1984)}]{Danielewicz:1982kk}%
  \BibitemOpen
  \bibfield  {author} {\bibinfo {author} {\bibfnamefont {P.}~\bibnamefont
  {Danielewicz}},\ }\bibfield  {title} {\enquote {\bibinfo {title} {{Quantum
  Theory of Nonequilibrium Processes. 1.}}}\ }\href {\doibase
  10.1016/0003-4916(84)90092-7} {\bibfield  {journal} {\bibinfo  {journal}
  {Annals Phys.}\ }\textbf {\bibinfo {volume} {152}},\ \bibinfo {pages}
  {239--304} (\bibinfo {year} {1984})}\BibitemShut {NoStop}%
\bibitem [{\citenamefont {Niemi}\ and\ \citenamefont
  {Semenoff}(1984)}]{Niemi:1983nf}%
  \BibitemOpen
  \bibfield  {author} {\bibinfo {author} {\bibfnamefont {A.~J.}\ \bibnamefont
  {Niemi}}\ and\ \bibinfo {author} {\bibfnamefont {G.~W.}\ \bibnamefont
  {Semenoff}},\ }\bibfield  {title} {\enquote {\bibinfo {title} {{Finite
  Temperature Quantum Field Theory in Minkowski Space}},}\ }\href {\doibase
  10.1016/0003-4916(84)90082-4} {\bibfield  {journal} {\bibinfo  {journal}
  {Annals Phys.}\ }\textbf {\bibinfo {volume} {152}},\ \bibinfo {pages} {105}
  (\bibinfo {year} {1984})}\BibitemShut {NoStop}%
\bibitem [{\citenamefont {Landsman}\ and\ \citenamefont {van
  Weert}(1987)}]{Landsman:1986uw}%
  \BibitemOpen
  \bibfield  {author} {\bibinfo {author} {\bibfnamefont {N.~P.}\ \bibnamefont
  {Landsman}}\ and\ \bibinfo {author} {\bibfnamefont {C.~G.}\ \bibnamefont {van
  Weert}},\ }\bibfield  {title} {\enquote {\bibinfo {title} {{Real and
  Imaginary Time Field Theory at Finite Temperature and Density}},}\ }\href
  {\doibase 10.1016/0370-1573(87)90121-9} {\bibfield  {journal} {\bibinfo
  {journal} {Phys. Rept.}\ }\textbf {\bibinfo {volume} {145}},\ \bibinfo
  {pages} {141} (\bibinfo {year} {1987})}\BibitemShut {NoStop}%
\bibitem [{\citenamefont {Calzetta}\ and\ \citenamefont
  {Hu}(1988)}]{Calzetta:1986cq}%
  \BibitemOpen
  \bibfield  {author} {\bibinfo {author} {\bibfnamefont {E.}~\bibnamefont
  {Calzetta}}\ and\ \bibinfo {author} {\bibfnamefont {B.~L.}\ \bibnamefont
  {Hu}},\ }\bibfield  {title} {\enquote {\bibinfo {title} {{Nonequilibrium
  Quantum Fields: Closed Time Path Effective Action, Wigner Function and
  Boltzmann Equation}},}\ }\href {\doibase 10.1103/PhysRevD.37.2878} {\bibfield
   {journal} {\bibinfo  {journal} {Phys. Rev.}\ }\textbf {\bibinfo {volume}
  {D37}},\ \bibinfo {pages} {2878} (\bibinfo {year} {1988})}\BibitemShut
  {NoStop}%
\bibitem [{\citenamefont {Knoll}\ \emph {et~al.}(2001)\citenamefont {Knoll},
  \citenamefont {Ivanov},\ and\ \citenamefont {Voskresensky}}]{Knoll:2001jx}%
  \BibitemOpen
  \bibfield  {author} {\bibinfo {author} {\bibfnamefont {Joern}\ \bibnamefont
  {Knoll}}, \bibinfo {author} {\bibfnamefont {{\relax Yu}.~B.}\ \bibnamefont
  {Ivanov}}, \ and\ \bibinfo {author} {\bibfnamefont {D.~N.}\ \bibnamefont
  {Voskresensky}},\ }\bibfield  {title} {\enquote {\bibinfo {title} {{Exact
  conservation laws of the gradient expanded Kadanoff-Baym equations}},}\
  }\href {\doibase 10.1006/aphy.2001.6185} {\bibfield  {journal} {\bibinfo
  {journal} {Annals Phys.}\ }\textbf {\bibinfo {volume} {293}},\ \bibinfo
  {pages} {126--146} (\bibinfo {year} {2001})},\ \Eprint
  {http://arxiv.org/abs/nucl-th/0102044} {arXiv:nucl-th/0102044 [nucl-th]}
  \BibitemShut {NoStop}%
\bibitem [{\citenamefont {Blaizot}\ and\ \citenamefont
  {Iancu}(2002)}]{Blaizot:2001nr}%
  \BibitemOpen
  \bibfield  {author} {\bibinfo {author} {\bibfnamefont {Jean-Paul}\
  \bibnamefont {Blaizot}}\ and\ \bibinfo {author} {\bibfnamefont {Edmond}\
  \bibnamefont {Iancu}},\ }\bibfield  {title} {\enquote {\bibinfo {title} {{The
  Quark gluon plasma: Collective dynamics and hard thermal loops}},}\ }\href
  {\doibase 10.1016/S0370-1573(01)00061-8} {\bibfield  {journal} {\bibinfo
  {journal} {Phys. Rept.}\ }\textbf {\bibinfo {volume} {359}},\ \bibinfo
  {pages} {355--528} (\bibinfo {year} {2002})},\ \Eprint
  {http://arxiv.org/abs/hep-ph/0101103} {arXiv:hep-ph/0101103 [hep-ph]}
  \BibitemShut {NoStop}%
\bibitem [{\citenamefont {Calzetta}\ and\ \citenamefont
  {Hu}(2008)}]{Calzetta:2008iqa}%
  \BibitemOpen
  \bibfield  {author} {\bibinfo {author} {\bibfnamefont {Esteban~A.}\
  \bibnamefont {Calzetta}}\ and\ \bibinfo {author} {\bibfnamefont {Bei-Lok~B.}\
  \bibnamefont {Hu}},\ }\href {\doibase 10.1017/CBO9780511535123} {\emph
  {\bibinfo {title} {{Nonequilibrium Quantum Field Theory}}}},\ Cambridge
  Monographs on Mathematical Physics\ (\bibinfo  {publisher} {Cambridge
  University Press},\ \bibinfo {year} {2008})\BibitemShut {NoStop}%
\bibitem [{\citenamefont {Berges}(2015)}]{Berges:2015kfa}%
  \BibitemOpen
  \bibfield  {author} {\bibinfo {author} {\bibfnamefont {Jurgen}\ \bibnamefont
  {Berges}},\ }\bibfield  {title} {\enquote {\bibinfo {title} {{Nonequilibrium
  Quantum Fields: From Cold Atoms to Cosmology}},}\ }\href@noop {} {\
  (\bibinfo {year} {2015})},\ \Eprint {http://arxiv.org/abs/1503.02907}
  {arXiv:1503.02907 [hep-ph]} \BibitemShut {NoStop}%
\bibitem [{\citenamefont {Weinstock}\ \emph {et~al.}(2005)\citenamefont
  {Weinstock}, \citenamefont {Schmidt},\ and\ \citenamefont
  {Prokopec}}]{Weinstock:2005xlg}%
  \BibitemOpen
  \bibfield  {author} {\bibinfo {author} {\bibfnamefont {S.}~\bibnamefont
  {Weinstock}}, \bibinfo {author} {\bibfnamefont {M.~G.}\ \bibnamefont
  {Schmidt}}, \ and\ \bibinfo {author} {\bibfnamefont {T.}~\bibnamefont
  {Prokopec}},\ }\bibfield  {title} {\enquote {\bibinfo {title} {{Transport
  Equations for Chiral Fermions to Order $\hbar$ and Electroweak
  Baryogenesis}},}\ }in\ \href {\doibase 10.1142/9789812702159_0041} {\emph
  {\bibinfo {booktitle} {{Proceedings, 6th International Conference on Strong
  and Electroweak Matter (SEWM 2004): Helsinki, Finland, June 16-19, 2004}}}}\
  (\bibinfo {year} {2005})\ pp.\ \bibinfo {pages} {291--295}\BibitemShut
  {NoStop}%
\bibitem [{\citenamefont {Iso}\ and\ \citenamefont
  {Shimada}(2014)}]{Iso:2014afa}%
  \BibitemOpen
  \bibfield  {author} {\bibinfo {author} {\bibfnamefont {Satoshi}\ \bibnamefont
  {Iso}}\ and\ \bibinfo {author} {\bibfnamefont {Kengo}\ \bibnamefont
  {Shimada}},\ }\bibfield  {title} {\enquote {\bibinfo {title} {{Coherent
  Flavour Oscillation and CP Violating Parameter in Thermal Resonant
  Leptogenesis}},}\ }\href {\doibase 10.1007/JHEP08(2014)043} {\bibfield
  {journal} {\bibinfo  {journal} {JHEP}\ }\textbf {\bibinfo {volume} {08}},\
  \bibinfo {pages} {043} (\bibinfo {year} {2014})},\ \Eprint
  {http://arxiv.org/abs/1404.4816} {arXiv:1404.4816 [hep-ph]} \BibitemShut
  {NoStop}%
\bibitem [{\citenamefont {Blanchet}\ \emph {et~al.}(2013)\citenamefont
  {Blanchet}, \citenamefont {Di~Bari}, \citenamefont {Jones},\ and\
  \citenamefont {Marzola}}]{Blanchet:2011xq}%
  \BibitemOpen
  \bibfield  {author} {\bibinfo {author} {\bibfnamefont {Steve}\ \bibnamefont
  {Blanchet}}, \bibinfo {author} {\bibfnamefont {Pasquale}\ \bibnamefont
  {Di~Bari}}, \bibinfo {author} {\bibfnamefont {David~A.}\ \bibnamefont
  {Jones}}, \ and\ \bibinfo {author} {\bibfnamefont {Luca}\ \bibnamefont
  {Marzola}},\ }\bibfield  {title} {\enquote {\bibinfo {title} {{Leptogenesis
  with heavy neutrino flavours: from density matrix to Boltzmann equations}},}\
  }\href {\doibase 10.1088/1475-7516/2013/01/041} {\bibfield  {journal}
  {\bibinfo  {journal} {JCAP}\ }\textbf {\bibinfo {volume} {1301}},\ \bibinfo
  {pages} {041} (\bibinfo {year} {2013})},\ \Eprint
  {http://arxiv.org/abs/1112.4528} {arXiv:1112.4528 [hep-ph]} \BibitemShut
  {NoStop}%
\bibitem [{\citenamefont {Abada}\ \emph {et~al.}(2006)\citenamefont {Abada},
  \citenamefont {Davidson}, \citenamefont {Ibarra}, \citenamefont
  {Josse-Michaux}, \citenamefont {Losada},\ and\ \citenamefont
  {Riotto}}]{Abada:2006ea}%
  \BibitemOpen
  \bibfield  {author} {\bibinfo {author} {\bibfnamefont {A.}~\bibnamefont
  {Abada}}, \bibinfo {author} {\bibfnamefont {S.}~\bibnamefont {Davidson}},
  \bibinfo {author} {\bibfnamefont {A.}~\bibnamefont {Ibarra}}, \bibinfo
  {author} {\bibfnamefont {F.-X.}\ \bibnamefont {Josse-Michaux}}, \bibinfo
  {author} {\bibfnamefont {M.}~\bibnamefont {Losada}}, \ and\ \bibinfo {author}
  {\bibfnamefont {A.}~\bibnamefont {Riotto}},\ }\bibfield  {title} {\enquote
  {\bibinfo {title} {{Flavour Matters in Leptogenesis}},}\ }\href {\doibase
  10.1088/1126-6708/2006/09/010} {\bibfield  {journal} {\bibinfo  {journal}
  {JHEP}\ }\textbf {\bibinfo {volume} {09}},\ \bibinfo {pages} {010} (\bibinfo
  {year} {2006})},\ \Eprint {http://arxiv.org/abs/hep-ph/0605281}
  {arXiv:hep-ph/0605281} \BibitemShut {NoStop}%
\bibitem [{\citenamefont {De~Simone}\ and\ \citenamefont
  {Riotto}(2007{\natexlab{c}})}]{DeSimone:2006nrs}%
  \BibitemOpen
  \bibfield  {author} {\bibinfo {author} {\bibfnamefont {Andrea}\ \bibnamefont
  {De~Simone}}\ and\ \bibinfo {author} {\bibfnamefont {Antonio}\ \bibnamefont
  {Riotto}},\ }\bibfield  {title} {\enquote {\bibinfo {title} {{On the impact
  of flavour oscillations in leptogenesis}},}\ }\href {\doibase
  10.1088/1475-7516/2007/02/005} {\bibfield  {journal} {\bibinfo  {journal}
  {JCAP}\ }\textbf {\bibinfo {volume} {0702}},\ \bibinfo {pages} {005}
  (\bibinfo {year} {2007}{\natexlab{c}})},\ \Eprint
  {http://arxiv.org/abs/hep-ph/0611357} {arXiv:hep-ph/0611357 [hep-ph]}
  \BibitemShut {NoStop}%
\bibitem [{\citenamefont {Beneke}\ \emph {et~al.}(2011)\citenamefont {Beneke},
  \citenamefont {Garbrecht}, \citenamefont {Fidler}, \citenamefont {Herranen},\
  and\ \citenamefont {Schwaller}}]{Beneke:2010dz}%
  \BibitemOpen
  \bibfield  {author} {\bibinfo {author} {\bibfnamefont {Martin}\ \bibnamefont
  {Beneke}}, \bibinfo {author} {\bibfnamefont {Bjorn}\ \bibnamefont
  {Garbrecht}}, \bibinfo {author} {\bibfnamefont {Christian}\ \bibnamefont
  {Fidler}}, \bibinfo {author} {\bibfnamefont {Matti}\ \bibnamefont
  {Herranen}}, \ and\ \bibinfo {author} {\bibfnamefont {Pedro}\ \bibnamefont
  {Schwaller}},\ }\bibfield  {title} {\enquote {\bibinfo {title} {{Flavoured
  Leptogenesis in the CTP Formalism}},}\ }\href {\doibase
  10.1016/j.nuclphysb.2010.10.001} {\bibfield  {journal} {\bibinfo  {journal}
  {Nucl. Phys.}\ }\textbf {\bibinfo {volume} {B843}},\ \bibinfo {pages}
  {177--212} (\bibinfo {year} {2011})},\ \Eprint
  {http://arxiv.org/abs/1007.4783} {arXiv:1007.4783 [hep-ph]} \BibitemShut
  {NoStop}%
\bibitem [{\citenamefont {Millington}\ and\ \citenamefont
  {Pilaftsis}(2013{\natexlab{a}})}]{Millington:2012pf}%
  \BibitemOpen
  \bibfield  {author} {\bibinfo {author} {\bibfnamefont {Peter}\ \bibnamefont
  {Millington}}\ and\ \bibinfo {author} {\bibfnamefont {Apostolos}\
  \bibnamefont {Pilaftsis}},\ }\bibfield  {title} {\enquote {\bibinfo {title}
  {{Perturbative nonequilibrium thermal field theory}},}\ }\href {\doibase
  10.1103/PhysRevD.88.085009} {\bibfield  {journal} {\bibinfo  {journal} {Phys.
  Rev.}\ }\textbf {\bibinfo {volume} {D88}},\ \bibinfo {pages} {085009}
  (\bibinfo {year} {2013}{\natexlab{a}})},\ \Eprint
  {http://arxiv.org/abs/1211.3152} {arXiv:1211.3152 [hep-ph]} \BibitemShut
  {NoStop}%
\bibitem [{\citenamefont {Bhupal~Dev}\ \emph {et~al.}(2015)\citenamefont
  {Bhupal~Dev}, \citenamefont {Millington}, \citenamefont {Pilaftsis},\ and\
  \citenamefont {Teresi}}]{Dev:2014wsa}%
  \BibitemOpen
  \bibfield  {author} {\bibinfo {author} {\bibfnamefont {P.~S.}\ \bibnamefont
  {Bhupal~Dev}}, \bibinfo {author} {\bibfnamefont {Peter}\ \bibnamefont
  {Millington}}, \bibinfo {author} {\bibfnamefont {Apostolos}\ \bibnamefont
  {Pilaftsis}}, \ and\ \bibinfo {author} {\bibfnamefont {Daniele}\ \bibnamefont
  {Teresi}},\ }\bibfield  {title} {\enquote {\bibinfo {title} {{Kadanoff–Baym
  approach to flavour mixing and oscillations in resonant leptogenesis}},}\
  }\href {\doibase 10.1016/j.nuclphysb.2014.12.003} {\bibfield  {journal}
  {\bibinfo  {journal} {Nucl. Phys.}\ }\textbf {\bibinfo {volume} {B891}},\
  \bibinfo {pages} {128--158} (\bibinfo {year} {2015})},\ \Eprint
  {http://arxiv.org/abs/1410.6434} {arXiv:1410.6434 [hep-ph]} \BibitemShut
  {NoStop}%
\bibitem [{\citenamefont {Kartavtsev}\ \emph {et~al.}(2016)\citenamefont
  {Kartavtsev}, \citenamefont {Millington},\ and\ \citenamefont
  {Vogel}}]{Kartavtsev:2015vto}%
  \BibitemOpen
  \bibfield  {author} {\bibinfo {author} {\bibfnamefont {Alexander}\
  \bibnamefont {Kartavtsev}}, \bibinfo {author} {\bibfnamefont {Peter}\
  \bibnamefont {Millington}}, \ and\ \bibinfo {author} {\bibfnamefont
  {Hendrik}\ \bibnamefont {Vogel}},\ }\bibfield  {title} {\enquote {\bibinfo
  {title} {{Lepton asymmetry from mixing and oscillations}},}\ }\href {\doibase
  10.1007/JHEP06(2016)066} {\bibfield  {journal} {\bibinfo  {journal} {JHEP}\
  }\textbf {\bibinfo {volume} {06}},\ \bibinfo {pages} {066} (\bibinfo {year}
  {2016})},\ \Eprint {http://arxiv.org/abs/1601.03086} {arXiv:1601.03086
  [hep-ph]} \BibitemShut {NoStop}%
\bibitem [{\citenamefont {Racker}(2020)}]{Racker:2020avp}%
  \BibitemOpen
  \bibfield  {author} {\bibinfo {author} {\bibfnamefont {J.}~\bibnamefont
  {Racker}},\ }\bibfield  {title} {\enquote {\bibinfo {title} {{CP violation in
  mixing and oscillations for leptogenesis with quasi-degenerate neutrinos}},}\
  }\href@noop {} {\  (\bibinfo {year} {2020})},\ \Eprint
  {http://arxiv.org/abs/2012.05354} {arXiv:2012.05354 [hep-ph]} \BibitemShut
  {NoStop}%
\bibitem [{\citenamefont {Granelli}\ \emph {et~al.}(2020)\citenamefont
  {Granelli}, \citenamefont {Moffat},\ and\ \citenamefont
  {Petcov}}]{Granelli:2020ysj}%
  \BibitemOpen
  \bibfield  {author} {\bibinfo {author} {\bibfnamefont {A.}~\bibnamefont
  {Granelli}}, \bibinfo {author} {\bibfnamefont {K.}~\bibnamefont {Moffat}}, \
  and\ \bibinfo {author} {\bibfnamefont {S.~T.}\ \bibnamefont {Petcov}},\
  }\bibfield  {title} {\enquote {\bibinfo {title} {{Flavoured Resonant
  Leptogenesis at Sub-Tev Scales}},}\ }\href@noop {} {\  (\bibinfo {year}
  {2020})},\ \Eprint {http://arxiv.org/abs/2009.03166} {arXiv:2009.03166
  [hep-ph]} \BibitemShut {NoStop}%
\bibitem [{\citenamefont {Esteban}\ \emph {et~al.}(2020)\citenamefont
  {Esteban}, \citenamefont {Gonzalez-Garcia}, \citenamefont {Maltoni},
  \citenamefont {Schwetz},\ and\ \citenamefont {Zhou}}]{Esteban:2020cvm}%
  \BibitemOpen
  \bibfield  {author} {\bibinfo {author} {\bibfnamefont {Ivan}\ \bibnamefont
  {Esteban}}, \bibinfo {author} {\bibfnamefont {M.~C.}\ \bibnamefont
  {Gonzalez-Garcia}}, \bibinfo {author} {\bibfnamefont {Michele}\ \bibnamefont
  {Maltoni}}, \bibinfo {author} {\bibfnamefont {Thomas}\ \bibnamefont
  {Schwetz}}, \ and\ \bibinfo {author} {\bibfnamefont {Albert}\ \bibnamefont
  {Zhou}},\ }\bibfield  {title} {\enquote {\bibinfo {title} {{The Fate of
  Hints: Updated Global Analysis of Three-Flavor Neutrino Oscillations}},}\
  }\href {\doibase 10.1007/JHEP09(2020)178} {\bibfield  {journal} {\bibinfo
  {journal} {JHEP}\ }\textbf {\bibinfo {volume} {09}},\ \bibinfo {pages} {178}
  (\bibinfo {year} {2020})},\ \Eprint {http://arxiv.org/abs/2007.14792}
  {arXiv:2007.14792 [hep-ph]} \BibitemShut {NoStop}%
\bibitem [{\citenamefont {Giuliani}\ \emph {et~al.}(2019)\citenamefont
  {Giuliani}, \citenamefont {Gomez~Cadenas}, \citenamefont {Pascoli},
  \citenamefont {Previtali}, \citenamefont {Saakyan}, \citenamefont
  {Sch\"affner},\ and\ \citenamefont {Sch\"onert}}]{Giuliani:2019uno}%
  \BibitemOpen
  \bibfield  {author} {\bibinfo {author} {\bibfnamefont {A.}~\bibnamefont
  {Giuliani}}, \bibinfo {author} {\bibfnamefont {J.~J.}\ \bibnamefont
  {Gomez~Cadenas}}, \bibinfo {author} {\bibfnamefont {S.}~\bibnamefont
  {Pascoli}}, \bibinfo {author} {\bibfnamefont {E.}~\bibnamefont {Previtali}},
  \bibinfo {author} {\bibfnamefont {R.}~\bibnamefont {Saakyan}}, \bibinfo
  {author} {\bibfnamefont {K.}~\bibnamefont {Sch\"affner}}, \ and\ \bibinfo
  {author} {\bibfnamefont {S.}~\bibnamefont {Sch\"onert}} (\bibinfo
  {collaboration} {APPEC Committee}),\ }\bibfield  {title} {\enquote {\bibinfo
  {title} {{Double Beta Decay APPEC Committee Report}},}\ }\href@noop {} {\
  (\bibinfo {year} {2019})},\ \Eprint {http://arxiv.org/abs/1910.04688}
  {arXiv:1910.04688 [hep-ex]} \BibitemShut {NoStop}%
\bibitem [{\citenamefont {Abe}\ \emph {et~al.}(2020)\citenamefont {Abe} \emph
  {et~al.}}]{Abe:2019vii}%
  \BibitemOpen
  \bibfield  {author} {\bibinfo {author} {\bibfnamefont {K.}~\bibnamefont
  {Abe}} \emph {et~al.} (\bibinfo {collaboration} {T2K}),\ }\bibfield  {title}
  {\enquote {\bibinfo {title} {{Constraint on the matter–antimatter
  symmetry-violating phase in neutrino oscillations}},}\ }\href {\doibase
  10.1038/s41586-020-2177-0, 10.1038/s41586-020-2415-5} {\bibfield  {journal}
  {\bibinfo  {journal} {Nature}\ }\textbf {\bibinfo {volume} {580}},\ \bibinfo
  {pages} {339--344} (\bibinfo {year} {2020})},\ \bibinfo {note} {[Erratum:
  Nature583,no.7814,E16(2020)]},\ \Eprint {http://arxiv.org/abs/1910.03887}
  {arXiv:1910.03887 [hep-ex]} \BibitemShut {NoStop}%
\bibitem [{\citenamefont {Abi}\ \emph {et~al.}(2018)\citenamefont {Abi} \emph
  {et~al.}}]{Abi:2018dnh}%
  \BibitemOpen
  \bibfield  {author} {\bibinfo {author} {\bibfnamefont {B.}~\bibnamefont
  {Abi}} \emph {et~al.} (\bibinfo {collaboration} {DUNE}),\ }\bibfield  {title}
  {\enquote {\bibinfo {title} {{The Dune Far Detector Interim Design Report
  Volume 1: Physics, Technology and Strategies}},}\ }\href@noop {} {\
  (\bibinfo {year} {2018})},\ \Eprint {http://arxiv.org/abs/1807.10334}
  {arXiv:1807.10334 [physics.ins-det]} \BibitemShut {NoStop}%
\bibitem [{\citenamefont {Davidson}\ \emph {et~al.}(2007)\citenamefont
  {Davidson}, \citenamefont {Isidori},\ and\ \citenamefont
  {Strumia}}]{Davidson:2006tg}%
  \BibitemOpen
  \bibfield  {author} {\bibinfo {author} {\bibfnamefont {Sacha}\ \bibnamefont
  {Davidson}}, \bibinfo {author} {\bibfnamefont {Gino}\ \bibnamefont
  {Isidori}}, \ and\ \bibinfo {author} {\bibfnamefont {Alessandro}\
  \bibnamefont {Strumia}},\ }\bibfield  {title} {\enquote {\bibinfo {title}
  {{The Smallest Neutrino Mass}},}\ }\href {\doibase
  10.1016/j.physletb.2007.01.015} {\bibfield  {journal} {\bibinfo  {journal}
  {Phys. Lett. B}\ }\textbf {\bibinfo {volume} {646}},\ \bibinfo {pages}
  {100--104} (\bibinfo {year} {2007})},\ \Eprint
  {http://arxiv.org/abs/hep-ph/0611389} {arXiv:hep-ph/0611389} \BibitemShut
  {NoStop}%
\bibitem [{\citenamefont {Casas}\ and\ \citenamefont
  {Ibarra}(2001)}]{Casas:2001sr}%
  \BibitemOpen
  \bibfield  {author} {\bibinfo {author} {\bibfnamefont {J.~A.}\ \bibnamefont
  {Casas}}\ and\ \bibinfo {author} {\bibfnamefont {A.}~\bibnamefont {Ibarra}},\
  }\bibfield  {title} {\enquote {\bibinfo {title} {{Oscillating neutrinos and
  $\mu \to e, \gamma$}},}\ }\href {\doibase 10.1016/S0550-3213(01)00475-8}
  {\bibfield  {journal} {\bibinfo  {journal} {Nucl. Phys.}\ }\textbf {\bibinfo
  {volume} {B618}},\ \bibinfo {pages} {171--204} (\bibinfo {year} {2001})},\
  \Eprint {http://arxiv.org/abs/hep-ph/0103065} {arXiv:hep-ph/0103065 [hep-ph]}
  \BibitemShut {NoStop}%
\bibitem [{\citenamefont {Zyla}\ \emph {et~al.}(2020)\citenamefont {Zyla} \emph
  {et~al.}}]{Zyla:2020zbs}%
  \BibitemOpen
  \bibfield  {author} {\bibinfo {author} {\bibfnamefont {P.A.}\ \bibnamefont
  {Zyla}} \emph {et~al.} (\bibinfo {collaboration} {Particle Data Group}),\
  }\bibfield  {title} {\enquote {\bibinfo {title} {{Review of Particle
  Physics}},}\ }\href {\doibase 10.1093/ptep/ptaa104} {\bibfield  {journal}
  {\bibinfo  {journal} {PTEP}\ }\textbf {\bibinfo {volume} {2020}},\ \bibinfo
  {pages} {083C01} (\bibinfo {year} {2020})}\BibitemShut {NoStop}%
\bibitem [{\citenamefont {Nardi}\ \emph {et~al.}(2006)\citenamefont {Nardi},
  \citenamefont {Nir}, \citenamefont {Roulet},\ and\ \citenamefont
  {Racker}}]{Nardi:2006fx}%
  \BibitemOpen
  \bibfield  {author} {\bibinfo {author} {\bibfnamefont {Enrico}\ \bibnamefont
  {Nardi}}, \bibinfo {author} {\bibfnamefont {Yosef}\ \bibnamefont {Nir}},
  \bibinfo {author} {\bibfnamefont {Esteban}\ \bibnamefont {Roulet}}, \ and\
  \bibinfo {author} {\bibfnamefont {Juan}\ \bibnamefont {Racker}},\ }\bibfield
  {title} {\enquote {\bibinfo {title} {{The Importance of flavor in
  leptogenesis}},}\ }\href {\doibase 10.1088/1126-6708/2006/01/164} {\bibfield
  {journal} {\bibinfo  {journal} {JHEP}\ }\textbf {\bibinfo {volume} {01}},\
  \bibinfo {pages} {164} (\bibinfo {year} {2006})},\ \Eprint
  {http://arxiv.org/abs/hep-ph/0601084} {arXiv:hep-ph/0601084 [hep-ph]}
  \BibitemShut {NoStop}%
\bibitem [{\citenamefont {Buchmuller}\ and\ \citenamefont
  {Fredenhagen}(2000)}]{Buchmuller:2000nd}%
  \BibitemOpen
  \bibfield  {author} {\bibinfo {author} {\bibfnamefont {Wilfried}\
  \bibnamefont {Buchmuller}}\ and\ \bibinfo {author} {\bibfnamefont {Stefan}\
  \bibnamefont {Fredenhagen}},\ }\bibfield  {title} {\enquote {\bibinfo {title}
  {{Quantum mechanics of baryogenesis}},}\ }\href {\doibase
  10.1016/S0370-2693(00)00573-6} {\bibfield  {journal} {\bibinfo  {journal}
  {Phys. Lett.}\ }\textbf {\bibinfo {volume} {B483}},\ \bibinfo {pages}
  {217--224} (\bibinfo {year} {2000})},\ \Eprint
  {http://arxiv.org/abs/hep-ph/0004145} {arXiv:hep-ph/0004145 [hep-ph]}
  \BibitemShut {NoStop}%
\bibitem [{\citenamefont {Garny}\ \emph {et~al.}(2009)\citenamefont {Garny},
  \citenamefont {Hohenegger}, \citenamefont {Kartavtsev},\ and\ \citenamefont
  {Lindner}}]{Garny:2009rv}%
  \BibitemOpen
  \bibfield  {author} {\bibinfo {author} {\bibfnamefont {M.}~\bibnamefont
  {Garny}}, \bibinfo {author} {\bibfnamefont {A.}~\bibnamefont {Hohenegger}},
  \bibinfo {author} {\bibfnamefont {A.}~\bibnamefont {Kartavtsev}}, \ and\
  \bibinfo {author} {\bibfnamefont {M.}~\bibnamefont {Lindner}},\ }\bibfield
  {title} {\enquote {\bibinfo {title} {{Systematic approach to leptogenesis in
  nonequilibrium QFT: Vertex contribution to the CP-violating parameter}},}\
  }\href {\doibase 10.1103/PhysRevD.80.125027} {\bibfield  {journal} {\bibinfo
  {journal} {Phys. Rev.}\ }\textbf {\bibinfo {volume} {D80}},\ \bibinfo {pages}
  {125027} (\bibinfo {year} {2009})},\ \Eprint {http://arxiv.org/abs/0909.1559}
  {arXiv:0909.1559 [hep-ph]} \BibitemShut {NoStop}%
\bibitem [{\citenamefont {Anisimov}\ \emph {et~al.}(2010)\citenamefont
  {Anisimov}, \citenamefont {Buchmüller}, \citenamefont {Drewes},\ and\
  \citenamefont {Mendizabal}}]{Anisimov:2010aq}%
  \BibitemOpen
  \bibfield  {author} {\bibinfo {author} {\bibfnamefont {Alexey}\ \bibnamefont
  {Anisimov}}, \bibinfo {author} {\bibfnamefont {Wilfried}\ \bibnamefont
  {Buchmüller}}, \bibinfo {author} {\bibfnamefont {Marco}\ \bibnamefont
  {Drewes}}, \ and\ \bibinfo {author} {\bibfnamefont {Sebastián}\ \bibnamefont
  {Mendizabal}},\ }\bibfield  {title} {\enquote {\bibinfo {title}
  {{Leptogenesis from Quantum Interference in a Thermal Bath}},}\ }\href
  {\doibase 10.1103/PhysRevLett.104.121102} {\bibfield  {journal} {\bibinfo
  {journal} {Phys. Rev. Lett.}\ }\textbf {\bibinfo {volume} {104}},\ \bibinfo
  {pages} {121102} (\bibinfo {year} {2010})},\ \Eprint
  {http://arxiv.org/abs/1001.3856} {arXiv:1001.3856 [hep-ph]} \BibitemShut
  {NoStop}%
\bibitem [{\citenamefont {Beneke}\ \emph {et~al.}(2010)\citenamefont {Beneke},
  \citenamefont {Garbrecht}, \citenamefont {Herranen},\ and\ \citenamefont
  {Schwaller}}]{Beneke:2010wd}%
  \BibitemOpen
  \bibfield  {author} {\bibinfo {author} {\bibfnamefont {Martin}\ \bibnamefont
  {Beneke}}, \bibinfo {author} {\bibfnamefont {Bjorn}\ \bibnamefont
  {Garbrecht}}, \bibinfo {author} {\bibfnamefont {Matti}\ \bibnamefont
  {Herranen}}, \ and\ \bibinfo {author} {\bibfnamefont {Pedro}\ \bibnamefont
  {Schwaller}},\ }\bibfield  {title} {\enquote {\bibinfo {title} {{Finite
  Number Density Corrections to Leptogenesis}},}\ }\href {\doibase
  10.1016/j.nuclphysb.2010.05.003} {\bibfield  {journal} {\bibinfo  {journal}
  {Nucl. Phys.}\ }\textbf {\bibinfo {volume} {B838}},\ \bibinfo {pages} {1--27}
  (\bibinfo {year} {2010})},\ \Eprint {http://arxiv.org/abs/1002.1326}
  {arXiv:1002.1326 [hep-ph]} \BibitemShut {NoStop}%
\bibitem [{\citenamefont {Herranen}\ \emph {et~al.}(2010)\citenamefont
  {Herranen}, \citenamefont {Kainulainen},\ and\ \citenamefont
  {Rahkila}}]{Herranen:2010mh}%
  \BibitemOpen
  \bibfield  {author} {\bibinfo {author} {\bibfnamefont {Matti}\ \bibnamefont
  {Herranen}}, \bibinfo {author} {\bibfnamefont {Kimmo}\ \bibnamefont
  {Kainulainen}}, \ and\ \bibinfo {author} {\bibfnamefont {Pyry~Matti}\
  \bibnamefont {Rahkila}},\ }\bibfield  {title} {\enquote {\bibinfo {title}
  {{Coherent quantum Boltzmann equations from cQPA}},}\ }\href {\doibase
  10.1007/JHEP12(2010)072} {\bibfield  {journal} {\bibinfo  {journal} {JHEP}\
  }\textbf {\bibinfo {volume} {12}},\ \bibinfo {pages} {072} (\bibinfo {year}
  {2010})},\ \Eprint {http://arxiv.org/abs/1006.1929} {arXiv:1006.1929
  [hep-ph]} \BibitemShut {NoStop}%
\bibitem [{\citenamefont {Fidler}\ \emph {et~al.}(2012)\citenamefont {Fidler},
  \citenamefont {Herranen}, \citenamefont {Kainulainen},\ and\ \citenamefont
  {Rahkila}}]{Fidler:2011yq}%
  \BibitemOpen
  \bibfield  {author} {\bibinfo {author} {\bibfnamefont {Christian}\
  \bibnamefont {Fidler}}, \bibinfo {author} {\bibfnamefont {Matti}\
  \bibnamefont {Herranen}}, \bibinfo {author} {\bibfnamefont {Kimmo}\
  \bibnamefont {Kainulainen}}, \ and\ \bibinfo {author} {\bibfnamefont
  {Pyry~Matti}\ \bibnamefont {Rahkila}},\ }\bibfield  {title} {\enquote
  {\bibinfo {title} {{Flavoured quantum Boltzmann equations from cQPA}},}\
  }\href {\doibase 10.1007/JHEP02(2012)065} {\bibfield  {journal} {\bibinfo
  {journal} {JHEP}\ }\textbf {\bibinfo {volume} {02}},\ \bibinfo {pages} {065}
  (\bibinfo {year} {2012})},\ \Eprint {http://arxiv.org/abs/1108.2309}
  {arXiv:1108.2309 [hep-ph]} \BibitemShut {NoStop}%
\bibitem [{\citenamefont {Herranen}\ \emph {et~al.}(2012)\citenamefont
  {Herranen}, \citenamefont {Kainulainen},\ and\ \citenamefont
  {Rahkila}}]{Herranen:2011zg}%
  \BibitemOpen
  \bibfield  {author} {\bibinfo {author} {\bibfnamefont {Matti}\ \bibnamefont
  {Herranen}}, \bibinfo {author} {\bibfnamefont {Kimmo}\ \bibnamefont
  {Kainulainen}}, \ and\ \bibinfo {author} {\bibfnamefont {Pyry~Matti}\
  \bibnamefont {Rahkila}},\ }\bibfield  {title} {\enquote {\bibinfo {title}
  {{Flavour-coherent propagators and Feynman rules: Covariant cQPA
  formulation}},}\ }\href {\doibase 10.1007/JHEP02(2012)080} {\bibfield
  {journal} {\bibinfo  {journal} {JHEP}\ }\textbf {\bibinfo {volume} {02}},\
  \bibinfo {pages} {080} (\bibinfo {year} {2012})},\ \Eprint
  {http://arxiv.org/abs/1108.2371} {arXiv:1108.2371 [hep-ph]} \BibitemShut
  {NoStop}%
\bibitem [{\citenamefont {Millington}\ and\ \citenamefont
  {Pilaftsis}(2013{\natexlab{b}})}]{Millington:2013isa}%
  \BibitemOpen
  \bibfield  {author} {\bibinfo {author} {\bibfnamefont {Peter}\ \bibnamefont
  {Millington}}\ and\ \bibinfo {author} {\bibfnamefont {Apostolos}\
  \bibnamefont {Pilaftsis}},\ }\bibfield  {title} {\enquote {\bibinfo {title}
  {{Perturbative Non-Equilibrium Thermal Field Theory to all Orders in Gradient
  Expansion}},}\ }\href {\doibase 10.1016/j.physletb.2013.05.044} {\bibfield
  {journal} {\bibinfo  {journal} {Phys. Lett.}\ }\textbf {\bibinfo {volume}
  {B724}},\ \bibinfo {pages} {56--62} (\bibinfo {year} {2013}{\natexlab{b}})},\
  \Eprint {http://arxiv.org/abs/1304.7249} {arXiv:1304.7249 [hep-ph]}
  \BibitemShut {NoStop}%
\bibitem [{\citenamefont {Anisimov}\ \emph
  {et~al.}(2011{\natexlab{b}})\citenamefont {Anisimov}, \citenamefont
  {Buchmüller}, \citenamefont {Drewes},\ and\ \citenamefont
  {Mendizabal}}]{Anisimov:2010dk}%
  \BibitemOpen
  \bibfield  {author} {\bibinfo {author} {\bibfnamefont {A.}~\bibnamefont
  {Anisimov}}, \bibinfo {author} {\bibfnamefont {W.}~\bibnamefont
  {Buchmüller}}, \bibinfo {author} {\bibfnamefont {M.}~\bibnamefont {Drewes}},
  \ and\ \bibinfo {author} {\bibfnamefont {S.}~\bibnamefont {Mendizabal}},\
  }\bibfield  {title} {\enquote {\bibinfo {title} {{Quantum Leptogenesis I}},}\
  }\href {\doibase 10.1016/j.aop.2011.02.002, 10.1016/j.aop.2013.05.00}
  {\bibfield  {journal} {\bibinfo  {journal} {Annals Phys.}\ }\textbf {\bibinfo
  {volume} {326}},\ \bibinfo {pages} {1998--2038} (\bibinfo {year}
  {2011}{\natexlab{b}})},\ \bibinfo {note} {[Erratum: Annals
  Phys.338,376(2011)]},\ \Eprint {http://arxiv.org/abs/1012.5821}
  {arXiv:1012.5821 [hep-ph]} \BibitemShut {NoStop}%
\bibitem [{\citenamefont {Dev}\ \emph {et~al.}(2018)\citenamefont {Dev},
  \citenamefont {Garny}, \citenamefont {Klaric}, \citenamefont {Millington},\
  and\ \citenamefont {Teresi}}]{Dev:2017wwc}%
  \BibitemOpen
  \bibfield  {author} {\bibinfo {author} {\bibfnamefont {Bhupal}\ \bibnamefont
  {Dev}}, \bibinfo {author} {\bibfnamefont {Mathias}\ \bibnamefont {Garny}},
  \bibinfo {author} {\bibfnamefont {Juraj}\ \bibnamefont {Klaric}}, \bibinfo
  {author} {\bibfnamefont {Peter}\ \bibnamefont {Millington}}, \ and\ \bibinfo
  {author} {\bibfnamefont {Daniele}\ \bibnamefont {Teresi}},\ }\bibfield
  {title} {\enquote {\bibinfo {title} {{Resonant enhancement in
  leptogenesis}},}\ }\href {\doibase 10.1142/S0217751X18420034} {\bibfield
  {journal} {\bibinfo  {journal} {Int. J. Mod. Phys.}\ }\textbf {\bibinfo
  {volume} {A33}},\ \bibinfo {pages} {1842003} (\bibinfo {year} {2018})},\
  \Eprint {http://arxiv.org/abs/1711.02863} {arXiv:1711.02863 [hep-ph]}
  \BibitemShut {NoStop}%
\bibitem [{\citenamefont {Sigl}\ and\ \citenamefont
  {Raffelt}(1993)}]{Sigl:1992fn}%
  \BibitemOpen
  \bibfield  {author} {\bibinfo {author} {\bibfnamefont {G.}~\bibnamefont
  {Sigl}}\ and\ \bibinfo {author} {\bibfnamefont {G.}~\bibnamefont {Raffelt}},\
  }\bibfield  {title} {\enquote {\bibinfo {title} {{General Kinetic Description
  of Relativistic Mixed Neutrinos}},}\ }\href {\doibase
  10.1016/0550-3213(93)90175-O} {\bibfield  {journal} {\bibinfo  {journal}
  {Nucl. Phys. B}\ }\textbf {\bibinfo {volume} {406}},\ \bibinfo {pages}
  {423--451} (\bibinfo {year} {1993})}\BibitemShut {NoStop}%
\bibitem [{\citenamefont {B\"odeker}\ and\ \citenamefont
  {Schr\"oder}(2019)}]{Bodeker:2019ajh}%
  \BibitemOpen
  \bibfield  {author} {\bibinfo {author} {\bibfnamefont {Dietrich}\
  \bibnamefont {B\"odeker}}\ and\ \bibinfo {author} {\bibfnamefont {Dennis}\
  \bibnamefont {Schr\"oder}},\ }\bibfield  {title} {\enquote {\bibinfo {title}
  {{Equilibration of right-handed electrons}},}\ }\href {\doibase
  10.1088/1475-7516/2019/05/010} {\bibfield  {journal} {\bibinfo  {journal}
  {JCAP}\ }\textbf {\bibinfo {volume} {05}},\ \bibinfo {pages} {010} (\bibinfo
  {year} {2019})},\ \Eprint {http://arxiv.org/abs/1902.07220} {arXiv:1902.07220
  [hep-ph]} \BibitemShut {NoStop}%
\bibitem [{\citenamefont {Ghisoiu}\ and\ \citenamefont
  {Laine}(2014)}]{Ghisoiu:2014ena}%
  \BibitemOpen
  \bibfield  {author} {\bibinfo {author} {\bibfnamefont {I.}~\bibnamefont
  {Ghisoiu}}\ and\ \bibinfo {author} {\bibfnamefont {M.}~\bibnamefont
  {Laine}},\ }\bibfield  {title} {\enquote {\bibinfo {title} {{Right-handed
  neutrino production rate at $T > 160$ GeV}},}\ }\href {\doibase
  10.1088/1475-7516/2014/12/032} {\bibfield  {journal} {\bibinfo  {journal}
  {JCAP}\ }\textbf {\bibinfo {volume} {1412}},\ \bibinfo {pages} {032}
  (\bibinfo {year} {2014})},\ \Eprint {http://arxiv.org/abs/1411.1765}
  {arXiv:1411.1765 [hep-ph]} \BibitemShut {NoStop}%
\bibitem [{\citenamefont {Garbrecht}\ \emph {et~al.}(2013)\citenamefont
  {Garbrecht}, \citenamefont {Glowna},\ and\ \citenamefont
  {Schwaller}}]{Garbrecht:2013urw}%
  \BibitemOpen
  \bibfield  {author} {\bibinfo {author} {\bibfnamefont {Björn}\ \bibnamefont
  {Garbrecht}}, \bibinfo {author} {\bibfnamefont {Frank}\ \bibnamefont
  {Glowna}}, \ and\ \bibinfo {author} {\bibfnamefont {Pedro}\ \bibnamefont
  {Schwaller}},\ }\bibfield  {title} {\enquote {\bibinfo {title} {{Scattering
  Rates For Leptogenesis: Damping of Lepton Flavour Coherence and Production of
  Singlet Neutrinos}},}\ }\href {\doibase 10.1016/j.nuclphysb.2013.08.020}
  {\bibfield  {journal} {\bibinfo  {journal} {Nucl. Phys.}\ }\textbf {\bibinfo
  {volume} {B877}},\ \bibinfo {pages} {1--35} (\bibinfo {year} {2013})},\
  \Eprint {http://arxiv.org/abs/1303.5498} {arXiv:1303.5498 [hep-ph]}
  \BibitemShut {NoStop}%
\bibitem [{\citenamefont {Biondini}\ \emph {et~al.}(2018)\citenamefont
  {Biondini} \emph {et~al.}}]{Biondini:2017rpb}%
  \BibitemOpen
  \bibfield  {author} {\bibinfo {author} {\bibfnamefont {Simone}\ \bibnamefont
  {Biondini}} \emph {et~al.},\ }\bibfield  {title} {\enquote {\bibinfo {title}
  {{Status of rates and rate equations for thermal leptogenesis}},}\ }\href
  {\doibase 10.1142/S0217751X18420046} {\bibfield  {journal} {\bibinfo
  {journal} {Int. J. Mod. Phys.}\ }\textbf {\bibinfo {volume} {A33}},\ \bibinfo
  {pages} {1842004} (\bibinfo {year} {2018})},\ \Eprint
  {http://arxiv.org/abs/1711.02864} {arXiv:1711.02864 [hep-ph]} \BibitemShut
  {NoStop}%
\bibitem [{\citenamefont {Jackson}\ and\ \citenamefont
  {Laine}(2020)}]{Jackson:2019tnr}%
  \BibitemOpen
  \bibfield  {author} {\bibinfo {author} {\bibfnamefont {G.}~\bibnamefont
  {Jackson}}\ and\ \bibinfo {author} {\bibfnamefont {M.}~\bibnamefont
  {Laine}},\ }\bibfield  {title} {\enquote {\bibinfo {title} {{A thermal
  neutrino interaction rate at NLO}},}\ }\href {\doibase
  10.1016/j.nuclphysb.2019.114870} {\bibfield  {journal} {\bibinfo  {journal}
  {Nucl. Phys.}\ }\textbf {\bibinfo {volume} {B950}},\ \bibinfo {pages}
  {114870} (\bibinfo {year} {2020})},\ \Eprint
  {http://arxiv.org/abs/1910.12880} {arXiv:1910.12880 [hep-ph]} \BibitemShut
  {NoStop}%
\bibitem [{\citenamefont {Kajantie}\ \emph {et~al.}(1996)\citenamefont
  {Kajantie}, \citenamefont {Laine}, \citenamefont {Rummukainen},\ and\
  \citenamefont {Shaposhnikov}}]{Kajantie:1996mn}%
  \BibitemOpen
  \bibfield  {author} {\bibinfo {author} {\bibfnamefont {K.}~\bibnamefont
  {Kajantie}}, \bibinfo {author} {\bibfnamefont {M.}~\bibnamefont {Laine}},
  \bibinfo {author} {\bibfnamefont {K.}~\bibnamefont {Rummukainen}}, \ and\
  \bibinfo {author} {\bibfnamefont {Mikhail~E.}\ \bibnamefont {Shaposhnikov}},\
  }\bibfield  {title} {\enquote {\bibinfo {title} {{Is there a hot electroweak
  phase transition at m(H) larger or equal to m(W)?}}}\ }\href {\doibase
  10.1103/PhysRevLett.77.2887} {\bibfield  {journal} {\bibinfo  {journal}
  {Phys. Rev. Lett.}\ }\textbf {\bibinfo {volume} {77}},\ \bibinfo {pages}
  {2887--2890} (\bibinfo {year} {1996})},\ \Eprint
  {http://arxiv.org/abs/hep-ph/9605288} {arXiv:hep-ph/9605288 [hep-ph]}
  \BibitemShut {NoStop}%
\bibitem [{\citenamefont {D'Onofrio}\ and\ \citenamefont
  {Rummukainen}(2016)}]{DOnofrio:2015gop}%
  \BibitemOpen
  \bibfield  {author} {\bibinfo {author} {\bibfnamefont {Michela}\ \bibnamefont
  {D'Onofrio}}\ and\ \bibinfo {author} {\bibfnamefont {Kari}\ \bibnamefont
  {Rummukainen}},\ }\bibfield  {title} {\enquote {\bibinfo {title} {{Standard
  model cross-over on the lattice}},}\ }\href {\doibase
  10.1103/PhysRevD.93.025003} {\bibfield  {journal} {\bibinfo  {journal} {Phys.
  Rev.}\ }\textbf {\bibinfo {volume} {D93}},\ \bibinfo {pages} {025003}
  (\bibinfo {year} {2016})},\ \Eprint {http://arxiv.org/abs/1508.07161}
  {arXiv:1508.07161 [hep-ph]} \BibitemShut {NoStop}%
\bibitem [{\citenamefont {Carrington}\ \emph {et~al.}(2005)\citenamefont
  {Carrington}, \citenamefont {Kunstatter},\ and\ \citenamefont
  {Zaraket}}]{Carrington:2003ut}%
  \BibitemOpen
  \bibfield  {author} {\bibinfo {author} {\bibfnamefont {M.~E.}\ \bibnamefont
  {Carrington}}, \bibinfo {author} {\bibfnamefont {G.}~\bibnamefont
  {Kunstatter}}, \ and\ \bibinfo {author} {\bibfnamefont {H.}~\bibnamefont
  {Zaraket}},\ }\bibfield  {title} {\enquote {\bibinfo {title} {{2PI effective
  action and gauge invariance problems}},}\ }\href {\doibase
  10.1140/epjc/s2005-02277-x} {\bibfield  {journal} {\bibinfo  {journal} {Eur.
  Phys. J.}\ }\textbf {\bibinfo {volume} {C42}},\ \bibinfo {pages} {253--259}
  (\bibinfo {year} {2005})},\ \Eprint {http://arxiv.org/abs/hep-ph/0309084}
  {arXiv:hep-ph/0309084 [hep-ph]} \BibitemShut {NoStop}%
\bibitem [{\citenamefont {Garbrecht}\ and\ \citenamefont
  {Konstandin}(2009)}]{Garbrecht:2008cb}%
  \BibitemOpen
  \bibfield  {author} {\bibinfo {author} {\bibfnamefont {Bjorn}\ \bibnamefont
  {Garbrecht}}\ and\ \bibinfo {author} {\bibfnamefont {Thomas}\ \bibnamefont
  {Konstandin}},\ }\bibfield  {title} {\enquote {\bibinfo {title} {{Separation
  of Equilibration Time-Scales in the Gradient Expansion}},}\ }\href {\doibase
  10.1103/PhysRevD.79.085003} {\bibfield  {journal} {\bibinfo  {journal} {Phys.
  Rev.}\ }\textbf {\bibinfo {volume} {D79}},\ \bibinfo {pages} {085003}
  (\bibinfo {year} {2009})},\ \Eprint {http://arxiv.org/abs/0810.4016}
  {arXiv:0810.4016 [hep-ph]} \BibitemShut {NoStop}%
\bibitem [{\citenamefont {Abada}\ \emph {et~al.}(2019)\citenamefont {Abada},
  \citenamefont {Arcadi}, \citenamefont {Domcke}, \citenamefont {Drewes},
  \citenamefont {Klaric},\ and\ \citenamefont {Lucente}}]{Abada:2018oly}%
  \BibitemOpen
  \bibfield  {author} {\bibinfo {author} {\bibfnamefont {Asmaa}\ \bibnamefont
  {Abada}}, \bibinfo {author} {\bibfnamefont {Giorgio}\ \bibnamefont {Arcadi}},
  \bibinfo {author} {\bibfnamefont {Valerie}\ \bibnamefont {Domcke}}, \bibinfo
  {author} {\bibfnamefont {Marco}\ \bibnamefont {Drewes}}, \bibinfo {author}
  {\bibfnamefont {Juraj}\ \bibnamefont {Klaric}}, \ and\ \bibinfo {author}
  {\bibfnamefont {Michele}\ \bibnamefont {Lucente}},\ }\bibfield  {title}
  {\enquote {\bibinfo {title} {{Low-Scale Leptogenesis with Three Heavy
  Neutrinos}},}\ }\href {\doibase 10.1007/JHEP01(2019)164} {\bibfield
  {journal} {\bibinfo  {journal} {JHEP}\ }\textbf {\bibinfo {volume} {01}},\
  \bibinfo {pages} {164} (\bibinfo {year} {2019})},\ \Eprint
  {http://arxiv.org/abs/1810.12463} {arXiv:1810.12463 [hep-ph]} \BibitemShut
  {NoStop}%
\bibitem [{\citenamefont {Kersten}\ and\ \citenamefont
  {Smirnov}(2007)}]{Kersten:2007vk}%
  \BibitemOpen
  \bibfield  {author} {\bibinfo {author} {\bibfnamefont {Jörn}\ \bibnamefont
  {Kersten}}\ and\ \bibinfo {author} {\bibfnamefont {Alexei~{\relax Yu}.}\
  \bibnamefont {Smirnov}},\ }\bibfield  {title} {\enquote {\bibinfo {title}
  {{Right-Handed Neutrinos at CERN LHC and the Mechanism of Neutrino Mass
  Generation}},}\ }\href {\doibase 10.1103/PhysRevD.76.073005} {\bibfield
  {journal} {\bibinfo  {journal} {Phys. Rev.}\ }\textbf {\bibinfo {volume}
  {D76}},\ \bibinfo {pages} {073005} (\bibinfo {year} {2007})},\ \Eprint
  {http://arxiv.org/abs/0705.3221} {arXiv:0705.3221 [hep-ph]} \BibitemShut
  {NoStop}%
\bibitem [{\citenamefont {Antusch}\ \emph {et~al.}(2002)\citenamefont
  {Antusch}, \citenamefont {Kersten}, \citenamefont {Lindner},\ and\
  \citenamefont {Ratz}}]{Antusch:2002rr}%
  \BibitemOpen
  \bibfield  {author} {\bibinfo {author} {\bibfnamefont {Stefan}\ \bibnamefont
  {Antusch}}, \bibinfo {author} {\bibfnamefont {Jörn}\ \bibnamefont
  {Kersten}}, \bibinfo {author} {\bibfnamefont {Manfred}\ \bibnamefont
  {Lindner}}, \ and\ \bibinfo {author} {\bibfnamefont {Michael}\ \bibnamefont
  {Ratz}},\ }\bibfield  {title} {\enquote {\bibinfo {title} {{Neutrino mass
  matrix running for nondegenerate seesaw scales}},}\ }\href {\doibase
  10.1016/S0370-2693(02)01960-3} {\bibfield  {journal} {\bibinfo  {journal}
  {Phys. Lett.}\ }\textbf {\bibinfo {volume} {B538}},\ \bibinfo {pages}
  {87--95} (\bibinfo {year} {2002})},\ \Eprint
  {http://arxiv.org/abs/hep-ph/0203233} {arXiv:hep-ph/0203233 [hep-ph]}
  \BibitemShut {NoStop}%
\bibitem [{\citenamefont {Lin}\ \emph {et~al.}(2010)\citenamefont {Lin},
  \citenamefont {Merlo},\ and\ \citenamefont {Paris}}]{Lin:2009sq}%
  \BibitemOpen
  \bibfield  {author} {\bibinfo {author} {\bibfnamefont {Y.}~\bibnamefont
  {Lin}}, \bibinfo {author} {\bibfnamefont {L.}~\bibnamefont {Merlo}}, \ and\
  \bibinfo {author} {\bibfnamefont {A.}~\bibnamefont {Paris}},\ }\bibfield
  {title} {\enquote {\bibinfo {title} {{Running Effects on Lepton Mixing Angles
  in Flavour Models with Type I Seesaw}},}\ }\href {\doibase
  10.1016/j.nuclphysb.2010.04.007} {\bibfield  {journal} {\bibinfo  {journal}
  {Nucl. Phys.}\ }\textbf {\bibinfo {volume} {B835}},\ \bibinfo {pages}
  {238--261} (\bibinfo {year} {2010})},\ \Eprint
  {http://arxiv.org/abs/0911.3037} {arXiv:0911.3037 [hep-ph]} \BibitemShut
  {NoStop}%
\bibitem [{\citenamefont {Ibarra}\ \emph {et~al.}(2020)\citenamefont {Ibarra},
  \citenamefont {Strobl},\ and\ \citenamefont {Toma}}]{Ibarra:2020eia}%
  \BibitemOpen
  \bibfield  {author} {\bibinfo {author} {\bibfnamefont {Alejandro}\
  \bibnamefont {Ibarra}}, \bibinfo {author} {\bibfnamefont {Patrick}\
  \bibnamefont {Strobl}}, \ and\ \bibinfo {author} {\bibfnamefont {Takashi}\
  \bibnamefont {Toma}},\ }\bibfield  {title} {\enquote {\bibinfo {title}
  {{Two-loop renormalization group equations for right-handed neutrino masses
  and phenomenological implications}},}\ }\href {\doibase
  10.1103/PhysRevD.102.055011} {\bibfield  {journal} {\bibinfo  {journal}
  {Phys. Rev.}\ }\textbf {\bibinfo {volume} {D102}},\ \bibinfo {pages} {055011}
  (\bibinfo {year} {2020})},\ \Eprint {http://arxiv.org/abs/2006.13584}
  {arXiv:2006.13584 [hep-ph]} \BibitemShut {NoStop}%
\bibitem [{\citenamefont {Roy}\ and\ \citenamefont
  {Shaposhnikov}(2010)}]{Roy:2010xq}%
  \BibitemOpen
  \bibfield  {author} {\bibinfo {author} {\bibfnamefont {Ananda}\ \bibnamefont
  {Roy}}\ and\ \bibinfo {author} {\bibfnamefont {Mikhail}\ \bibnamefont
  {Shaposhnikov}},\ }\bibfield  {title} {\enquote {\bibinfo {title} {{Resonant
  production of the sterile neutrino dark matter and fine-tunings in the
  $\nu$MSM}},}\ }\href {\doibase 10.1103/PhysRevD.82.056014} {\bibfield
  {journal} {\bibinfo  {journal} {Phys. Rev.}\ }\textbf {\bibinfo {volume}
  {D82}},\ \bibinfo {pages} {056014} (\bibinfo {year} {2010})},\ \Eprint
  {http://arxiv.org/abs/1006.4008} {arXiv:1006.4008 [hep-ph]} \BibitemShut
  {NoStop}%
\end{thebibliography}%

\end{document}